\documentclass{aa}  

\usepackage{graphicx}
\usepackage{pdflscape}
\usepackage{txfonts}
\usepackage{natbib}
\usepackage{color}
\usepackage{amssymb}
\usepackage{mathrsfs}

\begin{document}

   \title{Sparse aperture masking interferometry survey of transitional discs  \thanks{Based on observations made with the Keck observatory 
(NASA program IDs N104N2 and N121N2).}
    }

   \subtitle{Search for substellar-mass companions and asymmetries in their parent discs}

   \author{M.\ Willson,
          \inst{1}
            S.\ Kraus,
          \inst{1}
          J.\ Kluska,
          \inst{1}
          J.\ D.\ Monnier,
          \inst{2}
          M.\ Ireland,
          \inst{3}
          A.\ Aarnio,
          \inst{2}
          M.\ L.\ Sitko,
          \inst{4,5,6}
          N.\ Calvet,
          \inst{2}
          C.\ Espaillat,
          \inst{7}
          D.\ J.\ Wilner
          \inst{8}
}
    \authorrunning{Willson et al.}

   \institute{
    $^{1}$~University of Exeter, Astrophysics Group, School of Physics, Stocker Road, Exeter EX4 4QL, UK\\
    $^{2}$~Department of Astronomy, University of Michigan, 311 West Hall, 1085 South University Ave, Ann Arbor, MI 48109, USA\\
    $^{3}$~Research School of Astronomy \& Astrophysics, Mount Stromlo Observatory Cotter Road, Weston Creek, ACT 2611, Australia\\
    $^{4}$~Department of Physics, University of Cincinnati, Cincinnati, OH 45221, USA\\
    $^{5}$~Space Science Institute, 475 Walnut St., Suite 205, Boulder, CO 80301, USA\\
    $^{6}$~Visiting Astronomer, NASA Infrared Telescope Facility\\ 
    $^{7}$~Department of Astronomy, Boston University, 725 Commonwealth Avenue, Boston, MA 02215, USA\\
    $^{8}$~Harvard-Smithsonian Center for Astrophysics, 60 Garden Street, MS-78, Cambridge, MA 02138, USA
             }

   \date{Received ; accepted }

 
  \abstract
   {
   Transitional discs are a class of circumstellar discs around young stars with extensive clearing of dusty material within their inner regions on 10s of au scales. One of the primary candidates for this kind of clearing is the formation of planet(s) within the disc that then accrete or clear their immediate area as they migrate through the disc.
   }
   {
   The goal of this survey was to search for asymmetries in the brightness distribution around a selection of transitional disc targets. We then aimed to determine whether these asymmetries trace dynamically-induced structures in the disc or the gap-opening planets themselves.
   }
   {
   Our sample included eight transitional discs. Using the Keck/NIRC2 instrument we utilised the Sparse Aperture Masking (SAM) interferometry technique to search for asymmetries indicative of ongoing planet formation. We searched for close-in companions using both model fitting and interferometric image reconstruction techniques. Using simulated data, we derived diagnostics that helped us to distinguish between point sources and extended asymmetric disc emission. In addition, we investigated the degeneracy between the contrast and separation that appear for marginally resolved companions.
   }
   {
   We found FP\,Tau to contain a previously unseen disc wall, and DM\,Tau, LkH$\alpha$330, and TW\,Hya to contain an asymmetric signal indicative of point source-like emission. 
   We placed upper limits on the contrast of a companion in RXJ1842.9-3532 and V2246\,Oph. We ruled the asymmetry signal in RXJ1615.3-3255 and V2062\,Oph to be false positives. 
    In the cases where our data indicated a potential companion we computed estimates for the value of $ M_{c}\dot M_{c}$ and found values in the range of $10^{-5}-10^{-3} M^2 _J yr^{-1}$.
   }
   {
   We found significant asymmetries in four targets. Of these, three were consistent with companions. We resolved a previously unseen gap in the disc of FP\,Tau extending inwards from approximately 10\,au.
   }

   \keywords{Techniques: interferometric --
                Planetary systems --
                Planets and satellites: formation --
                Planets and satellites: detection --
                Protoplanetary disks -- 
                Stars: pre-main sequence
               }

   \maketitle
%

\section{Introduction}
\label{sec:intro}

It is thought that planet formation is a direct result of aggregation and growth of dust particles within the protoplanetary discs that form around accreting protostars during the star formation process \citep{1996Icar..124...62P}. Within the spectral energy distributions (SEDs) of some more evolved discs there are dramatic drops in the near-infrared (NIR) to mid-infrared (MIR) flux from the disc compared to a classical T Tauri-type disc. This drop in flux is typically interpreted as being caused by the clearing of dust grains through mechanisms such as grain growth \citep{2008A&A...480..859B, 2011A&A...525A..11B}, photo-evaporation of dust grains by the stellar radiation field \citep{2011ARA&A..49..195A}, or interactions between a forming giant planet and the disc. These objects are classified as transitional discs (in case of discs with an inner dust-cleared cavity) or pre-transitional discs (in case of a gapped disc structure) and thought to be the sites of ongoing planet formation \citep{2014prpl.conf..497E}.

During the earliest stages of their lives, planetary cores are highly challenging to detect as they are deeply embedded within the dusty material of their parent discs. However, once they have gained sufficient mass to clear a gap (at the transitional or pre-transitional disc stage) they become accessible to high resolution imaging observations.  This phase likely coincides with the hydrodynamic collapse and oligarchic growth of proto-Jupiters and proto-brown dwarfs and is likely associated with the formation of an extended, hot circumplanetary disc that feeds material onto the accreting core \citep{1996Icar..124...62P,2012MNRAS.427.2597A}. Once protoplanetary cores have cleared most of their immediate disc environment, they can continue to accrete significant amounts of mass from material flowing through the gap \citep[$10^{-9}\,M_{\odot}yr^{-1}$,][]{2007MNRAS.378..369N, 2006ApJ...640.1110V}. Therefore, it is expected that protoplanets would appear as strong NIR sources within cleared gap regions.

Spatially resolving such systems proves a challenge as the close angular separation between the protoplanets and their parent stars and that the parent star is likely to be substantially brighter than even a rapidly accreting protoplanet. Spatially resolved evidence for protoplanetary companions could only be obtained for a small sample of objects so far: Coronagraphic imaging has revealed ring-like and spiral-like structures on scales of 50-200\,au \citep[e.g.\ Subaru/SEEDS survey;][]{2011ApJ...729L..17H,2013ApJ...762...48G} and sparse aperture masking interferometry (SAM) has resulted in the detection of small-scale asymmetries in the brightness distribution around young stars that have been interpreted as low-mass companions (T\,Cha: \citealt{2011A&A...528L...7H}; LkCa\,15: \citealt{2012ApJ...745....5K}; HD\,142527: \citealt{2012ApJ...753L..38B}) or as disc emission of a heated wall in a centro-symmetric disc seen under intermediate inclination (FL\,Cha: \citealt{2013ApJ...762L..12C}; T\,Cha: \citealt{2013A&A...552A...4O}). Only in the case of LkCa\,15 has this continuum detection been confirmed as an accreting companion, with subsequent observations using a combination of SAM and H$\alpha$ spectral differential imaging performed by \citet{2015Natur.527..342S} to demonstrate for the first time unambiguous evidence for the accreting nature of the companion.

Besides the emission associated with the protoplanets themselves and their associated circumplanetary discs, asymmetries can also be caused by dynamically-induced disc features such as spiral arms, disc warps \citep{2010A&A...519A..88A, 2009ApJ...704L..15M}, or disc physics-related processes such as gravitational instabilities, and density waves \citep{2007A&A...463.1017B}. Additionally a highly inclined disc can induce strong asymmetries in an axial-symmetric disc owing to forward-scattered light from the illuminated inner rim of disc \citep{2013A&A...552A...4O,2015MNRAS.450L...1C}.
Most of the evidence for these processes comes from photometric or spectroscopic monitoring investigations. For instance, it was found that the variability shows an anti-correlated behaviour at NIR and MIR wavelengths. In order to explain both the timescale and spectral behaviour of the variability, \citet{2011ApJ...728...49E} proposed shadowing effects from co-rotating disc warps at the inner dust rim triggered by orbiting planets. Such warps are also predicted by hydrodynamic simulations of discs with embedded planets \citep[e.g.][]{2010A&A...518A..16F} and would result in a highly asymmetric brightness distribution. The warp emission will be extended and more complex in geometry than a companion point source.

Here we report on a survey of eight transitional disc targets observed using the Keck-II/NIRC2 instrument over the course of three years of observations. We  discuss the observations in detail within Chapter~\ref{sec:observations}; in Chapters~\ref{sec:modelling} and ~\ref{sec:imgrec} the methodology of fitting for a companion in the closure phase data and the procedure for producing reconstructed images is explained; we show a series of simulations in Chapter~\ref{sec:simulations} representing different scenarios and use these to develop criteria for classifying non-detections, detections and potential disc features; in Chapter~\ref{sec:results} we present the results and classifications for each of the data sets and finally in Chapter~\ref{sec:conclusions} we outline our conclusions.

%

\section{Observations}
\label{sec:observations}

\begin{table*}
\caption{Target list} 
\label{tab:targlist}
\centering
\vspace{0.1cm}
\begin{tabular}{c c c c c c c c c c c c}
\hline\hline
Name & Association & Distance & SpecType & A$_v$ & T$_{eff}$ & M$_*$ & L$_*$ & R$_*$ & References \\
	&	& [pc]	 & 	 &  	& [K]	& [M$_{\odot}$] & [L$_{\odot} $]	& [R$_{\odot}$] &	 \\
\hline
\object{DM\,Tau} & Taurus & 140 & M1 & 0.0 & 3705 & 0.47 & 0.37 & 1.3 & 6,8 \\
\object{FP\,Tau} & Taurus & 140 & M5 & 0.3 & 3125 & 0.22 & 0.62 & 2.5 & 6,8 \\
\object{LkH$\alpha$\,330} & Perseus & 250 & G3 & 1.8 & 5830 & 1.25 & 2.78 & 1.5 & 2,5 \\
\object{RXJ1615.3-3255} & Oph & 185 & K4 & 1.00 & 4590 & 1.28 & 1.01 & 1.59 & 11,13 \\
\object{RXJ1842.9-3532} & CrA & 136 & K2 & 1.1 & 4900 & 1.33 & 1.29 & 1.5 & 14,15 \\
\object{TW\,Hya} & TWHya & 56 & M0 & 1.0 & 3850 & 0.57 & 0.64 & 1.7 & 3,12 \\
\object{V2062\,Oph} & Oph & 125 & K3 & 2.3 & 4730 & 1.4 & 1.3 & 1.7 & 1 \\
\object{V2246\,Oph} & Oph & 121.9 & K0 & 6.2 & 5016 & 2.2 & 20.5 & --- & 10,1,4,7 \\
\end{tabular}
\tablefoot{Column 1: Target; Column 2: Association; Column 3: Distance; Column 4: Spectral Type; Column 5: Visual Extinction; Column 6: Effective Temperature; Column 7: Stellar Mass; Column 8: Stellar Luminosity; Column 9: Stellar Radius; Column 10: Literature References: (1) \citet{1992A&AS...92..481B}, (2) \citet{2009ApJ...704..496B}, (3) \citet{2004AJ....128.1294C}, (4) \citet{1995ApJ...445..377C}, (5) \citet{1995A&AS..114..439F}, (6) \citet{2006ApJS..165..568F}, (7) \citet{2009ApJ...703..252J}, (8) \citet{1998AJ....115.2491K}, (9) \citet{2013ApJ...768...80K}, (10) \citet{2008ApJ...675L..29L}, (11) \citet{2010ApJ...718.1200M}, (12) \citet{1996A&AS..120..229R}, (13) \citet{2011ApJ...732...42A}, (14) \citet{2006ApJ...639.1138S}, (15) \citet{2007AJ....133.2524W}}
\end{table*}

\begin{table*}
\centering
\caption{Log for our Keck/NIRC2 observations}
\label{tab:obslog}
\begin{tabular}{c c c c c c c}
\hline\hline
Target & Filter & Date & N$_{visits}$ & On Sky Rotation & Calibrator \\
	&	& [dd/mm/yy] &  & [$^{\circ}$] & \\
\hline

DM\,Tau &  K' & 08/01/12 & 3 & 1 & HD\,285938 \\
FP\,Tau &  K' & 20/10/13 & 3 & 24 & HD\,283420, HD\,283477 \\
	    &  L' & 20/10/13 & 2 & 63 & HD\,283420, HD\,283477 \\
LkH$\alpha$\,330 &  CH4s & 16/11/13 & 3 & 32 & HD\,22781, HD\,281309\\
	    &  K' & 08/01/12 & 3 & 18 & HD\,22781, HD\,281309 \\
	    &     &          &  &  & 2M034+3304, 2M0400+3311 \\
RXJ1615.3-3255 &  K' & 09/06/14 & 4 & 15 & HD\,146569, HD\,146593, \\
	    &  	  & 	     &  &  & HD\,148806 \\
RXJ1842.9-3532 &  K' & 09/06/14 & 5 & 13 & HD\,171450, HD\,171574, \\
	    &  	  & 	     &  &  & HD\,176999 \\
TW\,Hya &  K' & 08/01/12 & 3 & 8 & HD\,94542, HD\,95105 \\
	    &  K' & 10/01/12 & 3 & 11 & HD\,94542, HD\,95105,  \\
	    &  	  & 	     &  &  & HD\,97507 \\
V2062\,Oph &  K'  & 09/06/14 & 5 & 16 & HD\,147681, HD\,148212, \\
	    &  	  & 	     &  &  & HD\,148562 \\
V2246\,Oph &  K'  & 09/06/14 & 1 & 2 & HD\,147742, HD\,148352 \\
\end{tabular}
\end{table*}

Our high-angular resolution observations were conducted using the NIRC2 instrument at the 10m Keck-II telescope located on the summit of Mauna Kea on Hawaii. We employed the sparse aperture masking technique, which allows us to remove atmospheric phase noise through the use of the closure phase technique. We employed the nine hole mask on NIRC2, which offers a good compromise between sensitivity and uv-coverage. The chosen wavebands were H, K' and L' band filters as we expect an accreting companion to emit strongly in these bands. A list of our target stars can be found in Table \ref{tab:targlist}.

Our data set was obtained during five nights between January 2012 and June 2014 (Table \ref{tab:obslog}). 
We observed most targets in a single epoch with the K' filter (2.124$\pm$0.351$\mu$m) to search for direct emission from any close protoplanetary candidates. We observed FP\,Tau and LkH$\alpha$\,330 again during the same epoch but in additional wavebands, L' (3.776$\pm$0.700$\mu$m) and H (CH4s; 1.633$\pm$0.33$\mu$m) respectively. We obtain an additional observation in the same epoch in the case of TW\,Hya.  

The NIRC2 data were reduced using the pipeline described previously in \citet{2008ApJ...678L..59I} 
and \citet{2013ApJ...768...80K}, providing calibrated closure phases. In order to record the instrument transfer function, we bracketed the science star observations with observations of two unresolved calibrators. We aimed to alternate between two (ideally three) different calibrator stars, which allows us to still calibrate our data even if a calibrator is found to be a previously unknown binary. A calibrator with spatially resolved structure will induce erroneous phase signals in our data, masking any companion signal or inducing a false signal. We test for multiplicity in our calibrators by calibrating them against each other. In the cases where we have three calibrators we can identify which calibrators are binaries by fitting a binary model. In this way we find a 5$\sigma$ binary signal in HD\,95105 during the observations of TW\,Hya on 2012-01-10. We additionally observed this calibrator on in the same epoch on 2012-01-08 and observed a binary signal from the same region of the sky, although substantially weaker during the first night at 3.5$\sigma$. The position angle for both nights was found to be 219$\pm$2$^\circ$ while the separations were found to be 113$\pm$4\,mas and 91$\pm$3\,mas and the contrasts vary between 5.4$\pm$0.3\,mag and 3.0$\pm$0.2\,mag over the first and second nights respectively. Therefore we remove this calibrator from the data reduction as a precaution.

As a diagnostic for identifying data sets adversely affected by degenerating factors such as short 
coherence times and vibration in the optical system, we plotted histograms of the raw closure 
phase measurements of the individual interferograms. We fit gaussians to the closure phase
distribution and derive the variance $\omega$.
For the calibrator stars, $\omega$ provides a measure for the residual phase noise
as these point sources should not exhibit an intrinsic non-zero phase signal. We list the measured variances in Table \ref{tab:cpFWHM}. The high variability seen in the L-band observations is related to the misalignment of the IR dichroic which has since been corrected.

Our observations represent "snapshots" of the targets with often little field rotation. This means that we are likely to suffer from hole aliasing problems creating strong artefacts. For this reason we only consider the most significant fit in the follow sections and do not discuss apparent additional asymmetric signals without complementary observations.

\begin{table*}
\centering
\caption{Phase noise variance $\omega$ for uncalibrated closures phases.}
\label{tab:cpFWHM}
\begin{tabular}{c c c c c}
\hline\hline
Science Target & Filter & Date & Calibrator & $\omega$\\
	 & 	 & [dd/mm/yy] &  & [$^\circ$]\\
\hline
DM\,Tau & K' & 08/01/12 & HD285938 & 4.51\\
\hline
FP\,Tau & K' & 20/10/13 & HD283420 & 5.85\\
 & K' & 20/10/13 & HD283477 & 5.86 \\
\hline
 & L' & 20/10/13 & HD283420 & 11.20\\
 & L' & 20/10/13 & HD283477 & 12.35\\
\hline
LkH$\alpha$\,330 &  H & 16/11/13 & 2M0340+3304 & 4.29\\
 &  H & 16/11/13 & 2M0400+3311 & 4.00\\
 &  H & 16/11/13 & HD22781 & 4.18\\
 &  H & 16/11/13 & HD23849 & 6.47\\
\hline
 & K' & 08/01/12 & HD22781 & 4.0\\
 & K' & 08/01/12 & HD281309 & 4.7\\
\hline
RXJ1615.3-3255 & K' & 09/06/14 & HD146369 & 10.92\\
 & K' & 09/06/14 & HD146593 & 10.78\\
 & K' & 09/06/14 & HD148806 & 9.01\\
\hline
RXJ1842.9-3532 & K' & 09/06/14 & HD171450 & 10.8\\	 
 & K' & 09/06/14 & HD171574 & 11.2\\
 & K' & 09/06/14 & HD176999 & 12.3\\
\hline
TW\,Hya & K' & 08/01/12 & HD94542 & 5.8\\
 & K' & 08/01/12 & HD95105 & 5.4\\
\hline
& K' & 10/01/12 & HD94542 & 9.3\\
& K' & 10/01/12 & HD95105 & 8.6\\
& K' & 10/01/12 & HD97507 & 9.3\\
\hline
V2062\,Oph & K' & 09/06/14 & HD147681 & 11.80\\
 & K' & 09/06/14 & HD148212 & 11.38\\
 & K' & 09/06/14 & HD148562 & 11.17\\
\hline
V2246\,Oph & K' & 09/06/14 & HD147742 & 18.4\\
 & K' & 09/06/14 & HD148352 & 15.4\\
\end{tabular}
\end{table*}
%

\section{Model fitting}
\label{sec:modelling}

\subsection{Binary model approach}

We fit a star+companion model \citep{2012ApJ...745....5K} to the measured closure phases. The free parameters are separation ($\rho$), positional angle (PA) and contrast ($f$). The results from this fitting are listed in Table \ref{tab:binresults}. 

\begin{figure}
\centering
$\begin{array}{@{\hspace{-6.0mm}} c @{\hspace{-4.8mm}} c}
\includegraphics[height=4cm, angle=0]{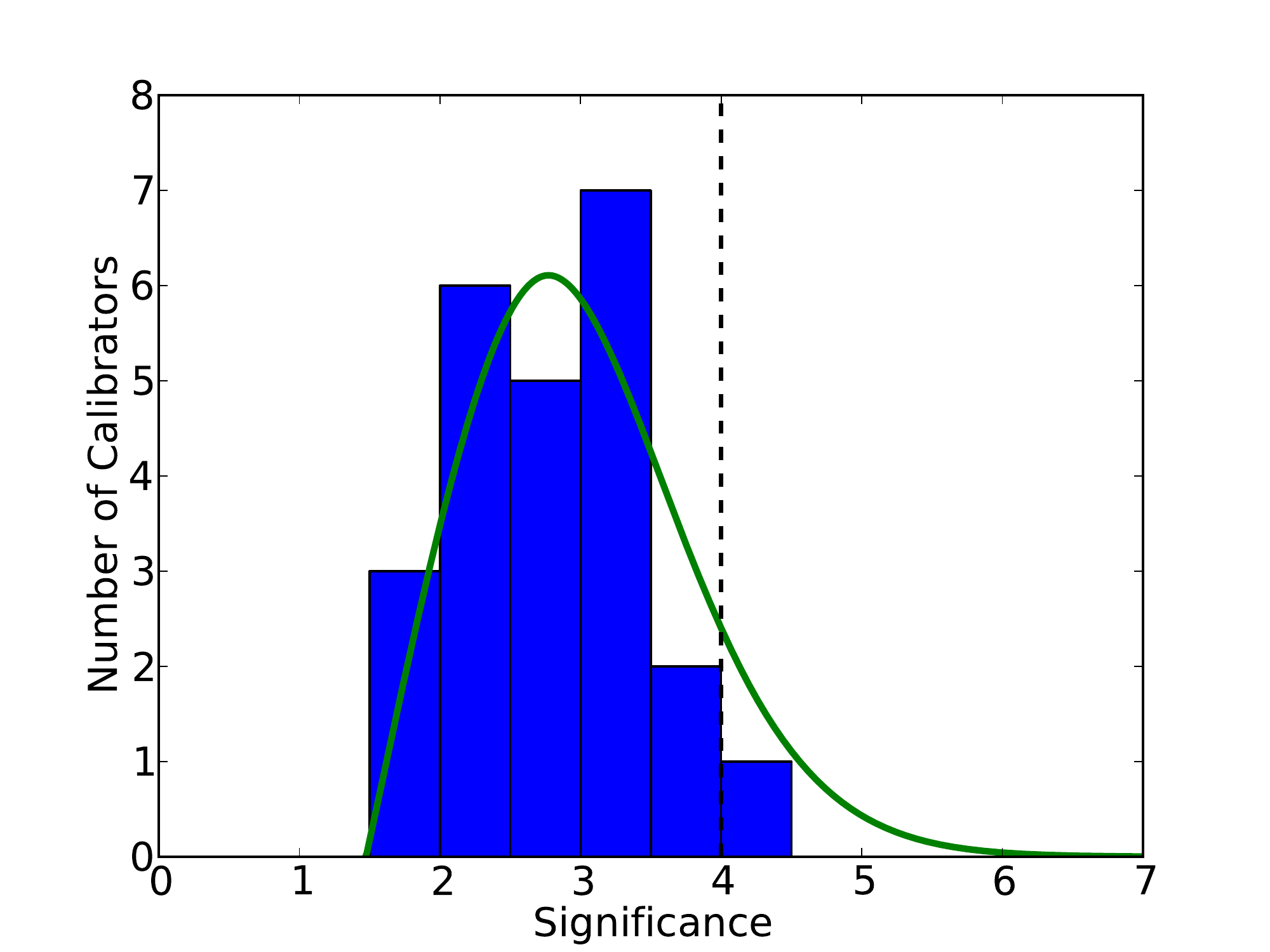} & \includegraphics[height=4cm, angle=0]{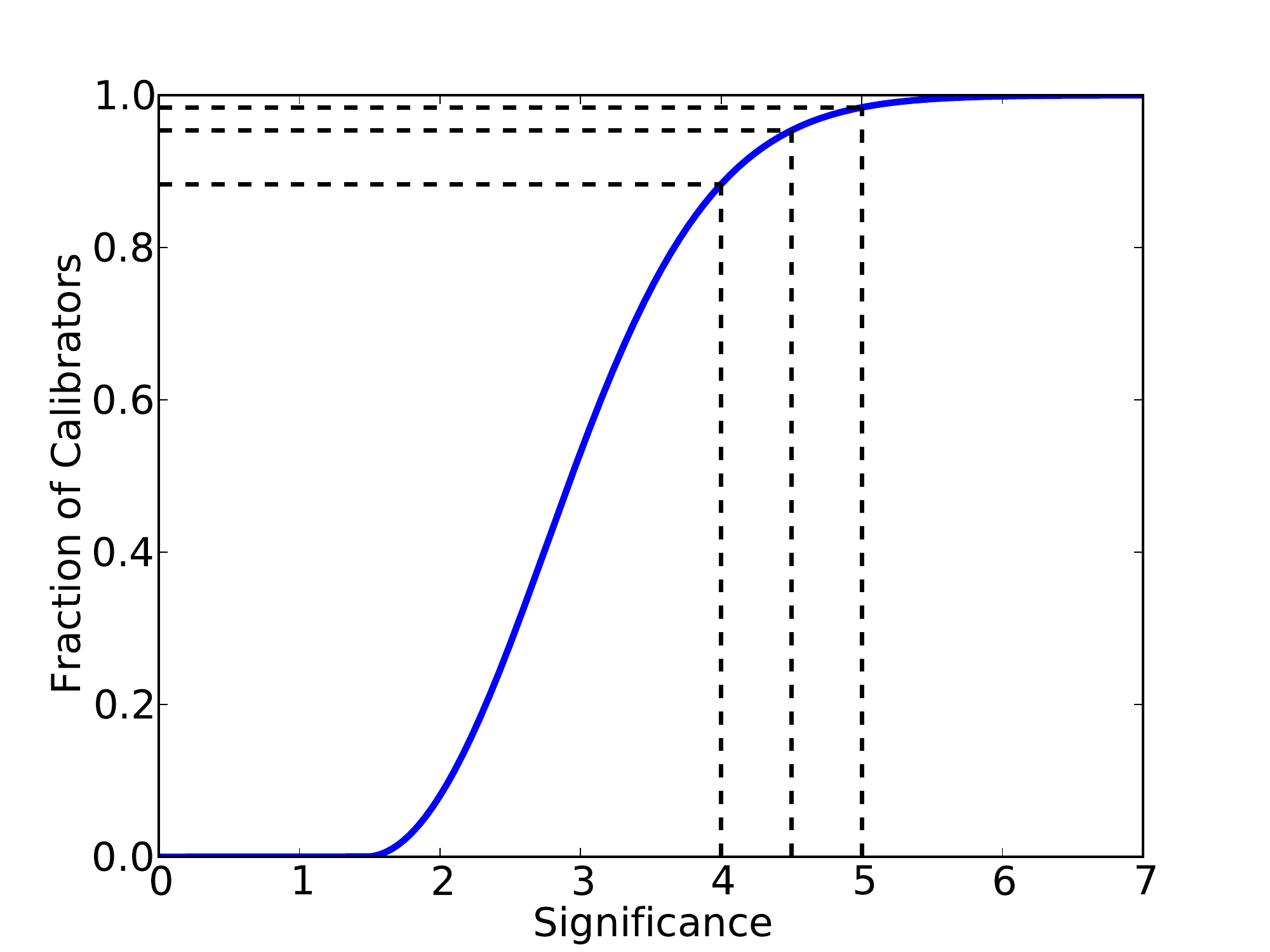}\\
 \end{array}$
\caption{
\textbf{Left: }Distribution of significances for binary model fits to our calibrators. \textbf{Right: }Cumulative distribution of the fitted poisson distribution. We expect no significant asymmetric signal within these stars so use them to form a crude test statistic for our threshold of 4$\sigma$. Fitting a poisson distribution we find a confidence level of 88\% for a 4$\sigma$ and more than 95\% for 4.5$\sigma$.
}
\label{CalibStats}
\end{figure}

We calibrate our detection threshold by investigating the best fit significance distribution in our sample of 24 calibrator star data sets. We fit binary models to the calibrated closure phases of the calibrator stars, using the other calibrators observed in the same area of the sky and close in time to build our sample. This leads to a correlation between the values obtained adversely affecting the shape of the distribution and the accuracy of our detection thresholds. Only one data set showed an asymmetric signal with a significance above 4.0$\sigma$, and none above 4.5$\sigma$, setting a simplistic confidence level of greater than 95\% for a significance of 4.0$\sigma$ and 99\% for significances above 4.5$\sigma$. Fitting a poisson distribution to the data set, we find more sophisticated confidence levels of 88.3\% for 4.0$\sigma$ and 95.4\% for 4.5$\sigma$ in agreement with our cruder approximation. Values $\geq$ $5.0\sigma$ represent confidence levels of greater than 98\%. This method is inaccurate as the low number of targets within each bin and the cross calibration of pairs or triples of calibrators results in values for the significance which are only semi-independent. We see no strong correlations between targets of similar R magnitudes or within the same night. This provides an intuitive interpretation for the meaning of the significance values we calculate (See Figure \ref{CalibStats}) and quantifies our ability to differentiate from false positives.

Additionally we consider the sample of 54 M-dwarfs observed with SAM as part of an investigation into M dwarf multiplicity by \citet{2016MNRAS.457.2877G} using Keck-II/NIRC2. They found approximately 25\,\% of science targets displayed asymmetric signals within their closure phases in the 4-5$\sigma$ range. Furthermore, 5\,\% were found to possess signals between 5-6$\sigma$. This places lower confidence levels on our 4$\sigma$ threshold but this sample of targets is likely to be strongly affected by systematics caused by the observation strategy of a single visit and the use of the Laser Guide Star. This will result in little on-sky rotation and inferior calibration so is not as applicable to our data set, except in a cases where we too have little on-sky rotation (i.e. DM\,Tau). Measured on-sky rotations are displayed in Table \ref{tab:obslog}. 

The fitting procedure outlined above is suitable if the brightness distribution resembles a binary, but it may be inadequate for more complex distributions such as triple systems, or complex disc features. To explore this we create maps of the significances ($\sigma$) produced in a binary fit. We use the following equation to produce complex visibilities for theoretical binary models from which we construct model closure phases to fit to our measured closure phases 

\begin{equation}
    \centering
    \label{eq:BinaryComplexVisibilities}
    V(u, v) = \frac{1 + f \exp(2 \pi i (u \alpha + v \beta))}{1 + f},
\end{equation}
Here, $f$ denotes the flux ratio of the model companion and the parent star, $u$ and $v$ are the Fourier plane coordinates and $\alpha$ and $\beta$ are the angular coordinates of the companion within the model. 
We then construct a grid of positions in RA and DEC with a resolution of 1\,mas, covering an area of 
400$\times$400\,mas with the parent star located in the centre of the field. At each position we fit 
for the best contrast and convert the calculated $\chi^2$s into a significance to form a map which enables us to make qualitative judgements about whether the detection resembles likely a companion or a more complex brightness distribution. The significance is estimated using:
\begin{equation}
    \centering
    \label{eq:significance}
    \sigma = \sqrt{\chi^2_{null} - \chi^2},
\end{equation}
where the $\chi^2_{null}$ is calculated using Eq.\,\ref{eq:BinaryComplexVisibilities} taking an unresolved single point source. 
We enforce within our fits positive flux and flux ratios less than 1.0, physically 
representing that the companion cannot be brighter than the parent star.

This modelling approach allows us to search for point source-like asymmetries consistent with a gap-clearing companion. However we are unable to distinguish companions from other potential sources of asymmetry which could mimic a point source in our data sets such as disc over-densities, accretion streams and other complex structure. For this reason we only consider significant detections to be companion candidates in need of further observation rather than confirmed companions. To establish their nature as protoplanets or substellar companions, evidence for orbital motion and ongoing accretion is required.

\subsection{Degeneracies}
\label{sec:degen}

\begin{figure}
\centering
$\begin{array}{@{\hspace{-4.2mm}} c @{\hspace{-1.0mm}} c}
     \includegraphics[height=4cm]{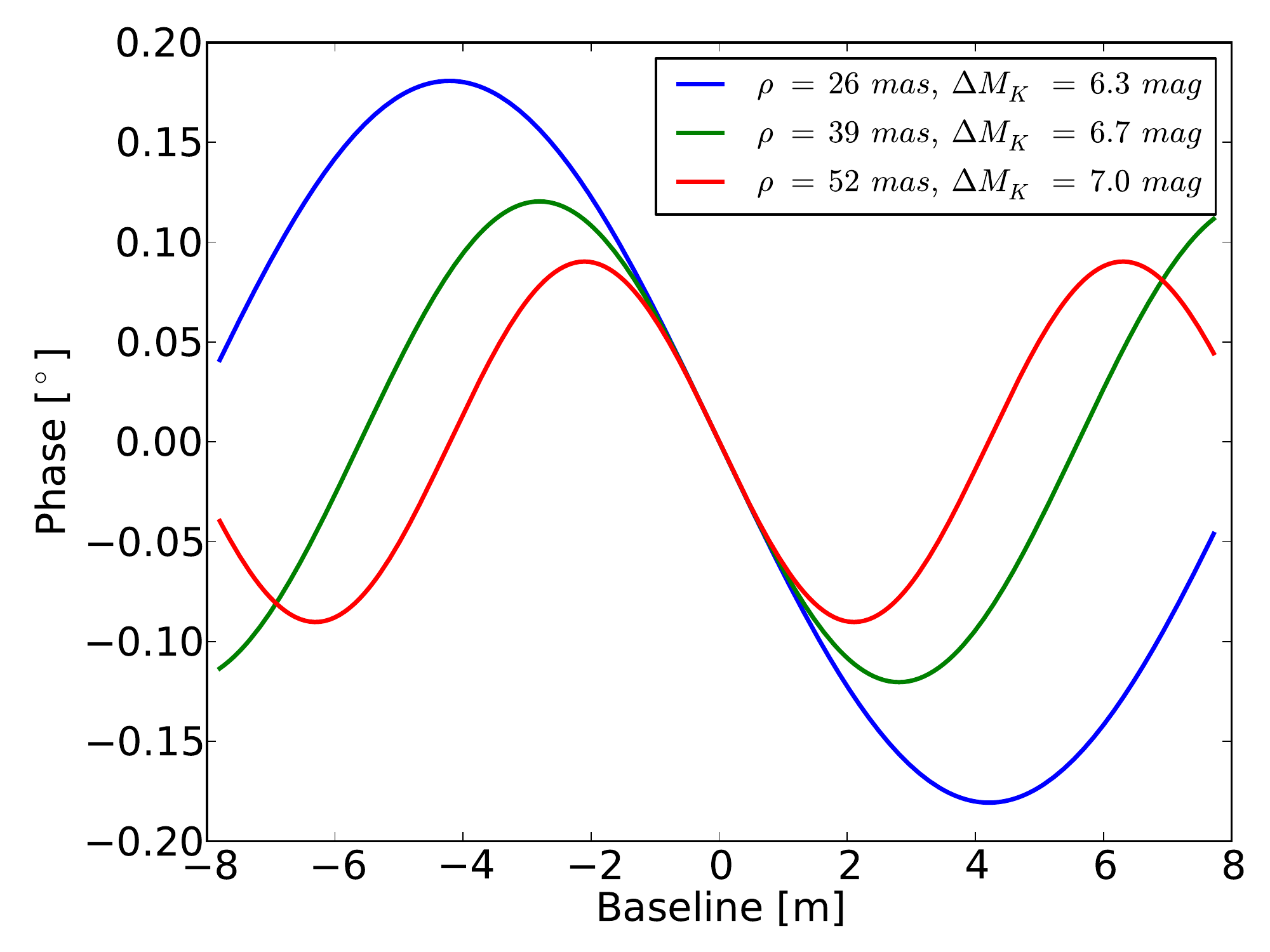} & 
     \includegraphics[height=4cm, angle=0, trim={0.5cm 0 0 0},clip]{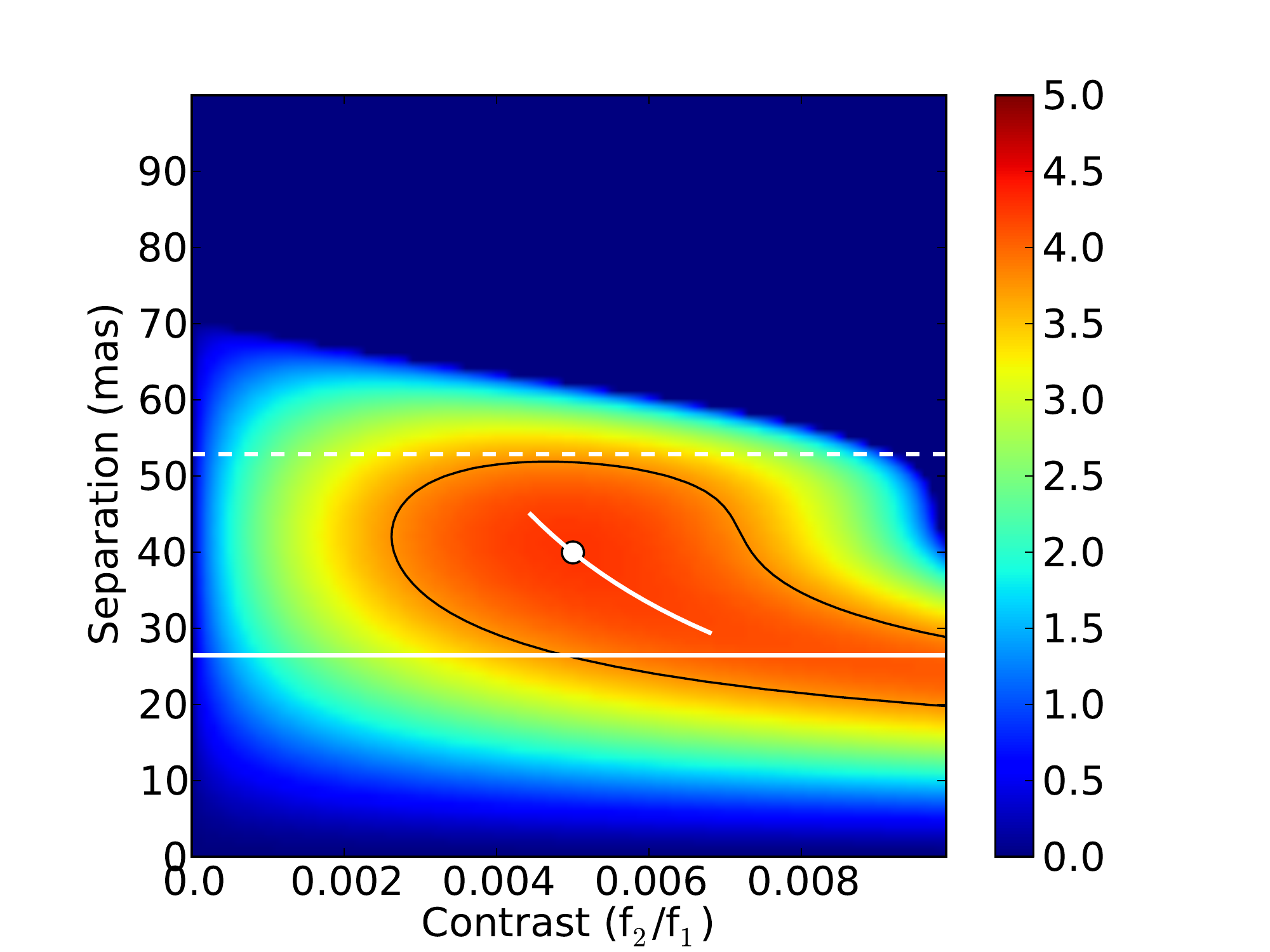} \\ 
 \end{array}$
\caption{Degeneracy plots of the detection in DM\,Tau. \textbf{Left: } Phases for three companions at three separations and their contrasts according to Eq.\,\ref{eq:BinaryContrastSmall}. \textbf{Right: } Fit of the degeneracy profile using the shortest projected baseline length. The fit using the longer baseline well describes the profile at larger separations but poorly describes the shorter separations as the contrast ratio asymptotically goes to 1.0 as the separation approaches $\lambda/2D$. The shortest baseline poorly follows the structure at larger separations but does follow closely the profile at closer separations as a result of its ability to probe the more SNR sensitive region close to the $\lambda/2D$ resolution limit.}
\label{FigMaxMin}
\end{figure}

Detections of companions with separations $\rho \lesssim \lambda/2D$ are problematic to fit because of a degeneracy that appears between the separation and contrast. In this separation regime, the phases do not sample the full sinusoidal modulation that is required to constrain the companion contrast and separation separately. This makes our fits highly sensitive to the signal-to-noise ratio (SNR) of the closure phases, resulting in a range of separations and contrasts that can reproduce the measured non-zero closure phases equally well (see Figure \ref{FigMaxMin}). We therefore find that a similarly good fit can be obtained for different separation/contrast pairings. This is most clearly seen within the significance maps themselves, producing lobe-like structures in the region between $\lambda/D$ and $\lambda/2D$. 

To explore this degeneracy and allow one to translate from one separation/contrast pair to another we take two approaches. The first approach is to plot the degeneracy directly. We plot a grid of contrasts against separations along the non-degenerate best-fit position angle and construct a significance map in the same manner as outlined above (see Figure \ref{FigMaxMin}).
In our second approach, we aim to derive an analytic expression for the separation/contrast degeneracy. For this purpose, we start from Eq.\,\ref{eq:BinaryComplexVisibilities} and retrieve the phase component, $\phi$,
\begin{equation}
\centering
\label{eq:BinaryPhase}
\tan \phi = \frac{f \sin (-2\pi b\rho)}{1 + f \cos (-2\pi b\rho)},
\end{equation}
where 
$\rho$ is the scalar companion separation for our best fit position and $b$ is the projected length of the baseline along the vector separation. Rearranging we find:
\begin{equation}
\centering
\label{eq:BinaryContrast}
f = \frac{\sin \phi}{\sin (-\phi -2\pi b\rho)},
\end{equation}
For small values of $\phi$ (i.e. values of $\phi < \pi /4$) this second equation can be further simplified using the small angle approximation:
\begin{equation}
\centering
\label{eq:BinaryContrastSmall}
f \approx -\frac{\phi}{\phi + 2\pi b\rho},
\end{equation}

To most accurately trace the profile of the degeneracy, we would need to use every $u$ projected baseline and weight according to their associated uncertainties. However, using simply the shortest projected baseline was found to be effective for tracing the degeneracy to smaller separations. Within our degeneracy plots, the physical degenerate region is shown by the black contour defining the $\Delta\sigma = 0.5$ region, while the white curve displays our analytical solution for the shortest projected baseline (see Figure \ref{FigMaxMin}, right).

In our data set, we find three cases for which the best-fit separation is within the degenerate region. In the case of DM\,Tau, the uncertainties in our closure phases are small enough to allow us to find a minimum. In this case the degeneracy results in enhanced uncertainties in the separation and contrast measurements. In the remaining cases (FP\,Tau, K-band; TW\,Hya, K-band) the SNR for the closure phases prevented our fitting algorithm from finding a minimum $\chi^2$. In these cases we set the separation to the resolution limit of our observations; $\lambda/2B$, where $\lambda$ is the wavelength of the observations and $B$ is the longest baseline from our mask. However the separations are not well constrained and solutions at larger separations and higher contrasts would result in good fits of similar significances. Our analytical solution enables us to calculate the contrast at a different separation from our fit here.

\begin{table*}
\centering
\caption{Binary fit results}
\label{tab:binresults}
\begin{tabular}{c c c c c c}
\hline\hline
Target & Filter & Date & Contrast & Significance & Comment\\
	 & 	 & [dd/mm/yy] & [mag] &  & \\
\hline

DM\,Tau & K' & 08/01/12 & 6.8$\pm$0.3 & 4.27 & Detection \\
FP\,Tau & K' & 20/10/13 & 4.4$\pm$1.2 & 5.99 & Disc Detection \\
	& L' & 20/10/13 & 4.1$\pm$0.3 & 3.55 & Non-Detection \\
LkH$\alpha$\,330 &  CH4s & 16/11/13 & 5.7$\pm$0.3 & 3.39 & Non-Detection \\
	& K' & 08/01/12 & 5.6$\pm$0.2 & 4.73 & Detection \\
RXJ1615.3-3255 & K' & 09/06/14 & 4.9$\pm$0.2 & 4.77 & False Positive\\
RXJ1842.9-3532 & K' & 09/06/14 & 5.5$\pm$0.2 & 3.72 & False Positive \\
TW\,Hya & K' & 08/01/12 & 5.6$\pm$0.2 & 4.46 & Detection \\
	& K' & 10/01/12 & 3.7$\pm$1.4 & 3.63 & Non-Detection \\
V2062\,Oph & K' & 09/06/14 & 5.00$\pm$0.2 & 5.52 & False Positive \\
V2246\,Oph & K' & 09/06/14 & 4.3$\pm$0.5 & 1.97 & Non-Detection \\
\end{tabular}
\tablefoot{
\tablefoottext{}{Criterion for detection or non-detection is based on a $\sigma$ of greater than 4.0 representing a confidence level greater than 88\%. However there are effects which can mimic a detection of between 4-5$\sigma$ so for these cases which look individually at each target and try attempt to rule them out through inspection of their reconstructed images, significance maps, uv-coverage and visibilities. Detections close to 6.0 $\sigma$ ($>$\,99.9\% confidence level) are considered strong enough that closer examination is not required. Classification as a disc detection is based upon identification of a strong double lobing. An asterisk denotes cases where we see a strong degeneracy.
}}
\end{table*}

%

\section{Image Reconstruction}
\label{sec:imgrec}

In order to retrieve the brightness distribution of the observed objects in a model-independent way, we use image reconstruction techniques developed for infrared long-baseline interferometry on our measured calibrated closure phases and visibilities.

The image estimation from the discrete points in the Fourier plane (the aperture masking measurements) can be considered as an inverse problem. Given that there are more pixels than measurements, the problem is ill-posed and solving it requires one to adopt a Bayesian approach. This amounts to minimising a global cost function ($\mathscr{F}$) defined as:
\begin{eqnarray}
\label{eqn:cost}
\mathscr{F} = \mathscr{F}_\mathrm{data} + \mu \mathscr{F}_\mathrm{rgl},
\end{eqnarray}
where $\mathscr{F}_\mathrm{data}$ is the likelihood term (here the $\chi^2$), $\mathscr{F}_\mathrm{rgl}$ is the regularisation term and $\mu$ the regularisation weight \citep[see][for more background information]{Thiebaut2008,Renard2011}. The likelihood term ensures that the image is reproducing the data whereas the regularisation term helps to fill the gaps in the Fourier space by interpolating it in a specific way defined by the user. This term helps also to converge to the most likely a-posteriori estimate of the image.

To perform our image reconstructions we have chosen the MiRA algorithm \citep{Thiebaut2008}. This algorithm is 
minimising the cost function ($\mathscr{F}$) with a downhill gradient method. In our objects the central star is spatially unresolved. In order to image its environment we have therefore modelled it as a point source 
and reconstruct an image of the environment only, using the approach outlined in \citet{Kluska2014}.

The images are defined to have 128$\times$128 pixels each. For the pixel size, we chose 5, 7 and 11\,mas for $H$, $K$ and $L$-bands respectively. We have chosen to use the quadratic smoothness regularisation \citep{Renard2011}.
We employed the L-curve method \citep[see][for more details]{Renard2011,Kluska2014} to determine the weight of the regularisation for all data sets and then used the average weight of all the L-curves which is $\mu=10^9$.

To define the fraction of the stellar flux in the parametric model, we made a grid of reconstructions with different flux ratios for the star. Because we are minimising the global cost function $\mathscr{F}$ we should have chosen the images having the minimum $\mathscr{F}$ value. Because of the regularisation effects, these images still have flux at the star position which is not physical. Therefore we decided to keep the flux ratio for which the image has 
the smaller likelihood term ($\mathscr{F}_\mathrm{data}$). These images do not differ significantly from the images with smaller $\mathscr{F}$ except in correcting this effect.

%

\section{Simulations - reference models}
\label{sec:simulations}

Within many of our reconstructed images and significance maps we see patterns or structures which are not consistent with simple point source companions. To aid our understanding of these structures we simulated a range of possible scenarios. We simulate companions with different separations, position angles and contrasts in order to understand potential effects that might be caused by the imperfect uv-coverage and to investigate how the structure of the significance maps changes within the fully resolved and partially resolved regimes described in Section \ref{sec:degen}. While we expect these scenarios to cover most structures likely to be seen, this is an incomplete set and other scenarios may occur.

For our simulations we model data sets that correspond to the K-band and the NIRC2 9-hole mask. We add phase noise with a variance of $\omega\,=\,4^\circ$, which resembles good conditions in our observations.

\subsection{Small-Separation/Unresolved Companion Scenario}
\label{sec:unresolved_companion_simulations}

\begin{figure}
\centering
$\begin{array}{@{\hspace{-2.0mm}} c @{\hspace{-7.0mm}} c}
     \includegraphics[height=4cm]{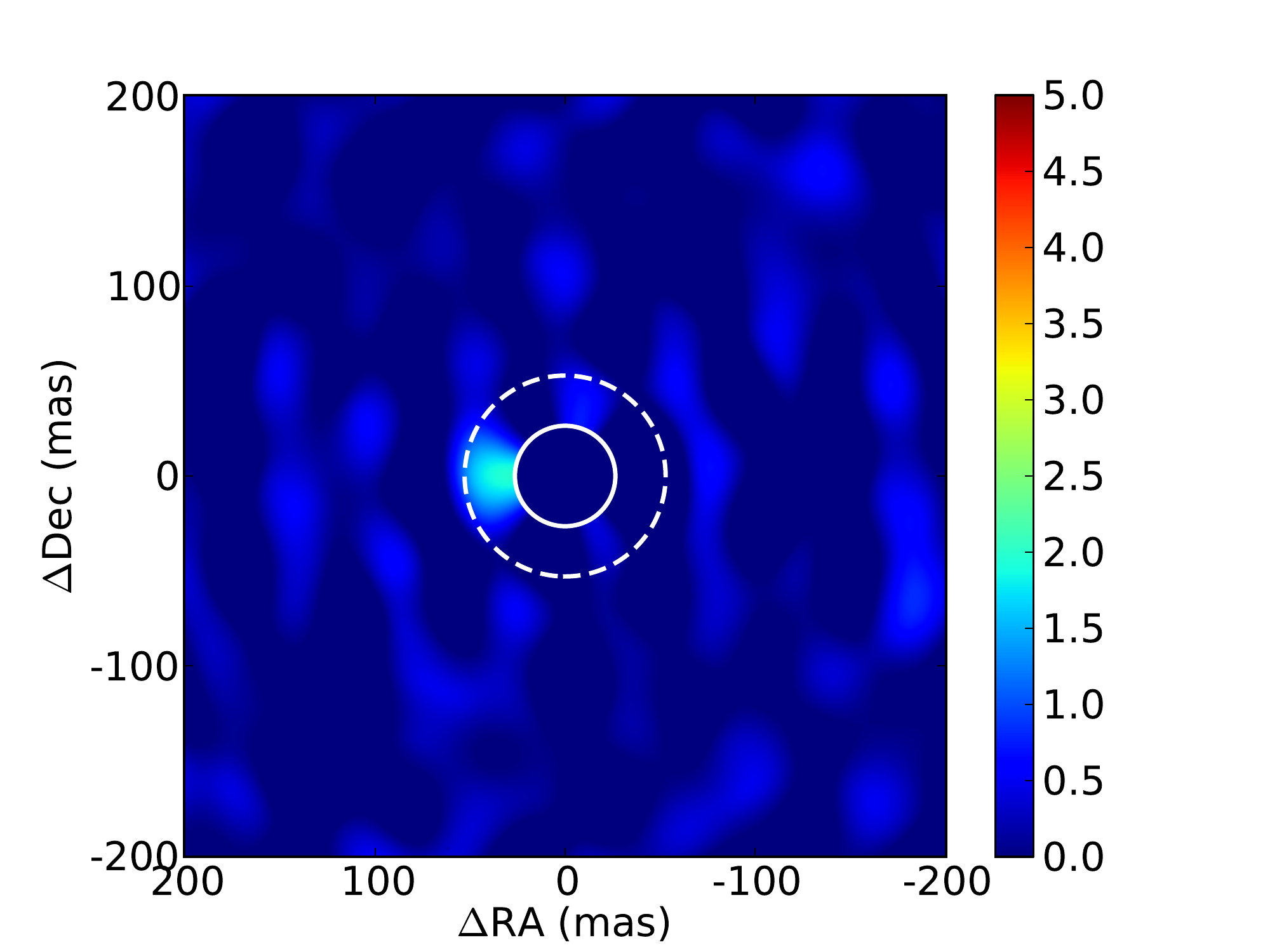}& \includegraphics[height=4cm, angle=0]{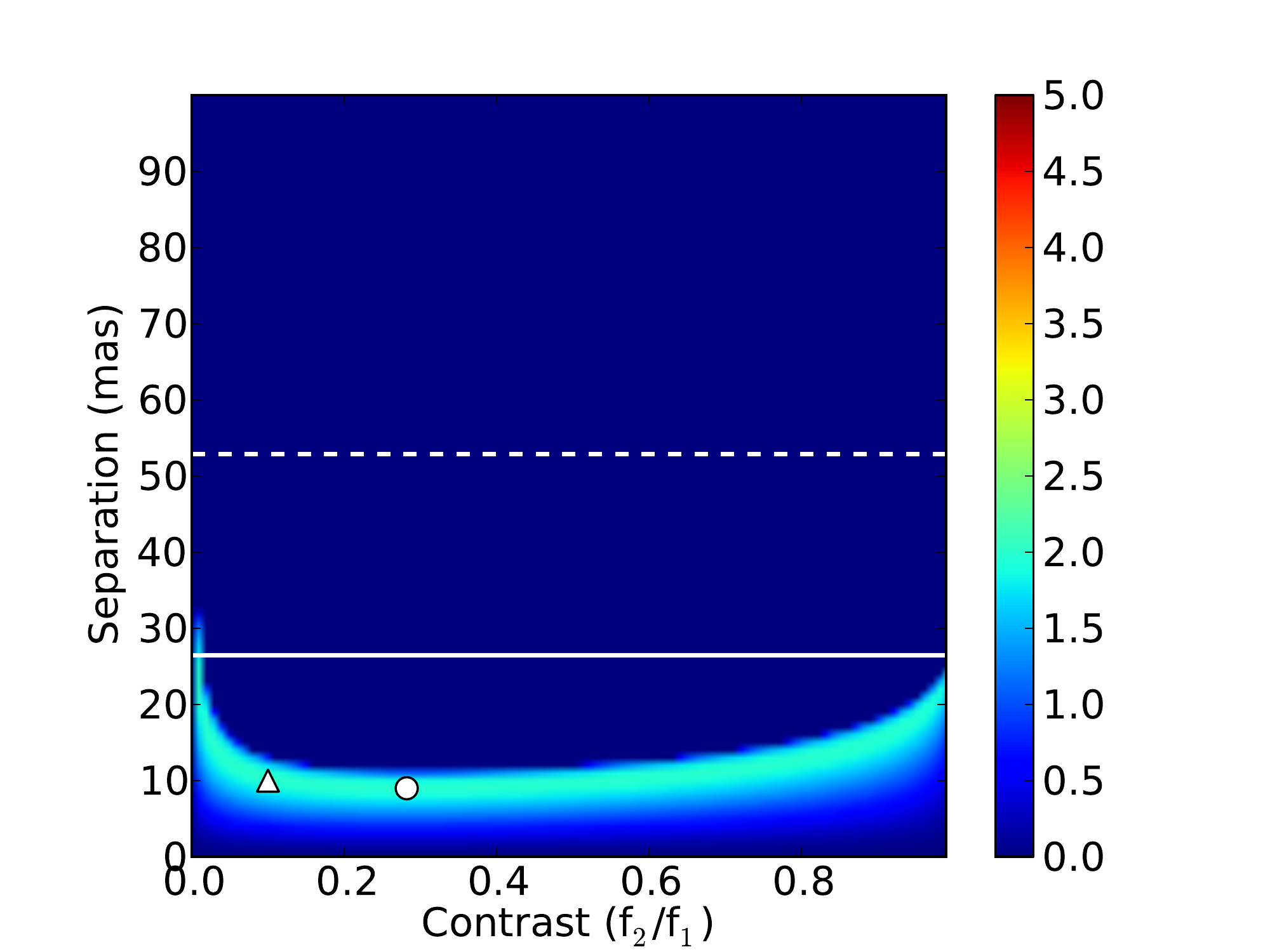} \\ 
 \end{array}$
\caption{ \textbf{Left: }Significance map. \textbf{Right: }Degeneracy plot. Simulated data of a a companion located at a separation of 10\,mas, a position angle of 90$^\circ$, and contributed 10\% of the total flux (white triangle). The white circle shows our best fit position. Within the background it is possible to see noise artefacts caused by holes within the uv-coverage. These holes create periodic signals within the background and may take on geometric patterns.
}
\label{FigNonDetcChi2}
\end{figure}

In data sets where the companion or disc wall was positioned at separations at or below $\lambda/2D$ we see that the images and significance maps become dominated by the gaussian noise placed in the models (see Figure~\ref{FigNonDetcChi2}). In all the cases shown, the artificial companion has a contrast of $f = 0.1$. We find a much reduced significance compared to a similar companion at larger separations. We therefore consider any companion with a separation below $\lambda/2D$ to be unresolved. In cases where our uv-coverage is sparser caused by the flagging of one or multiple holes during the data reduction process, we often see this noise as periodic signals in the background distribution. The strength of these periodic signals is dependent on the precise uv-coverage and the level of noise in the closure phases.

Within the reconstructed images, we find that data sets with an unresolved companion will simply be dominated by randomly distributed noise peaks (i.e.\ TW\,Hya, K'-band). We also encountered cases, where the image reconstruction algorithm attributed the flux elements of the companion to the central star (e.g.\ FP\,Tau, L'-band). Both are shown in Figure \ref{fig:NonDetections}. In these cases we are limited to placing lower limits on the possible contrast for a companion around these targets at separations within 200\,mas. This limit is set at the 99$\%$ confidence level which is determined by the individual noise properties of the data.

\subsection{Marginally Resolved Companion Scenario}
\label{sec:partially_companion_simulations}

\begin{figure}
\centering
$\begin{array}{@{\hspace{-2.0mm}} c @{\hspace{-7.0mm}} c}
     \includegraphics[height=4cm]{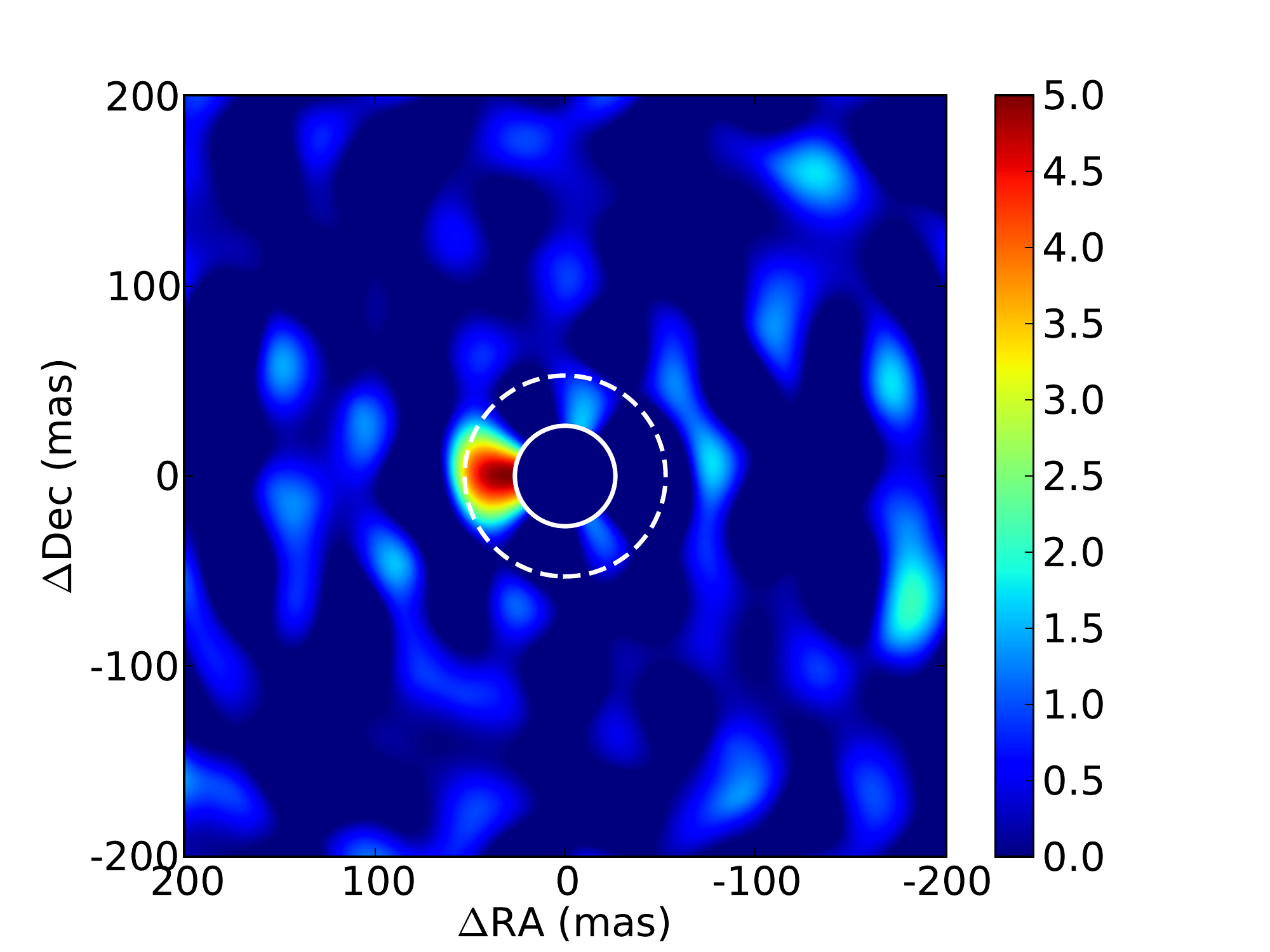}& \includegraphics[height=4cm, angle=0]{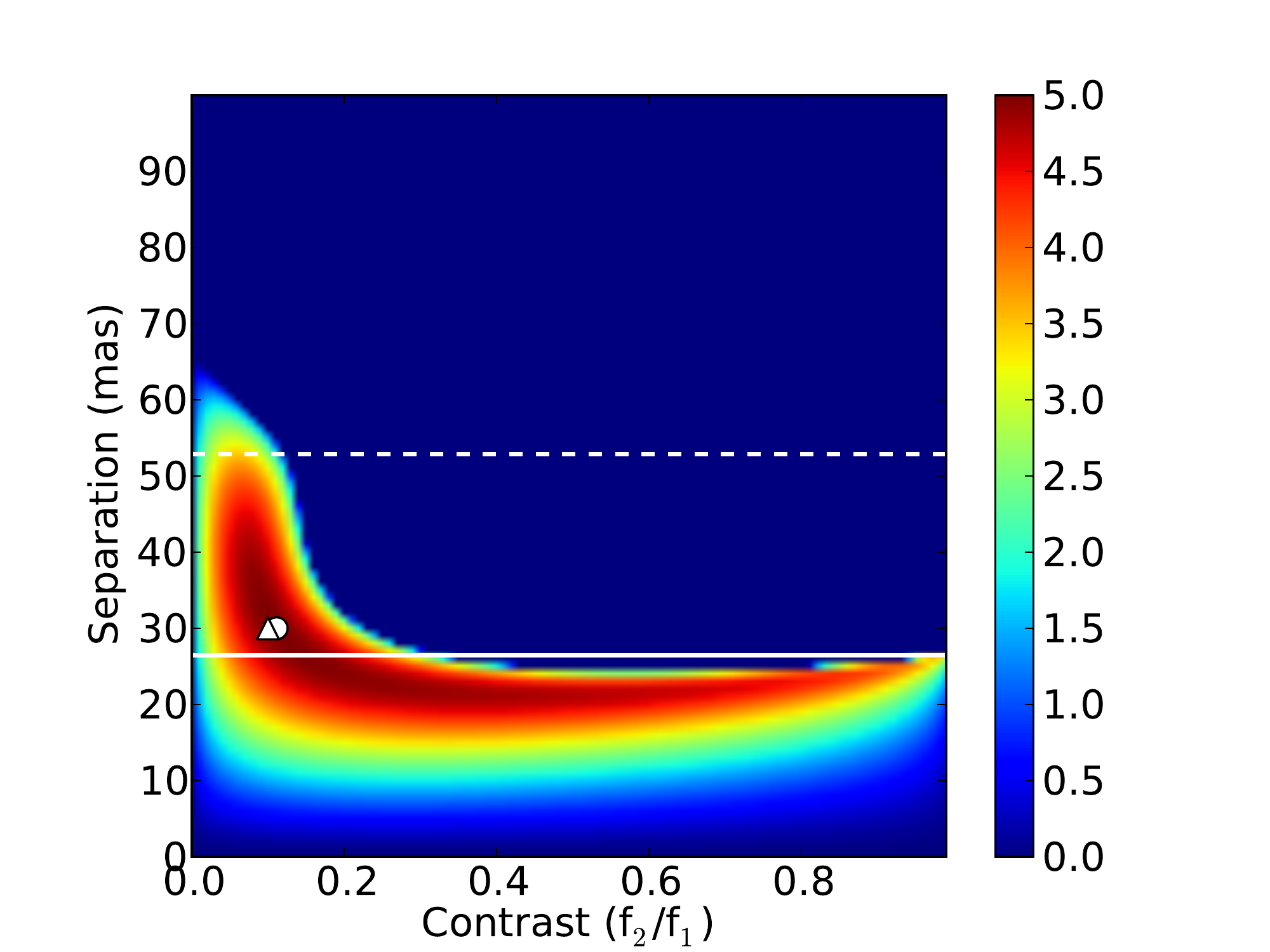} \\ 
 \end{array}$
\caption{\textbf{Left:} Significance map. \textbf{Right: }Degeneracy plot. Simulated data of a companion located at a separation of 30\,mas, a position angle of 90$^\circ$, and contributes 10\% of the total flux (white triangle). The white circle shows our best fit position. We see the distinctive lobing of a partially resolved companion. In this case with excellent SNR we are able to accurately identify the location of the companion but in practice this is not always the case.
}
\label{FigMargDetcChi2}
\end{figure}

To study a marginally resolved regime, we simulated data with a companion at a separation of 30\,mas. We observe the "strong lobe" structure characteristic of this regime (Figure \ref{FigMargDetcChi2}).  In the case of a low-contrast companion ($f=0.1$ in the simulation), the degenerate region is reasonably confined, while for higher-contrast companions the "lobes" are large and induce greater errors into estimations of both the position and contrast of any potential companion detection.

\subsection{Fully-Resolved Companion Scenario}
\label{sec:fully_companion_simulations}

\begin{figure}
\centering
$\begin{array}{@{\hspace{-4.0mm}} c @{\hspace{-7.0mm}} c}
\includegraphics[height=4cm]{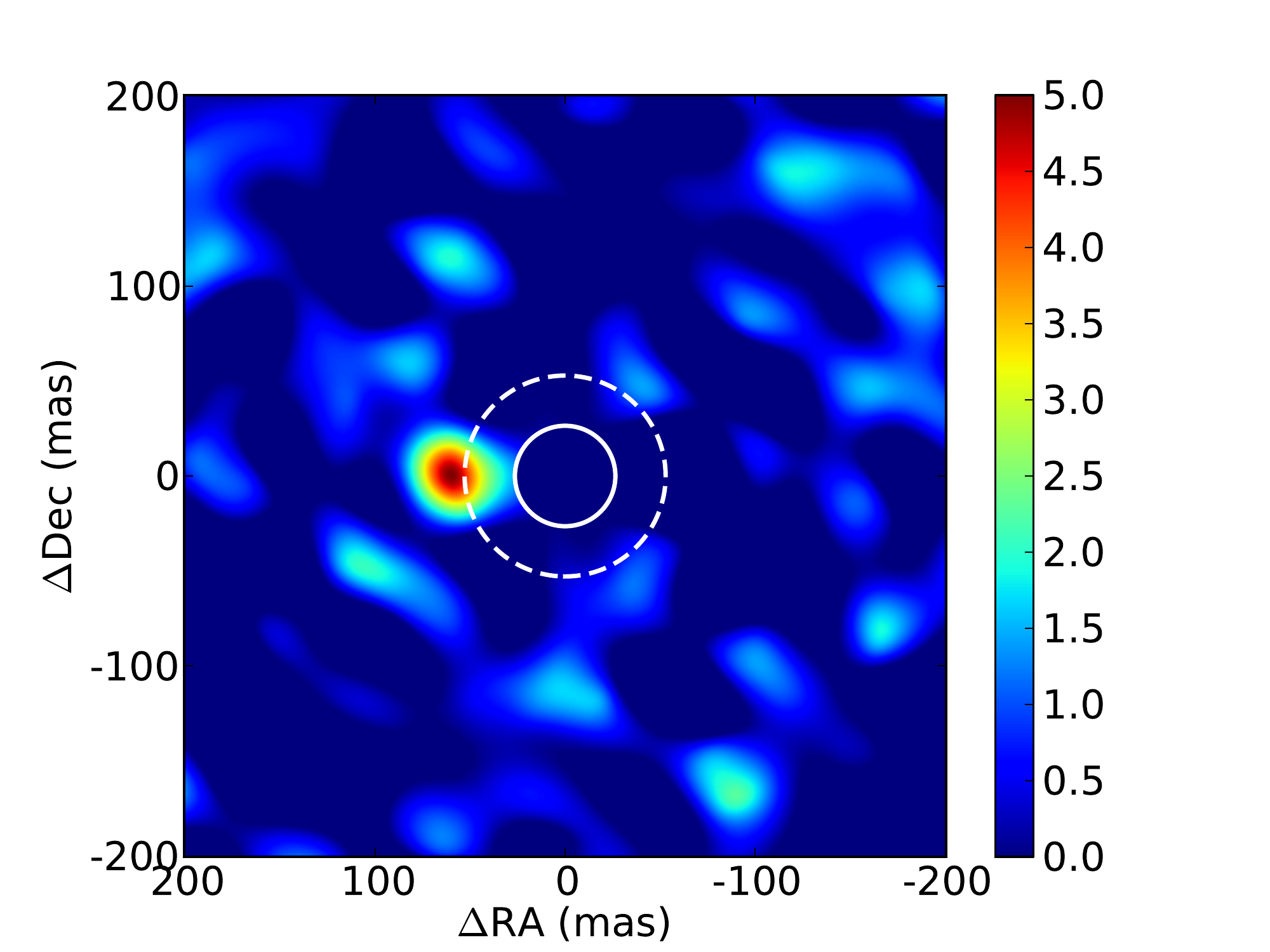} & \includegraphics[height=4cm]{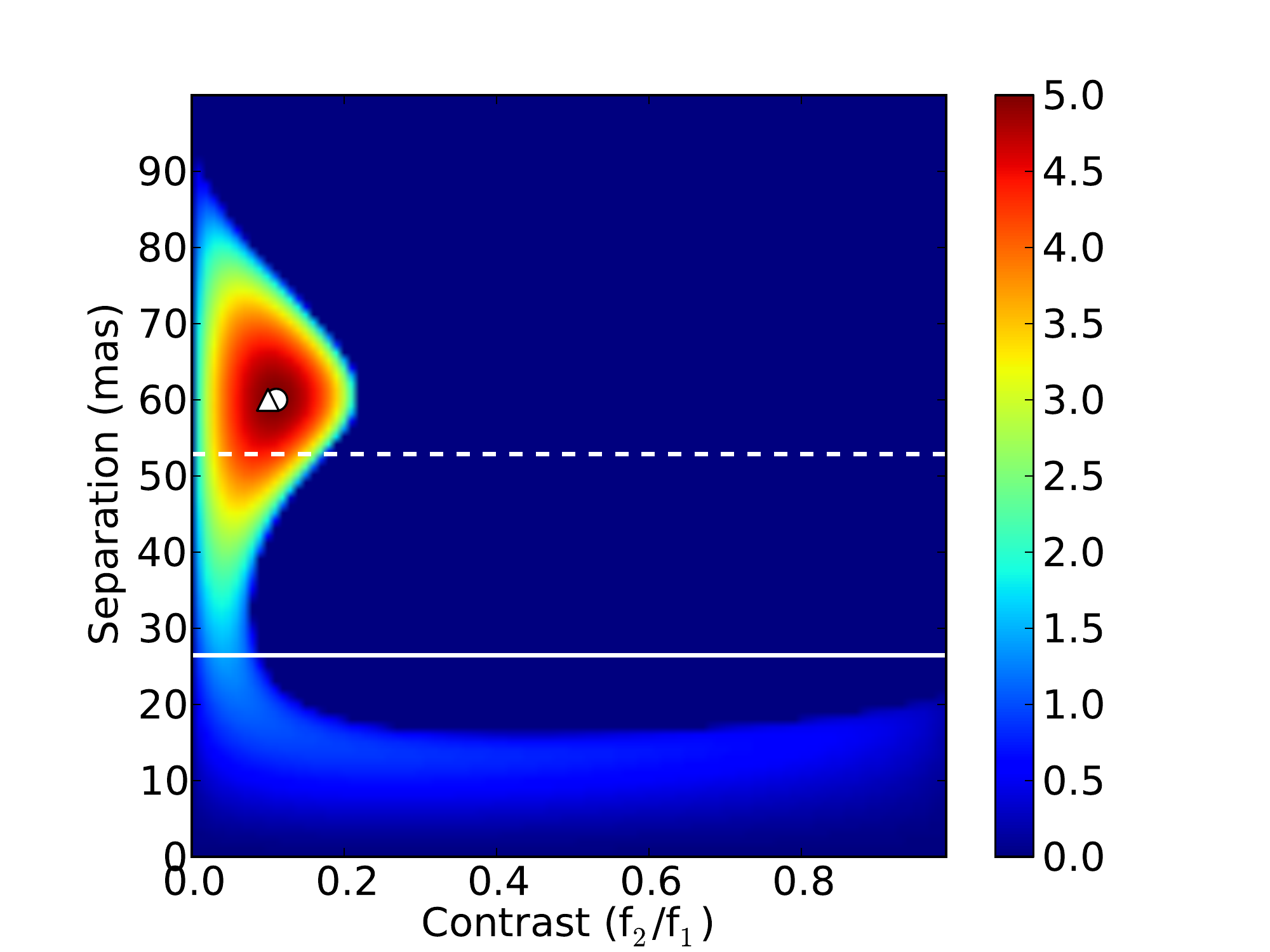} \\
 \end{array}$
\caption{\textbf{Left: }Significance map. \textbf{Right: }Degeneracy plot. Simulated data for a companion located at a separation of 60\,mas, a position angle of 90$^\circ$, and contributed 10\% of the total flux (white triangle). The white circle shows our best fit position. Here the companion is fully resolved and the position and contrast are well constrained. 
}
\label{FigFulDetcChi2}
\end{figure}

To simulate a fully-resolved companion we computed models with a companion located at a separation of 60\,mas, just beyond $\lambda/D$. 
At these separations we can see that the degenerate region has largely disappeared allowing the position of the companion to be well constrained (see Figure \ref{FigFulDetcChi2}). 

\subsection{Disc-feature scenario}
\label{sec:disk_feature_simulations}

Asymmetries in the brightness distribution can also be caused by disc-related structures, producing closure phase signals that might be difficult to discern from those produced by close-in companions \citep{2015MNRAS.450L...1C}. To investigate this scenario, we produced synthetic images that are intended to mimic the rim of a disc seen under intermediate inclination ($60^{\circ}$ from the face-on orientation) with a radius of 30\,mas, which corresponds to $\sim$\,$3\lambda/2D$. The image shown was produced by simulating a skewed ring with a gaussian profile and a width of 15\,mas, a skewness of 0.8, and whose major axis is oriented along position angle 0$^\circ$. The flux of the disc represents 1\% of the total flux in the frame. 

\begin{figure}
\centering
$\begin{array}{@{\hspace{-3.0mm}} c @{\hspace{-4.0mm}} c}
\includegraphics[height=4cm]{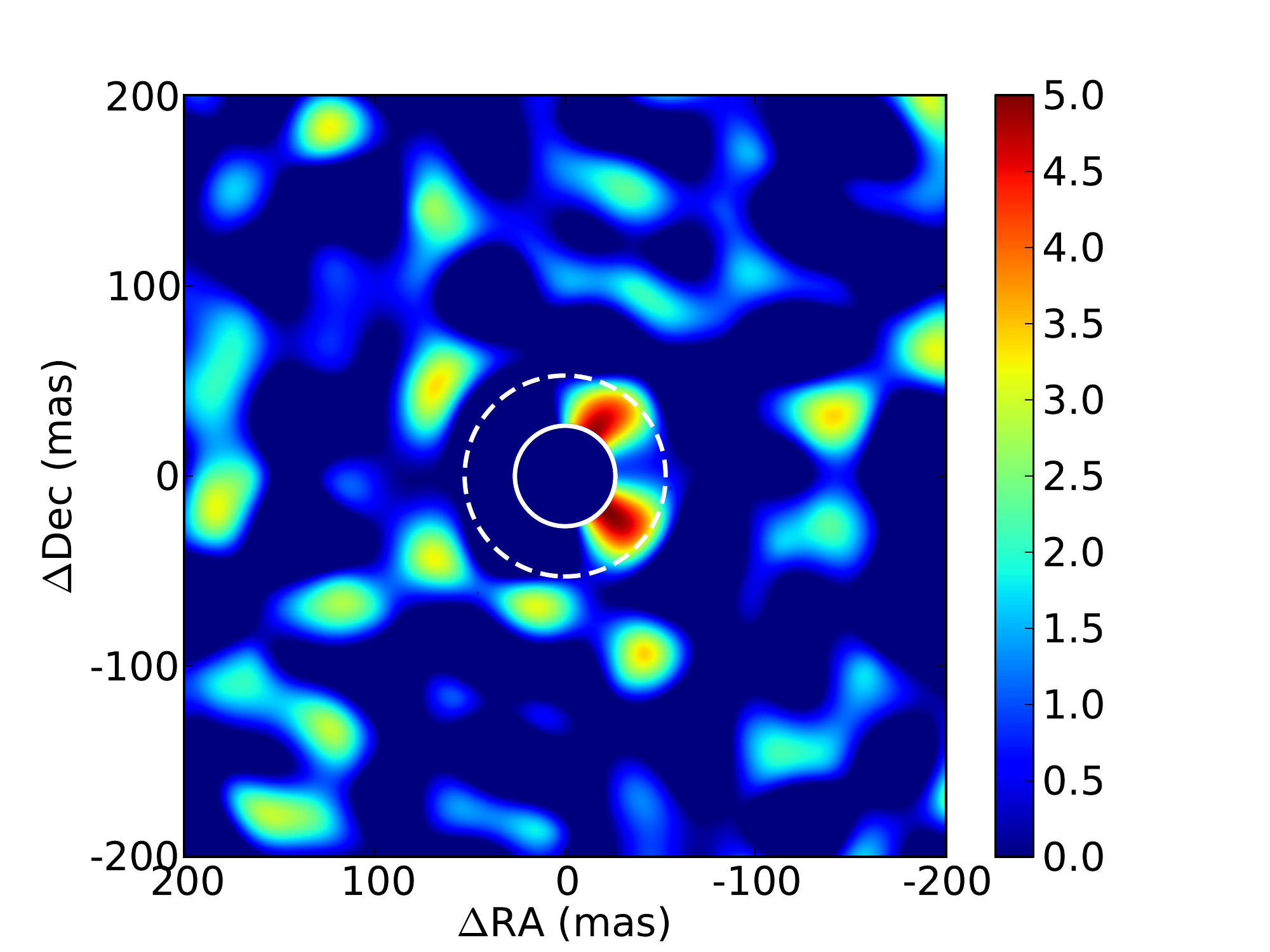} &
\includegraphics[height=4cm]{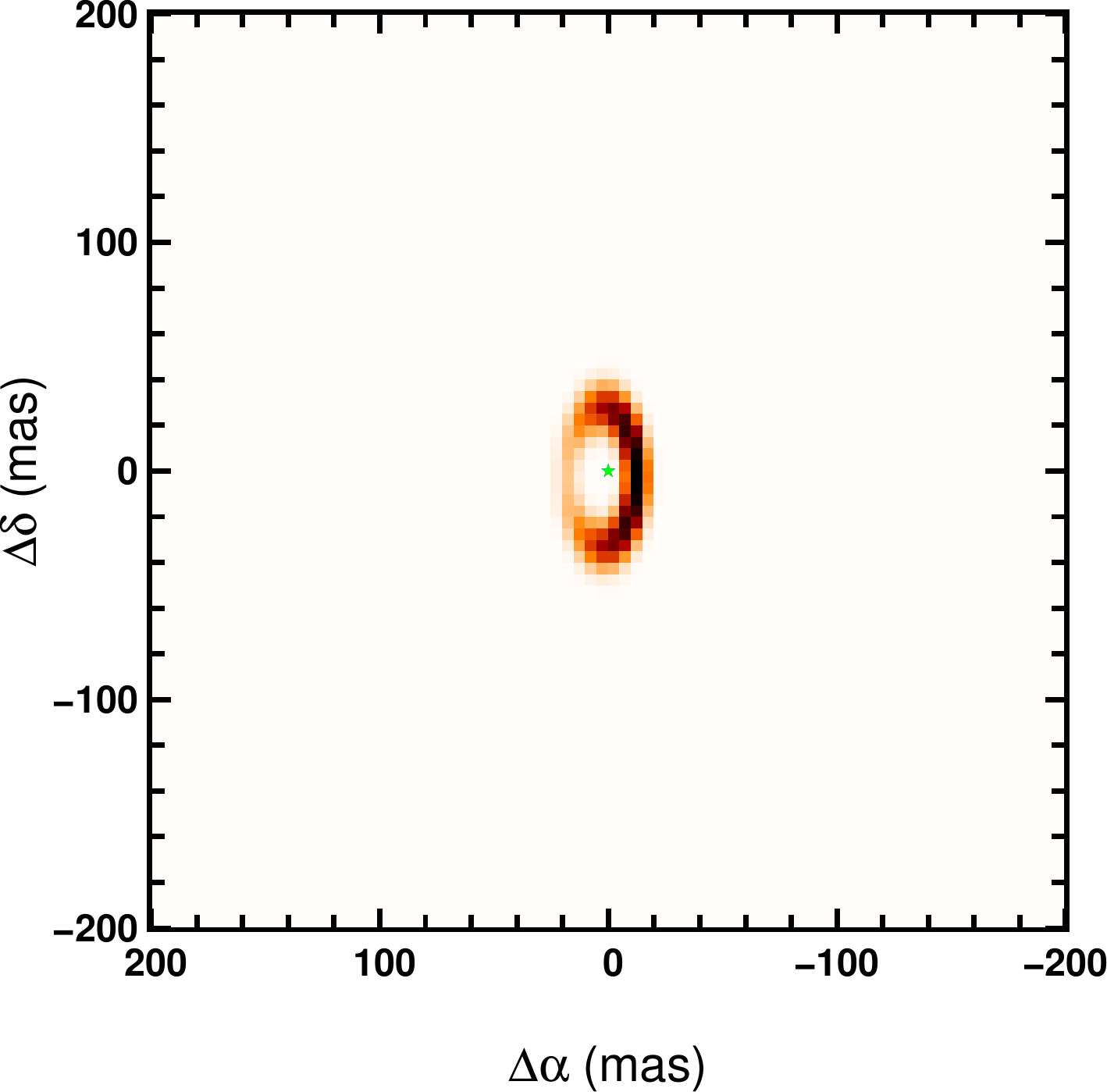} \\
 \end{array}$
\caption{\textbf{Left: }Significance map for a simulation with a partially resolved disc wall. The resulting significance maps show two strong detections located at the disc wall.  At increasing separations and resolution the two point sources begin to merge. \textbf{Right: }Input intensity distribution. Green star indicates the position of the parent star.
}
\label{FigDiskDetcChi2}
\end{figure}

Within all the significance maps from this scenario we observe that the significance contours take on double-lobed structures (Figure~\ref{FigDiskDetcChi2}). This is in agreement with previous work performed by \citet{2015MNRAS.450L...1C}, who showed that an inner wall of a optically thick disc will appear as two point-source like structures co-locational with the illuminated rim of the disc, bisected by the center of the disc wall. We find that these also appear within our significance maps and reconstructed images.

Extending the semi-major axis such that the ring appears outside of the degenerate region we begin to resolve the shape of the disc wall. This structure tends to be comparable in strength to the artefacts however and is unseen unless the flux contribution of the disc is not enhanced. This is the result of the flux becoming more spread out within the frame, inducing smaller phase signals.

\begin{figure}
\centering
$\begin{array}{@{\hspace{-3.0mm}} c @{\hspace{-4.0mm}} c}
\includegraphics[height=4cm]{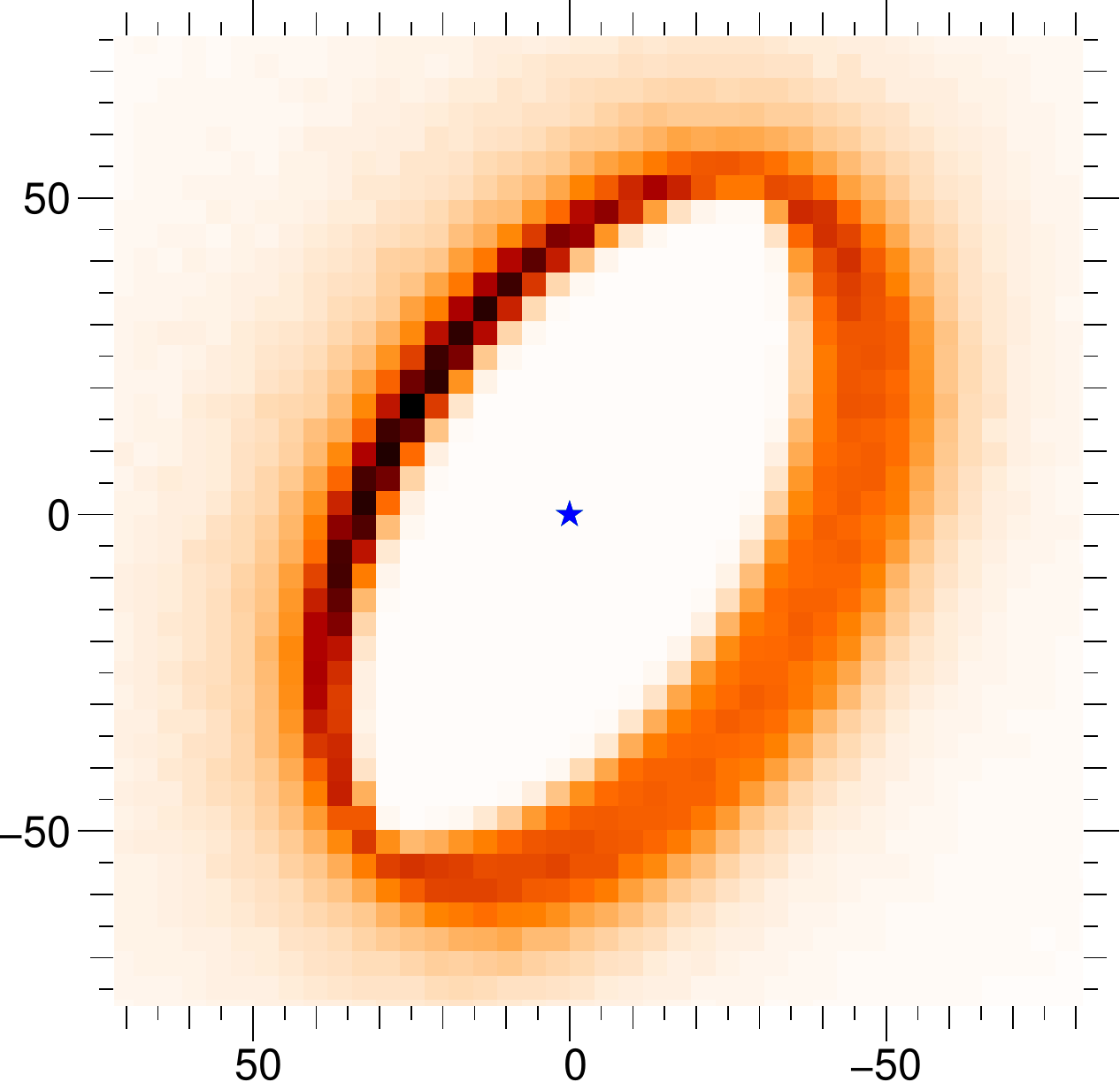} &
\includegraphics[height=4cm]{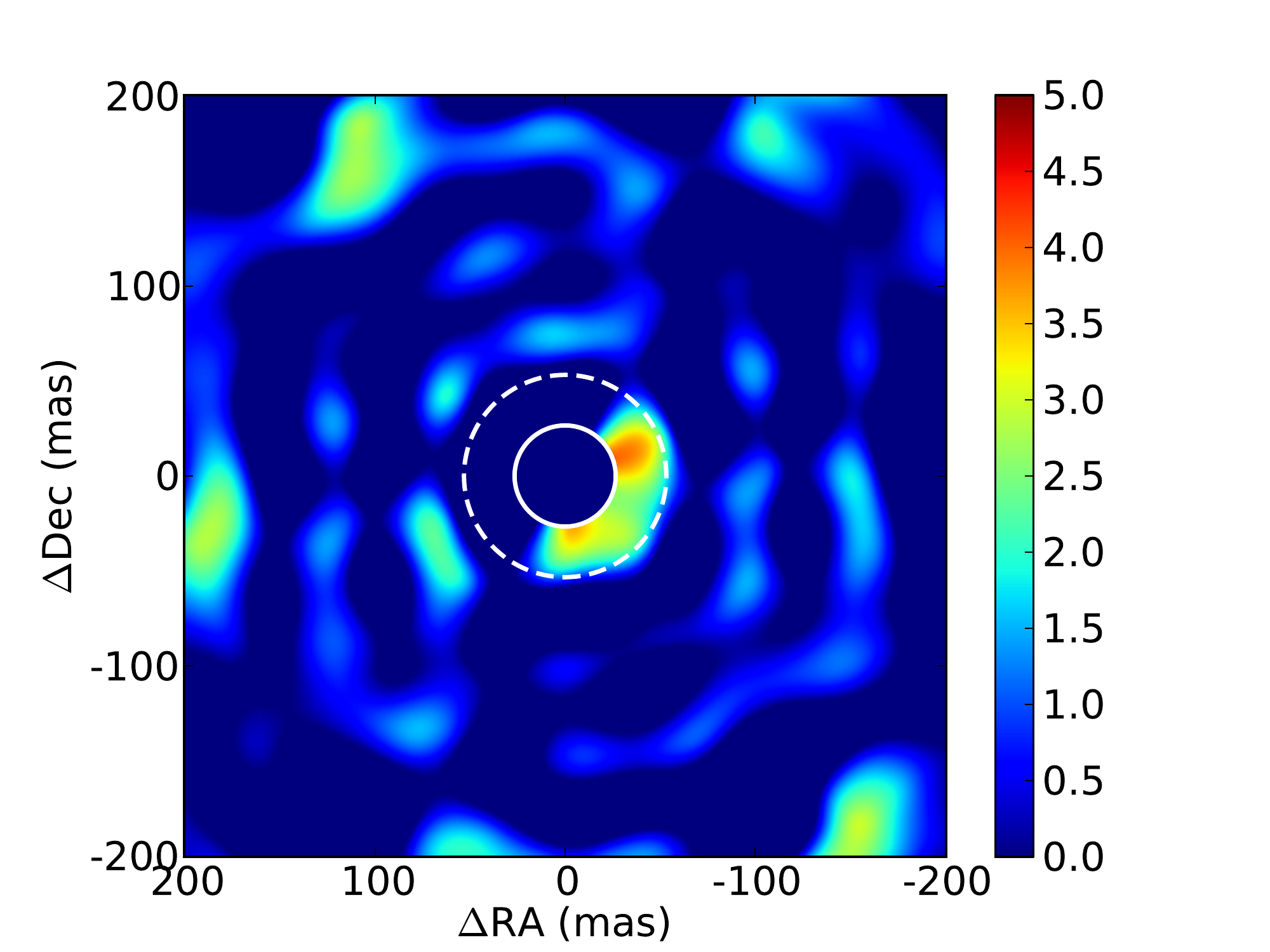} \\
\includegraphics[height=4cm]{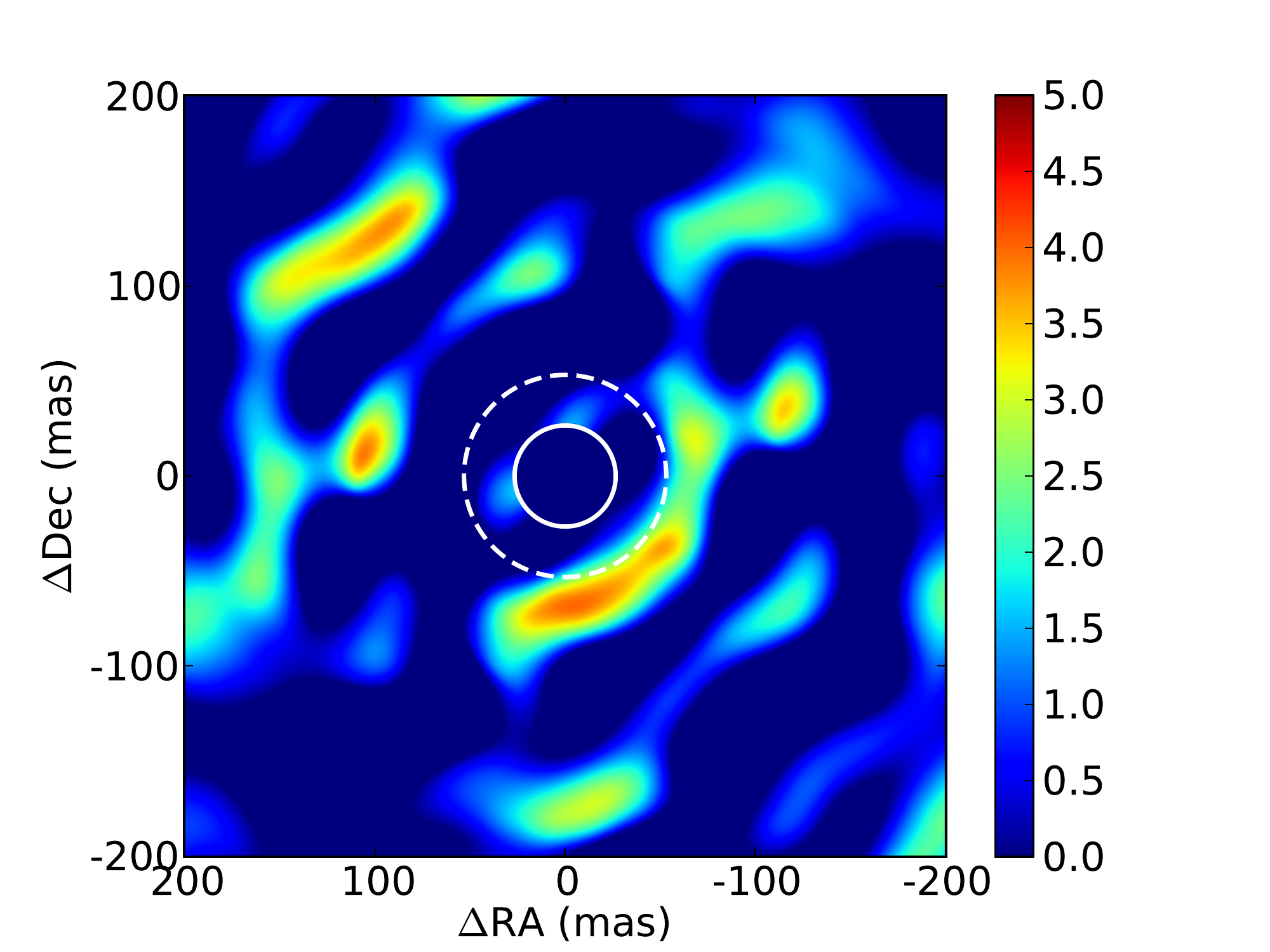} &
\includegraphics[height=4cm]{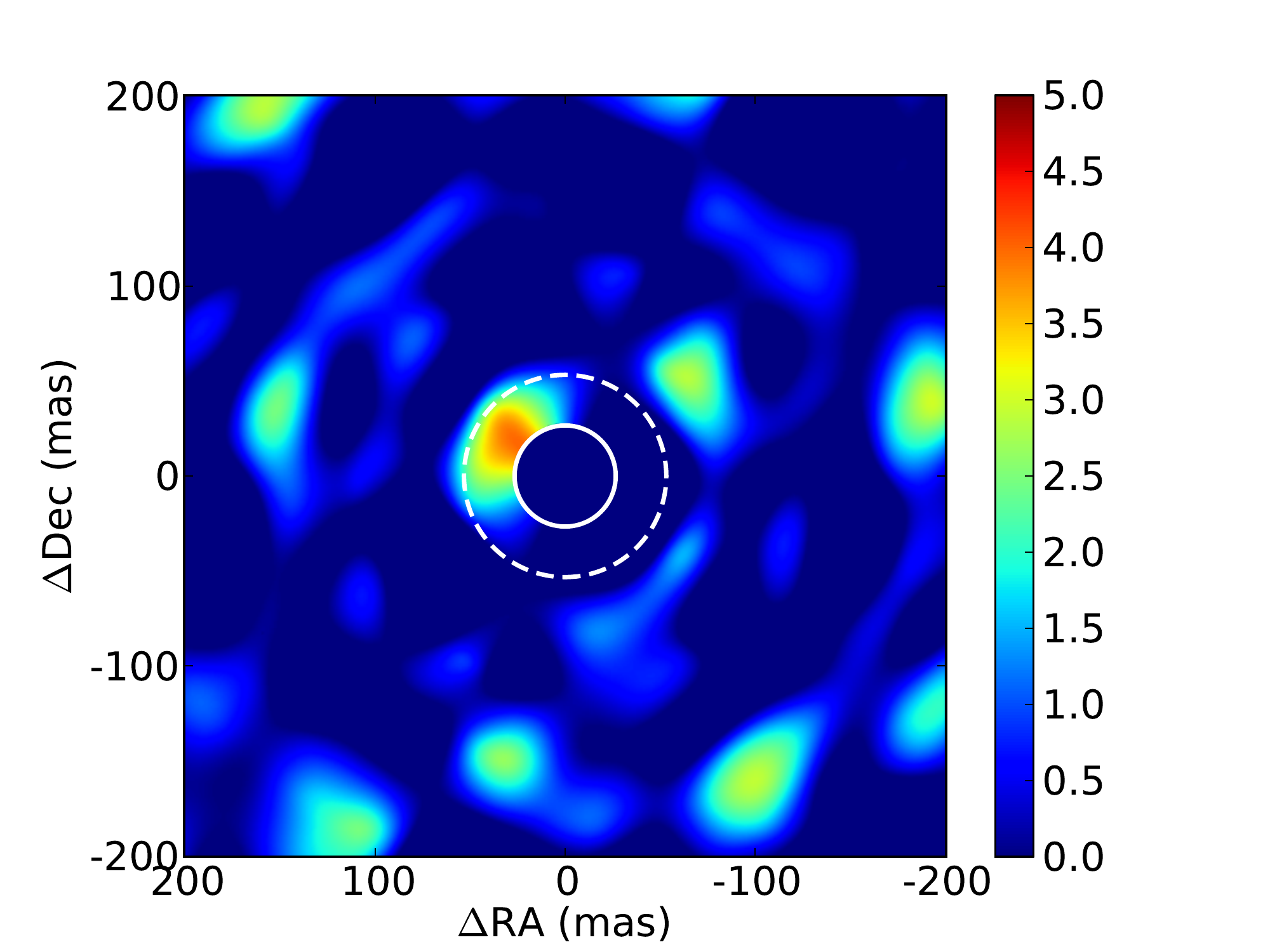} \\
 \end{array}$
\caption{\textbf{Top Left:} Base model for physical disc simulations. We scale this model for each semi-major axis case. Thermal emission from the far side of the disc is seen in the right side of the frame and forward scattered light in the top left. \textbf{Top Right: }30\,mas case. Here we see the double lobe structure caused by the far side of the inner disc wall. \textbf{Bottom Left: }60\,mas case. The arc of the far side of the disc wall is clearly seen but is comparable in strength to the artefacts within the frame and so would be difficult to identify in practice. \textbf{Bottom Right: }90\,mas case. The forward scattered component is now the dominant feature within the frame. It forms a single lobe owing to the greater concentration of flux in the centre of the arc than in the thermal emission from the opposite side of the disc wall.
}
\label{FigDiskDetcForward}
\end{figure}

To make a comparison to a more physical model we create a disc model using the radiative transfer code, TORUS \citep{2014ascl.soft04006H}. Here we can include effects such as forward scattering from the near edge of the disc. We scale this model to have semi-major axes of 30, 45, 60, 90, 120, and 180\,mas. The results are shown in Figure \ref{FigDiskDetcForward}. The forward scattered component, while containing more flux, is closer to the central star than the thermal component so only appears at larger separations. It also appears as a single lobe as a resut of flux being most concentrated at the centre of the arc whereas in the thermal case, the flux is more evenly spread across the disc wall. At larger separations this single lobe becomes more resolved, similar to the thermal emission seen in the bottom left frame of Figure \ref{FigDiskDetcForward} and similarly difficult to distinguish from artefacts.

\subsection{Disc-asymmetry scenario}
\label{sec:disk_asymmetry_simulations}

\begin{figure}
\centering
$\begin{array}{@{\hspace{-0.0mm}} c @{\hspace{-4.0mm}} c}
\includegraphics[height=4cm]{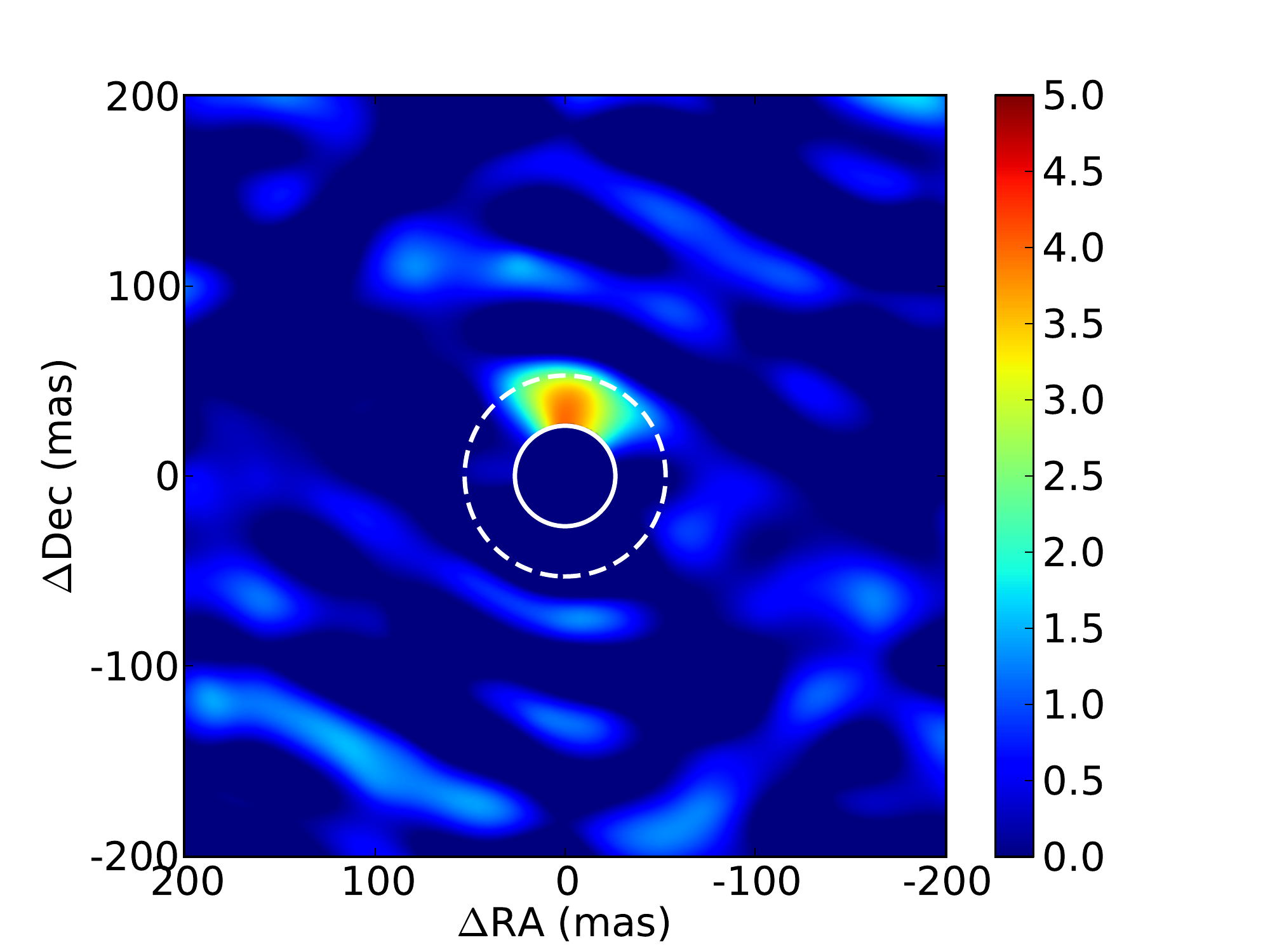} &
\includegraphics[height=4cm]{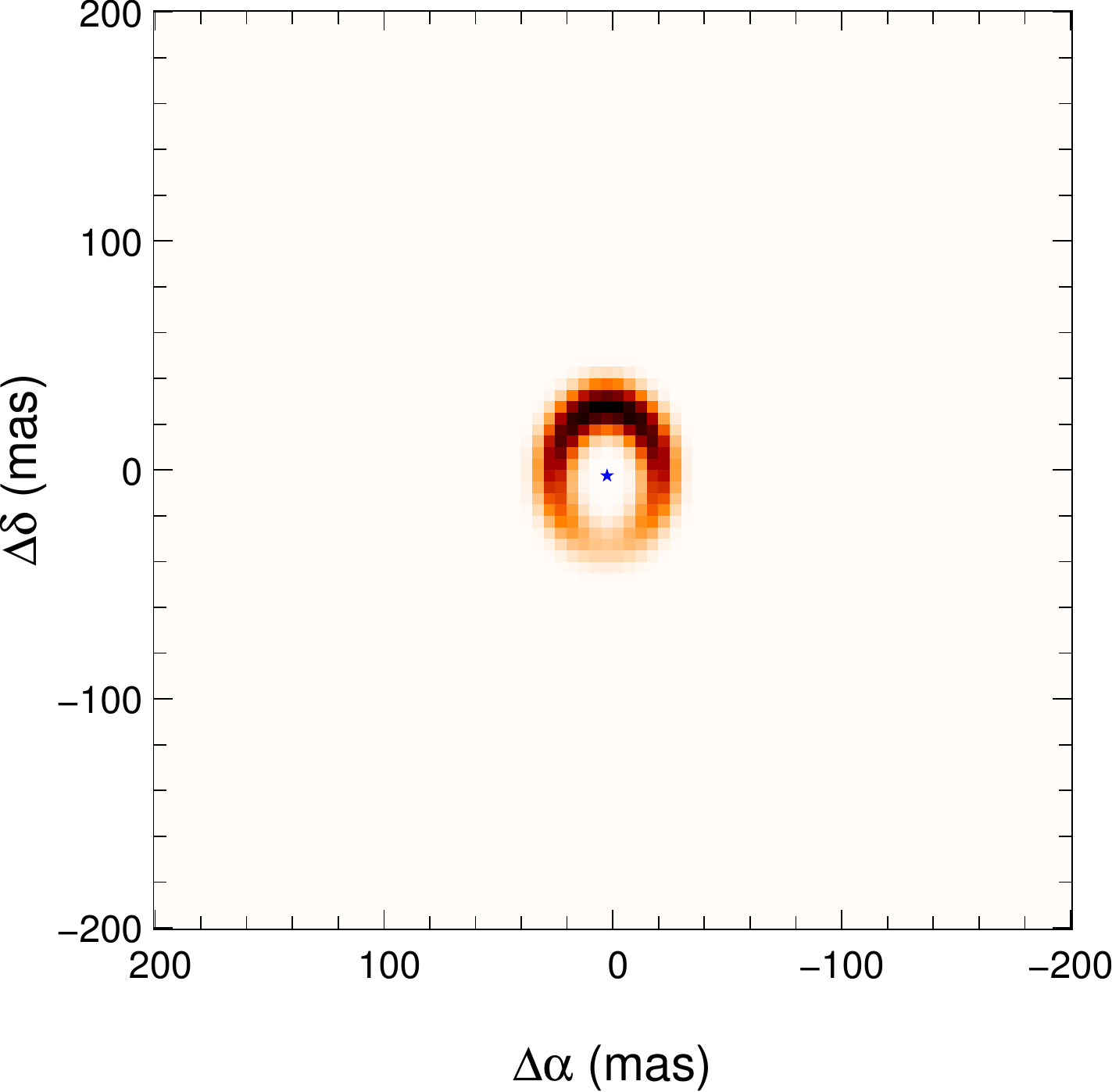} \\
\includegraphics[height=4cm]{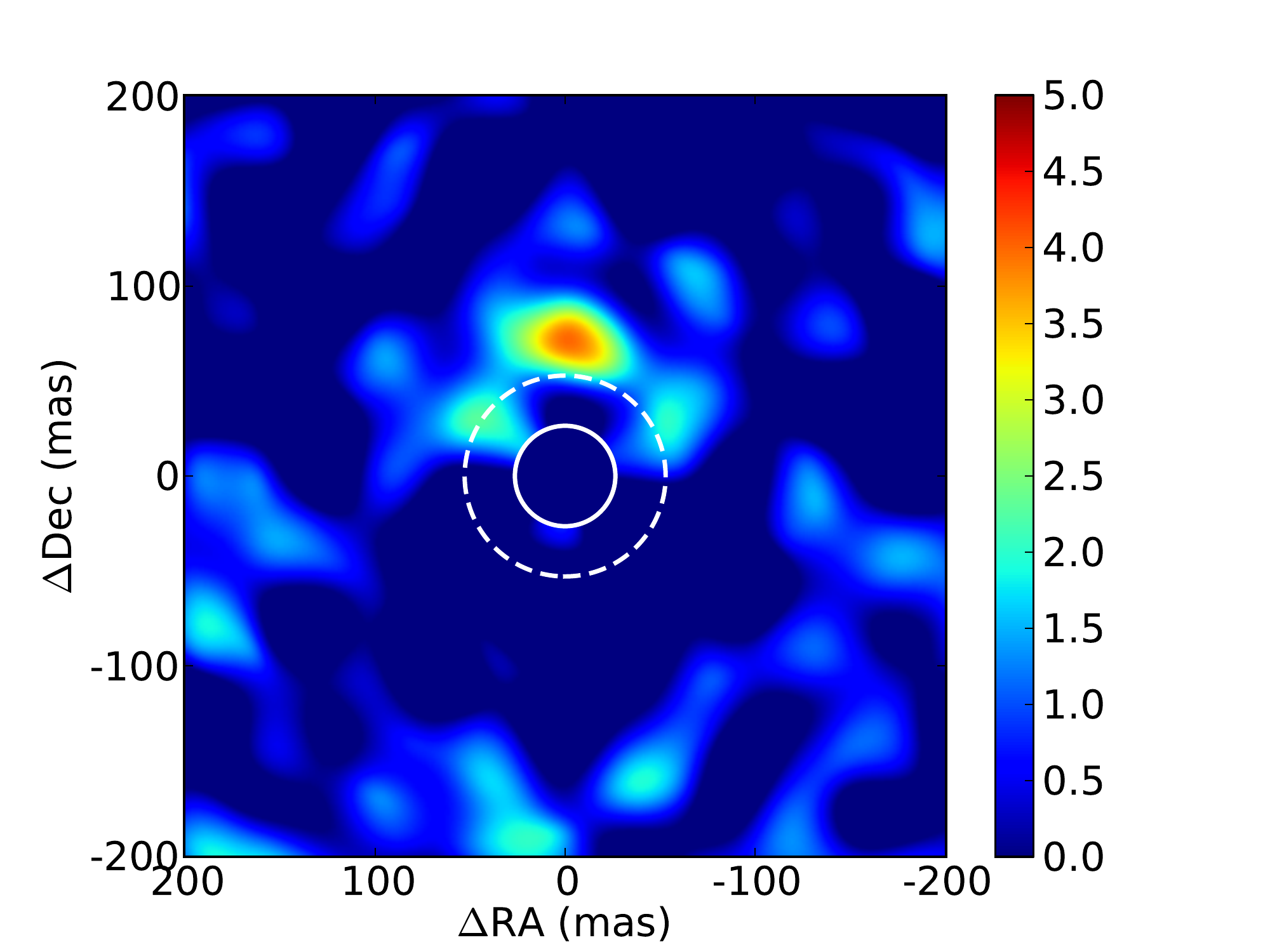} &
\includegraphics[height=4cm]{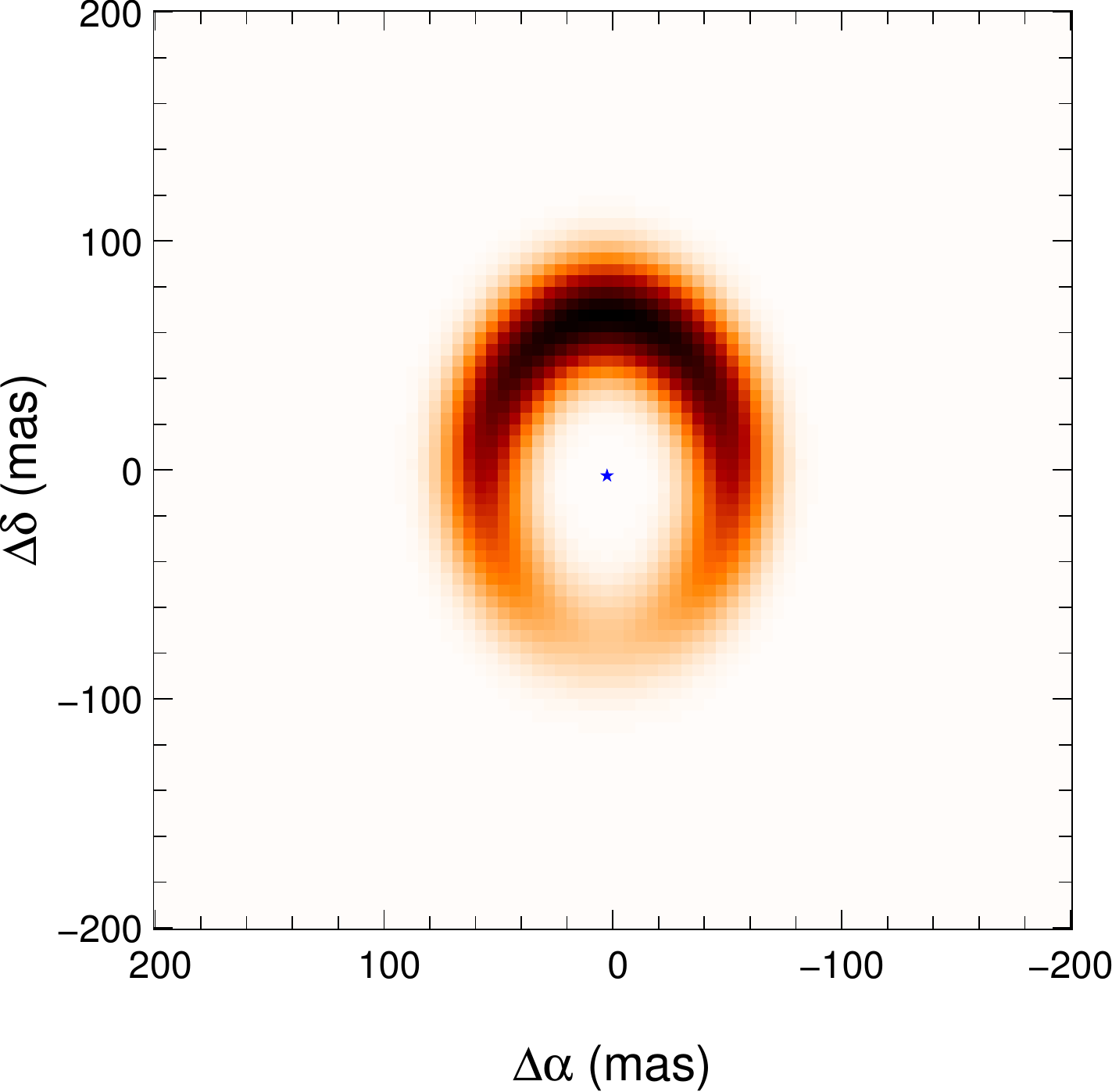} \\
 \end{array}$
\caption{\textbf{Left: }Significance map for a simulation with a partially resolved and fully resolved disc asymmetry. The resulting significance maps shows structure similar to a companion detection. \textbf{Right: }Input intensity distribution. Green star indicates the position of the parent star.
}
\label{FigDiskAsymChi2}
\end{figure}

To investigate scenarios in which an asymmetry is caused by an over density within the disc we simulate a disc feature with a contrast of 5:1 to the rest of the disc. We take a similar approach to Section \ref{sec:disk_feature_simulations} but skew the ring in such a way to resemble possible asymmetries such as those found in simulations by \citet{2007A&A...471.1043D}. These are extreme cases as it is difficult to physically create such a strong contrast particularly in continuum emission \citep{2015MNRAS.451.1147J}.

In Figure \ref{FigDiskAsymChi2} we show two cases representing partially resolved and fully resolved cases. Both strongly resemble the structures seen in our companion detection simulations in Sections \ref{sec:partially_companion_simulations} and \ref{sec:fully_companion_simulations}. This should be kept in mind when considering our companion detections without complementary multi-wavelength observations.

%

\section{Results}
\label{sec:results}
We identify potential candidates through a combination of setting a threshold on the significance of the binary fit and inspection of reconstructed images and significance maps.
In Table \ref{tab:companions} we list data sets in which we find significant closure phase asymmetries excluding false positives and each case is discussed individually in detail below. 

To calculate the semi-major axis for our candidates we assume circular orbits coplanar with the outer disc. Where disc inclination/position angle information is unavailable, we assume a face-on disc.  To estimate the companions' absolute magnitudes, we used the reddening law outlined in \citet{1989ApJ...345..245C}. From the dereddened absolute magnitudes, we then estimate values of $M_{c} \dot M_{c}$ using the accreting protoplanetary disc models described by \citet{0004-637X-799-1-16}. We match our dereddened absolute magnitudes to the table within \citet{0004-637X-799-1-16}, assuming an inner circumplanetary disc radius of 2R$_{\textrm{J}}$. This is a highly unknown quantity with a significant effect on the resultant values of $M_{c} \dot M_{c}$. We arbitrarily chose our inner disc radius to be the same value as that assumed by \citet{2015Natur.527..342S} for the purposes of comparison. When matching the absolute magntiudes are difficult to match we linearly interpolate. We are unable to directly estimate the mass of a potential companion as these objects are thought to likely possess extended, accreting circumplanetary discs that dominate the infrared excess emission; this prevents us from separating the mass $M_{c}$ and accretion rate $\dot M_{C}$.

\begin{table*}
\centering
\caption{Companion candidates}
\label{tab:companions}
\begin{tabular}{c c c c c c c c c}
\hline\hline

    Identifier & $\rho$ & PA & Contrast & Sig & Semi-Major Axis & R$_{In}$ & $M_K$ & $ M_{c} \dot M_{c}$ \tablefootmark{a} \\
	 & [mas] & [$^{\circ}$] & [mag] & [$\sigma$] & [AU] & [AU] & [mag] & [10$^{-6}$ M$^2 _J$yr$^{-1}$] \\
\hline

DM Tau & 43$\pm$7 & 121$\pm$6 & 6.8$\pm$0.3 & 4.27 & 6.5$\pm$1.7 ($\pm$1.0) & 3/19\tablefootmark{(b)} & 11.0$\pm$0.3 & 10 \\ 
LkH$\alpha$ 330 & 132$\pm$3 & 212.9$\pm$1.4 & 5.6$\pm$0.2 & 4.88 & 37$\pm$4 ($\pm$0.8) & 50 & 5.3$\pm$0.2 & 1000 \\ 
TW\,Hya & 101$\pm$4 & 283$\pm$2 & 5.6$\pm$0.2 & 4.46 & 5.5$\pm$0.8 ($\pm$0.2) & 4 & 9.0$\pm$0.2 & 10 \\ 
\\
\end{tabular}
\tablefoot{Columns are organised by Identifier, angular separation, position angle, significance, orbital separation based on previous observations of the disc inclination and position angle (error when distance error neglected), inner radius of optically thick disc, absolute magnitude of the companion, stellar accretion rate and companion mass. Dereddening was performed as described in \citet{1989ApJ...345..245C}
\tablefoottext{a}{Values derived from \citet{0004-637X-799-1-16} assuming circumplanetary disc radii extending from 2 $R_J$ outwards.}
\tablefoottext{b}{First value derived from fitting IR spectra, second from fitting to sub-mm data}
}
\end{table*}

\begin{figure*}
 \begin{center}
\scriptsize
$\begin{array}{ @{\hspace{-1.0mm}} c @{\hspace{-4.7mm}} c @{\hspace{-6.5mm}} c @{\hspace{-5.0mm}} c}
   \includegraphics[height=4cm, angle=0]{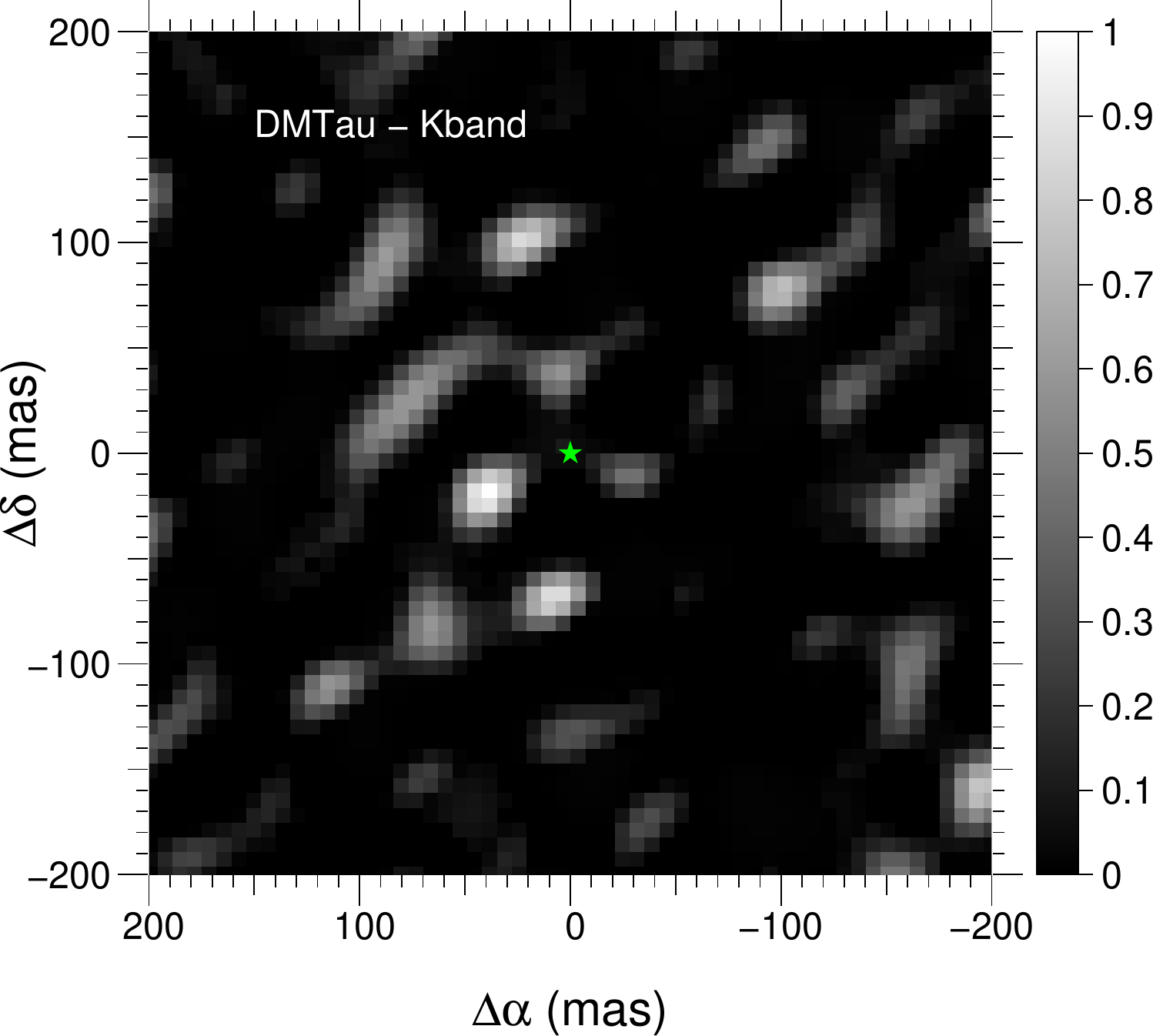} & \includegraphics[height=4cm, angle=0]{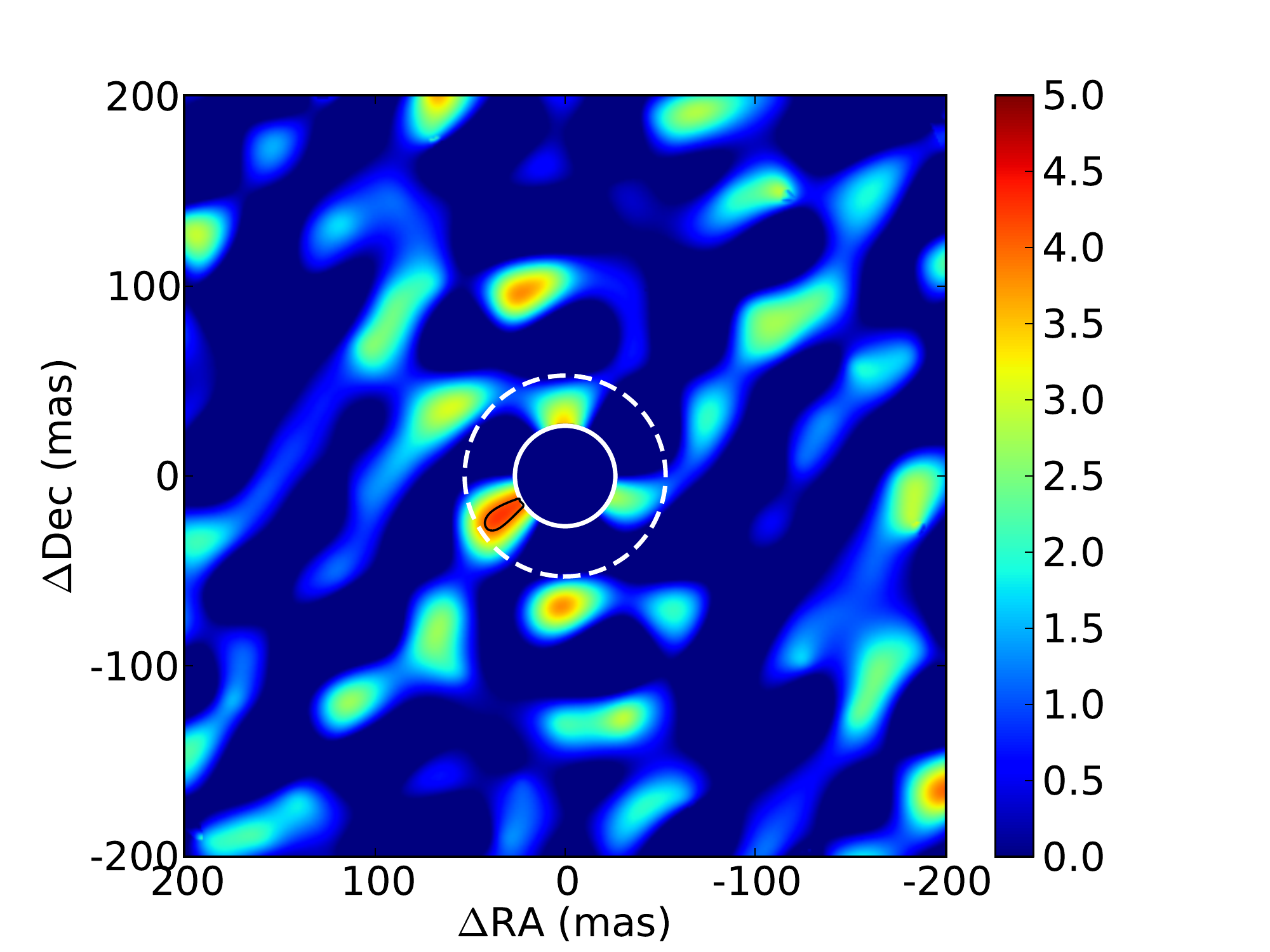} & \includegraphics[height=4cm, angle=0]{DMTau_KBand_120108_degeneracy_plot.pdf} &
   \includegraphics[height=4cm, angle=0]{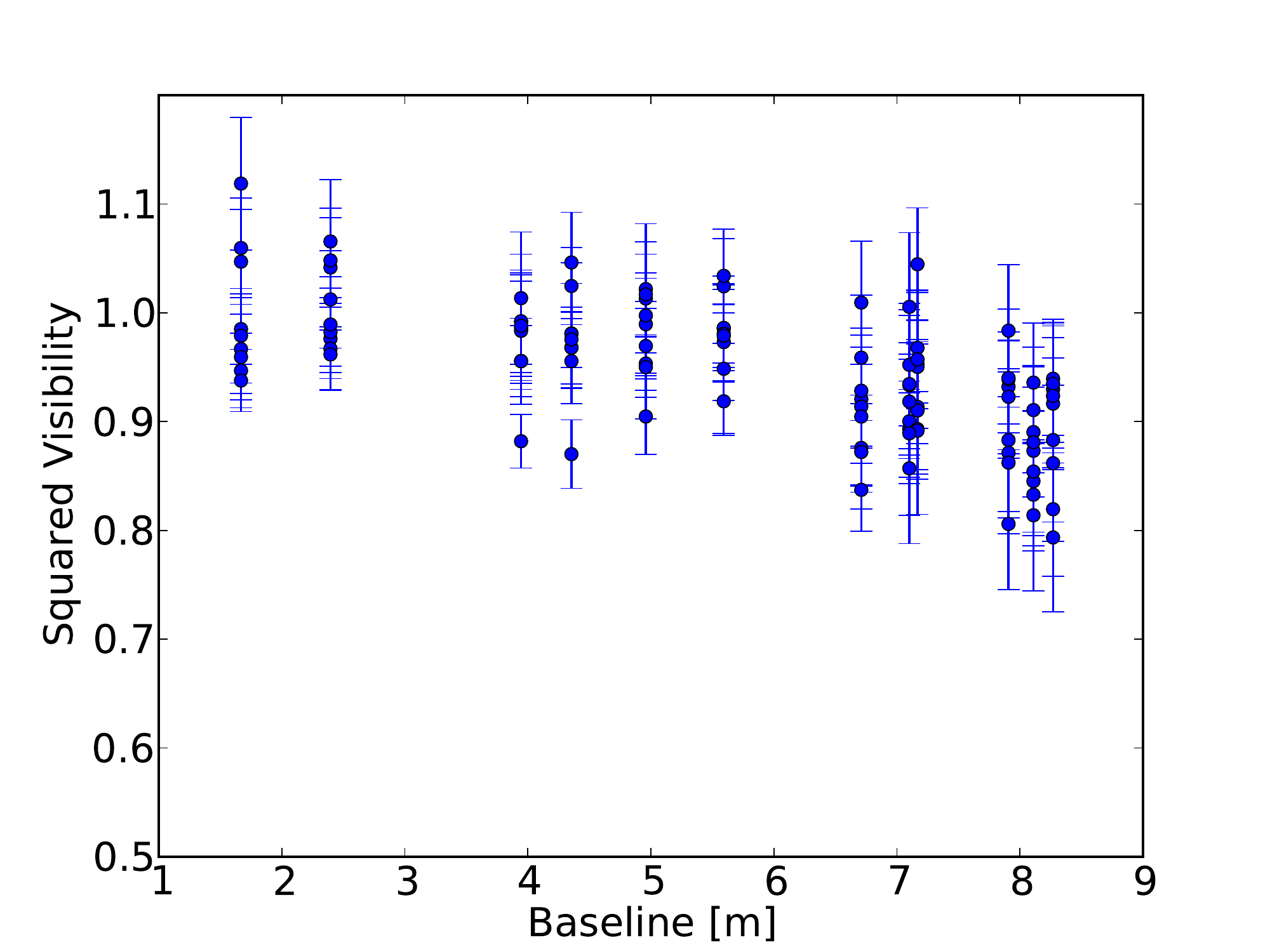}\\
    \includegraphics[height=4cm, angle=0]{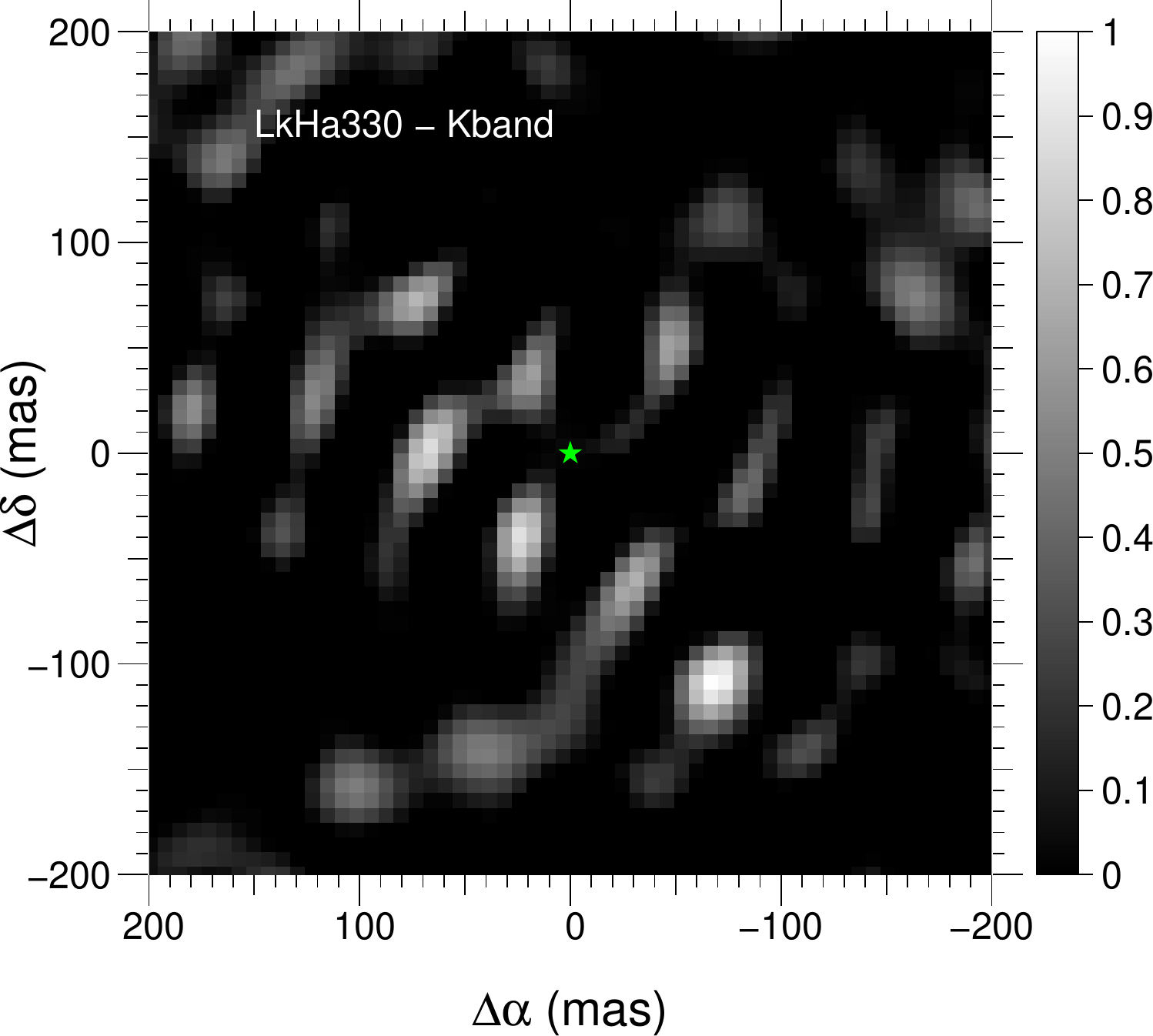} & \includegraphics[height=4cm, angle=0]{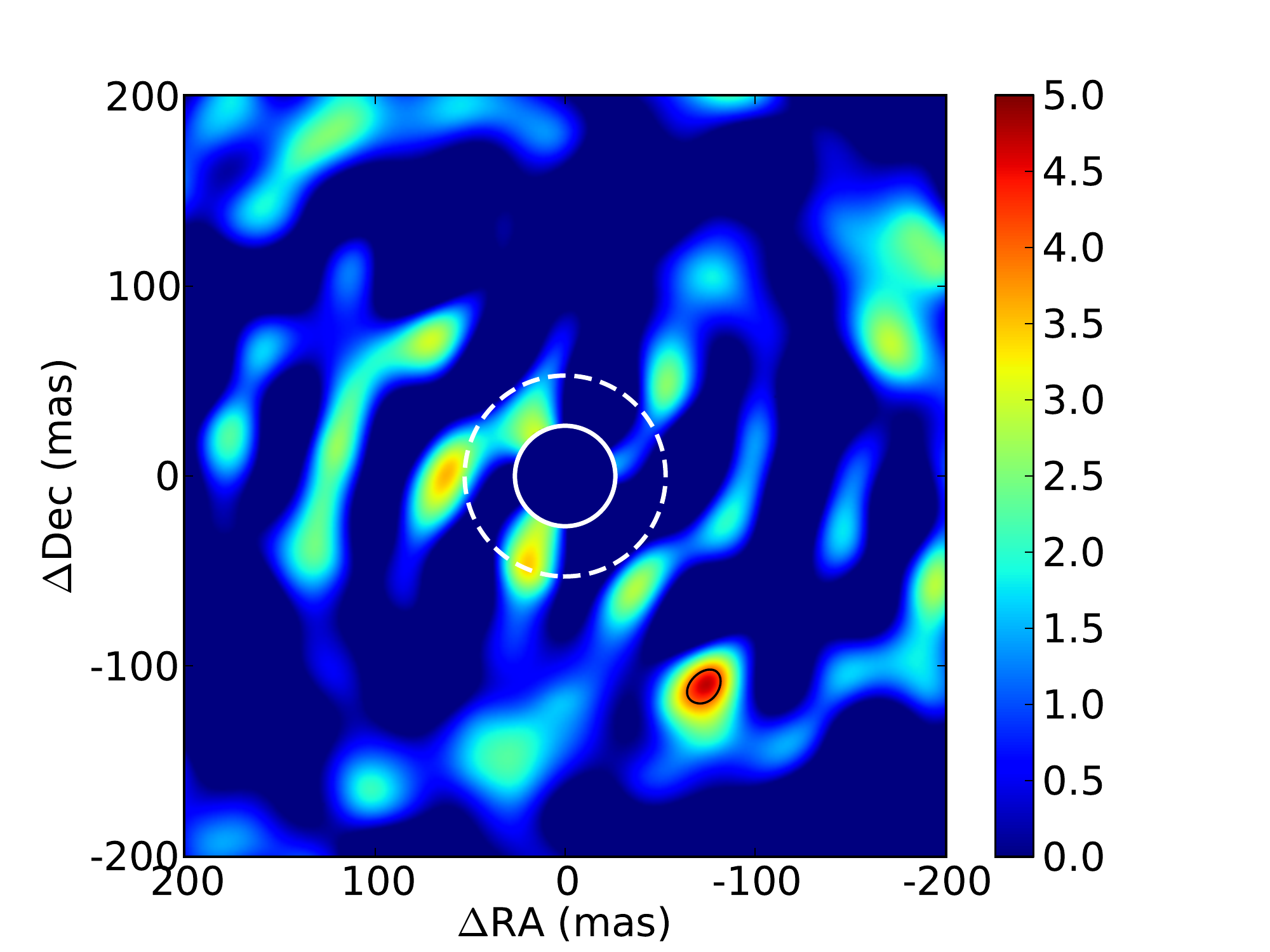} & \includegraphics[height=4cm, angle=0]{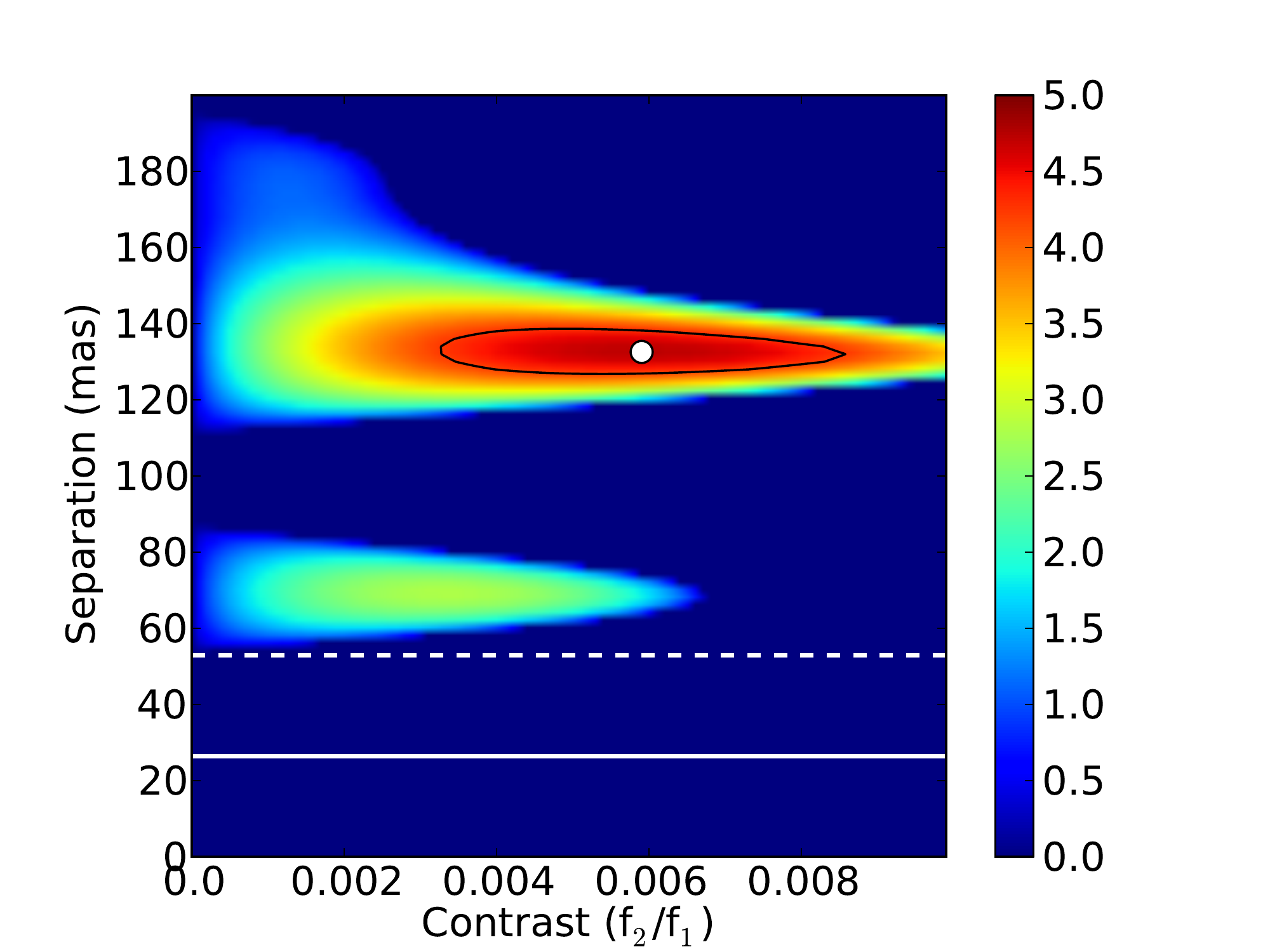} &
    \includegraphics[height=4cm, angle=0]{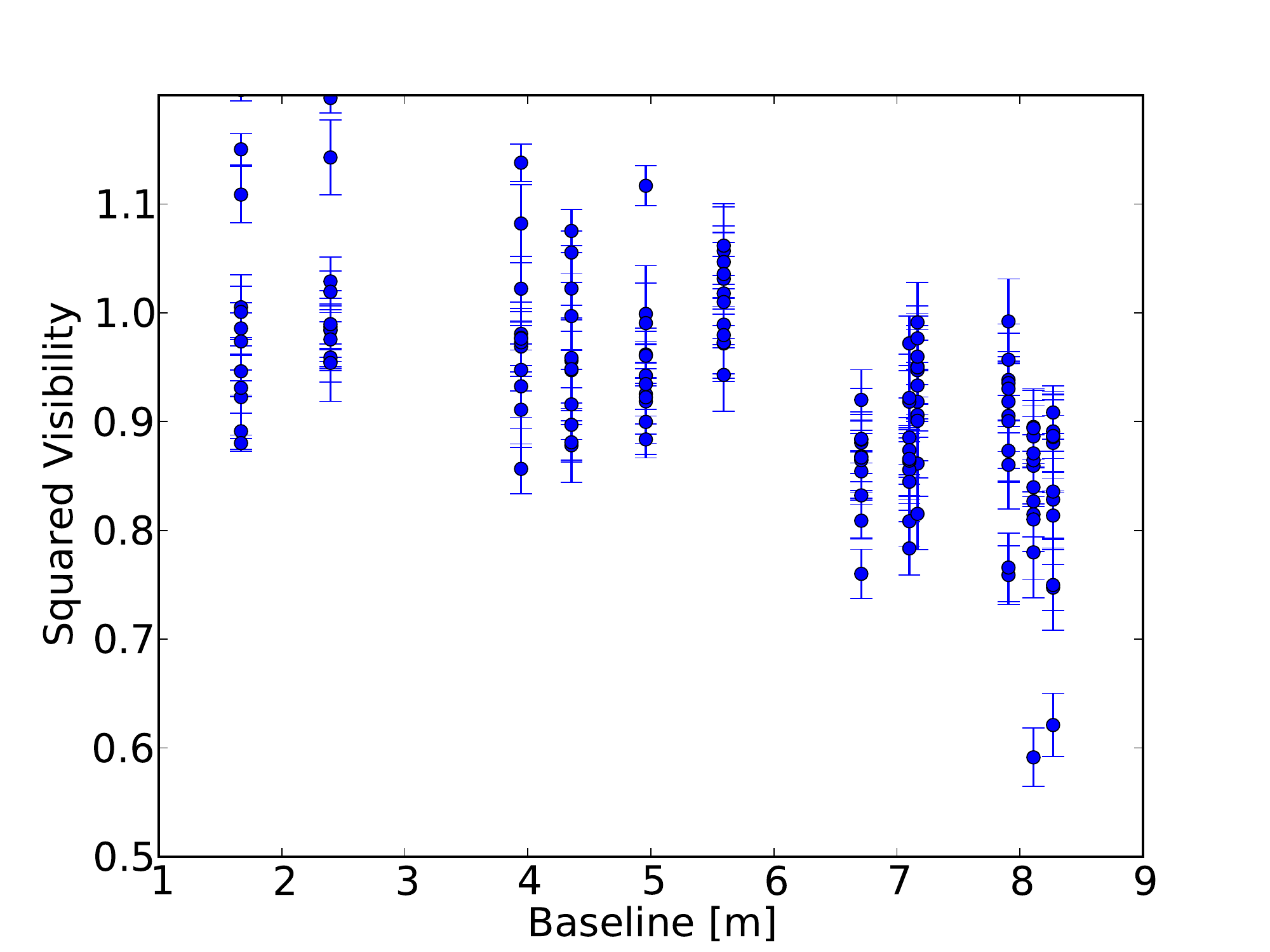}\\ 
    \includegraphics[height=4cm, angle=0]{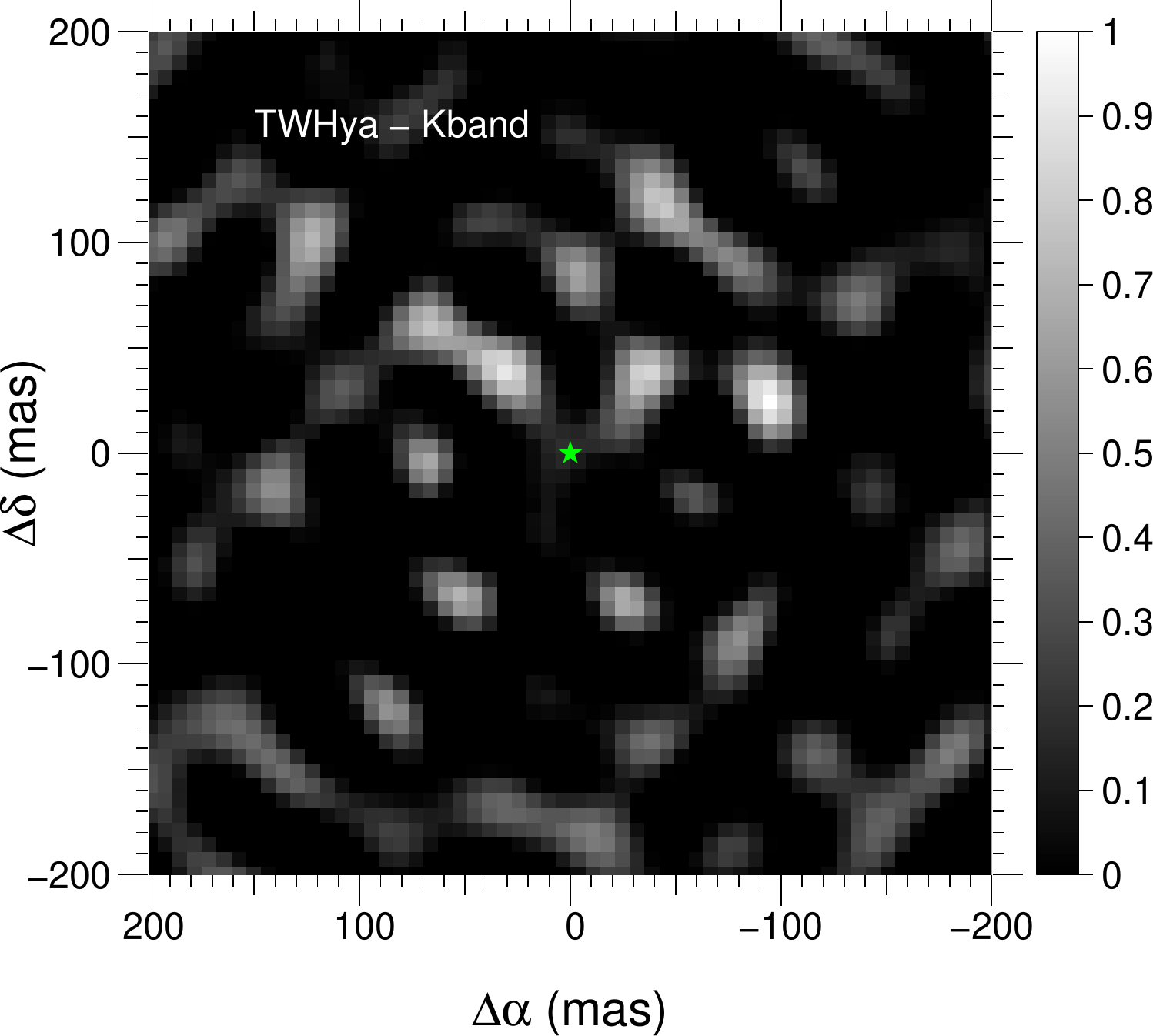} & \includegraphics[height=4cm, angle=0]{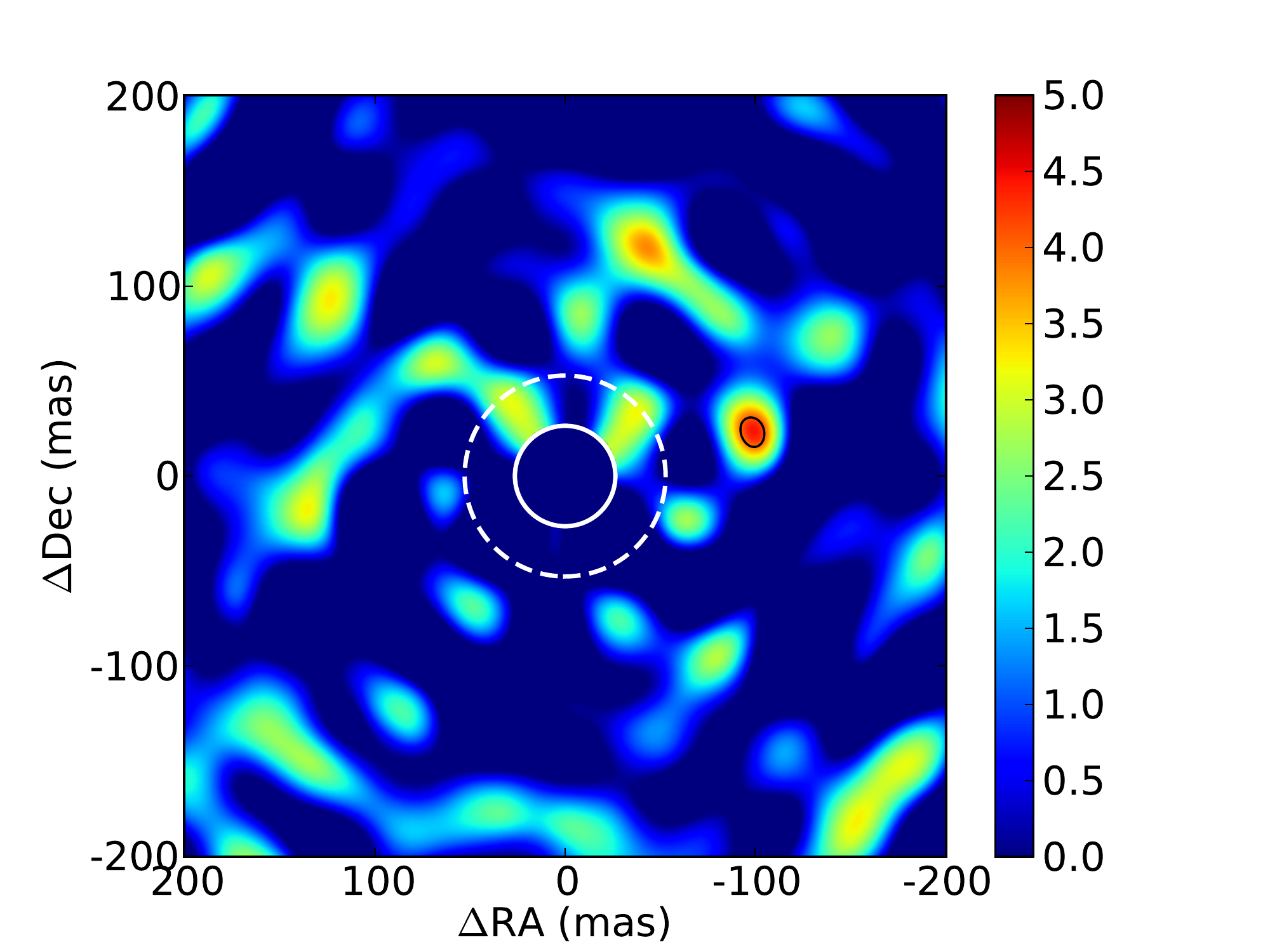} & \includegraphics[height=4cm, angle=0]{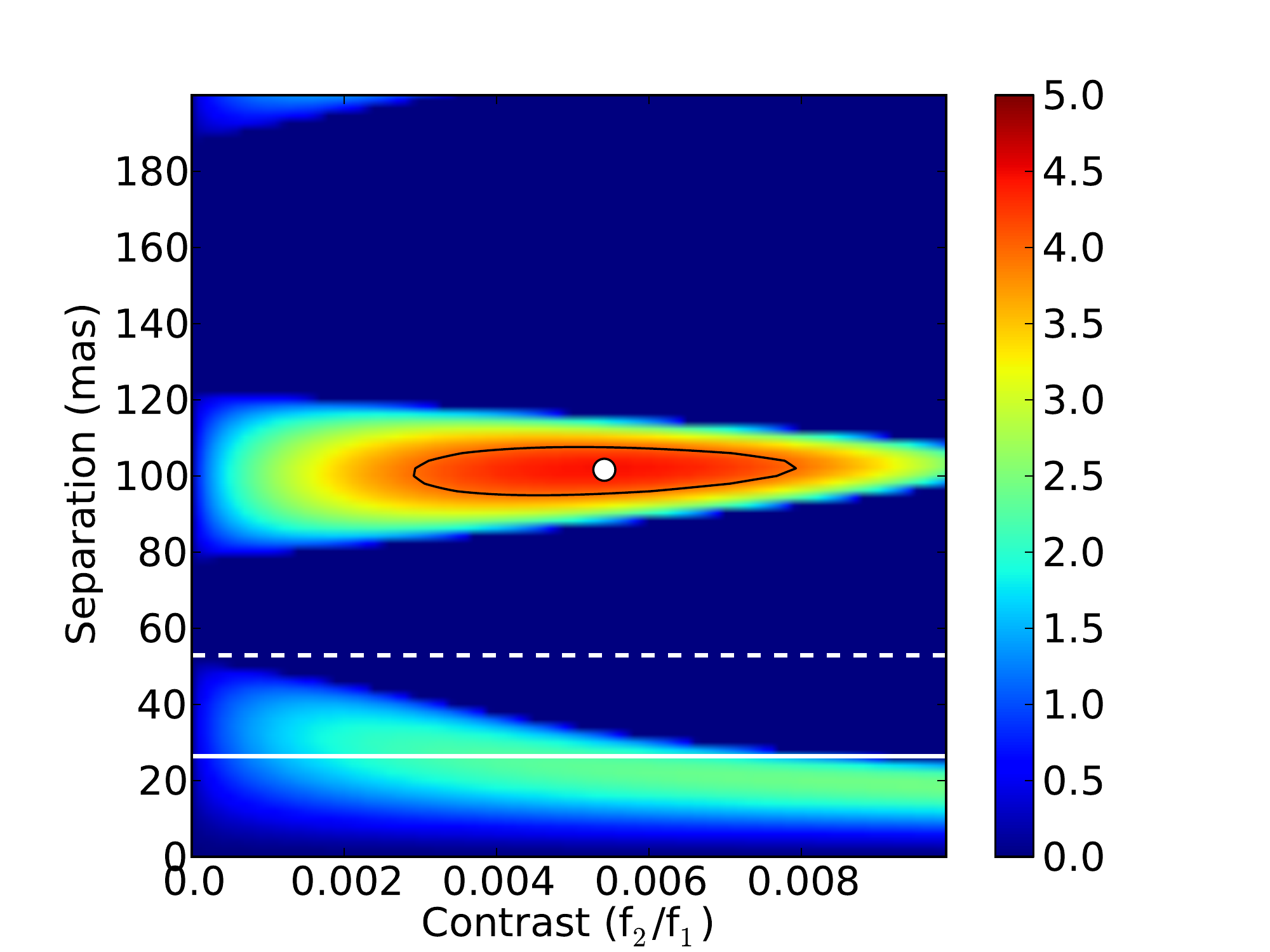} &
    \includegraphics[height=4cm, angle=0]{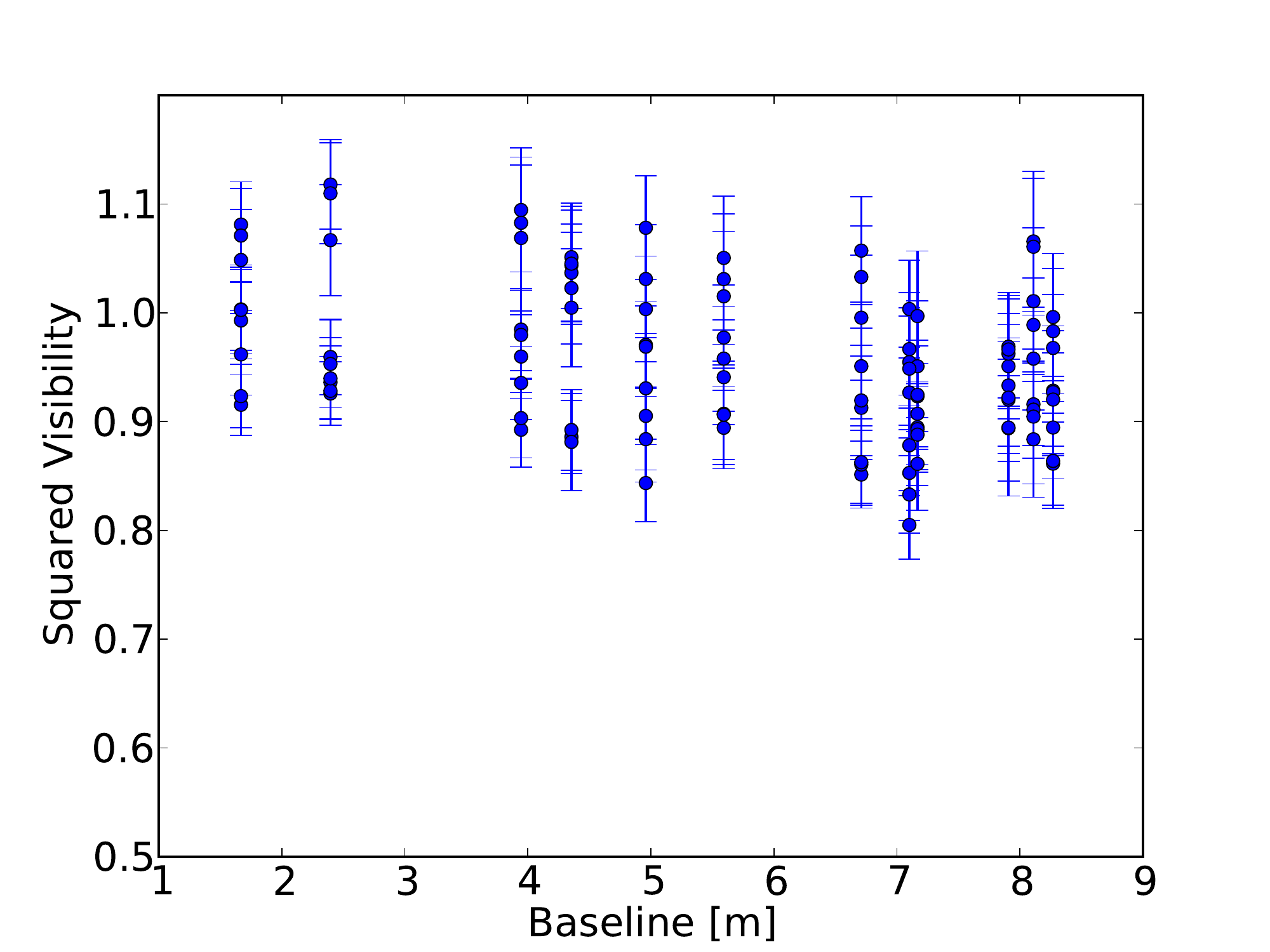}\\
    \end{array}$ 
\end{center}
\caption{Potential Candidate Detections: \textbf{Left:}  Reconstructed Image. \textbf{Middle Left:} Computed significance map. \textbf{Middle Right:} Degeneracy plot \textbf{Right:} V$^2$ plots. \textbf{First row:}DM Tau, K-band, \textbf{Second row:}, LkH$\alpha$330, K'-band, \textbf{Third row:}, TW\,Hya, K'-band}
\label{fig:PotCompanions}
\end{figure*}

\begin{figure*}
 \begin{center}
\scriptsize
$\begin{array}{ @{\hspace{-1.0mm}} c @{\hspace{-4.7mm}} c @{\hspace{-5.0mm}} c }
   \includegraphics[height=4cm, angle=0]{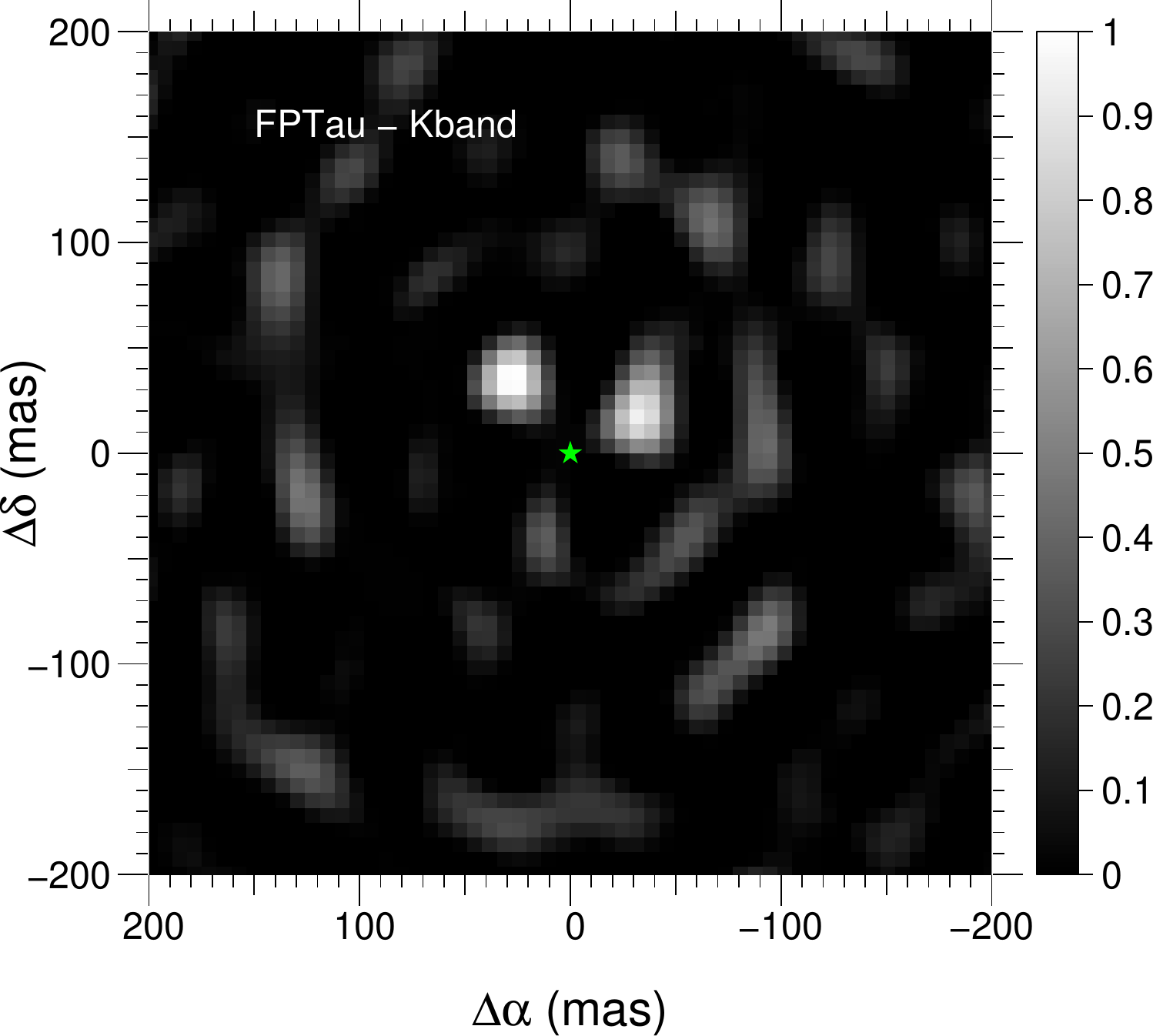} & \includegraphics[height=4cm, angle=0]{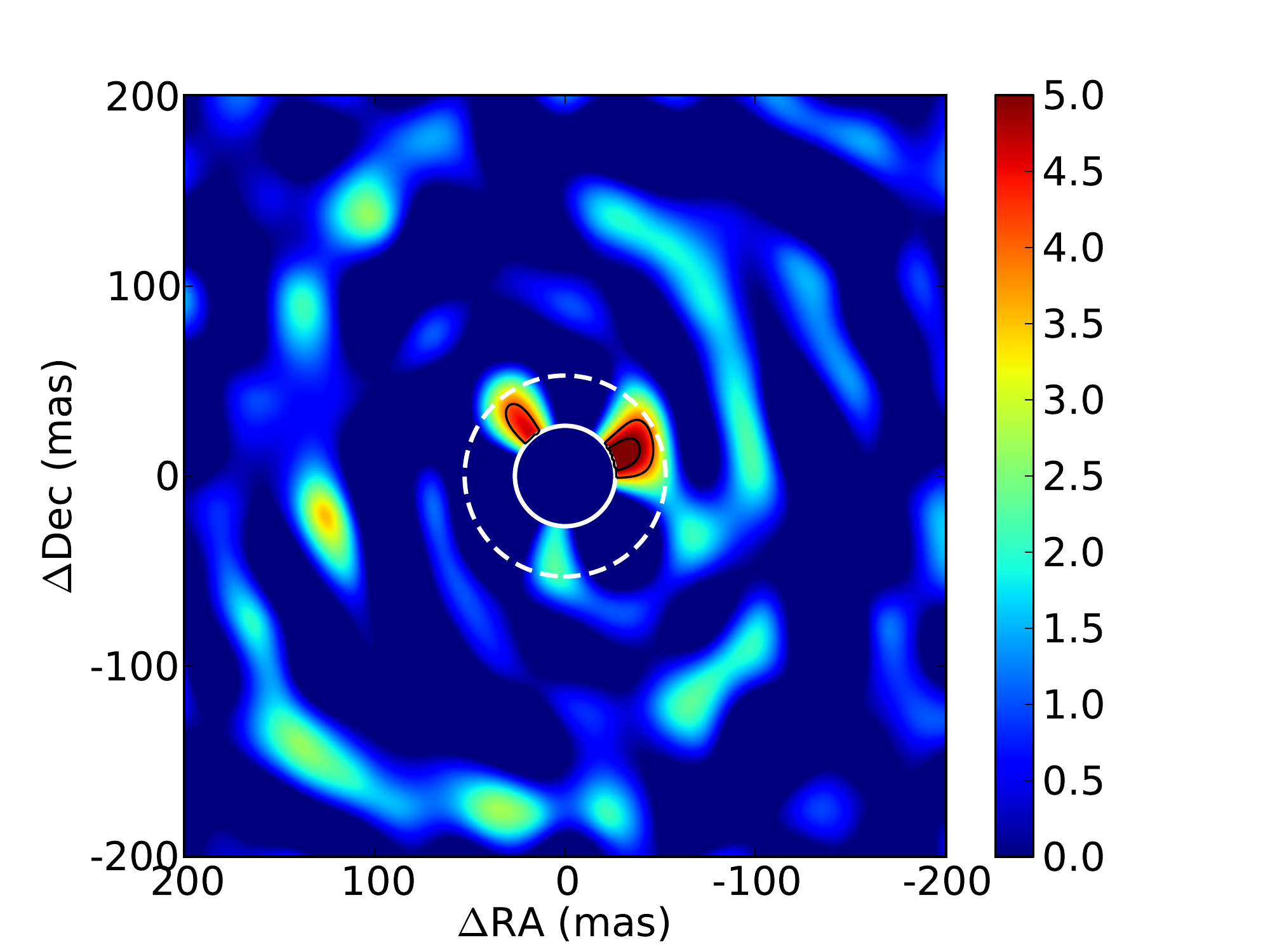} &
   \includegraphics[height=4cm, angle=0]{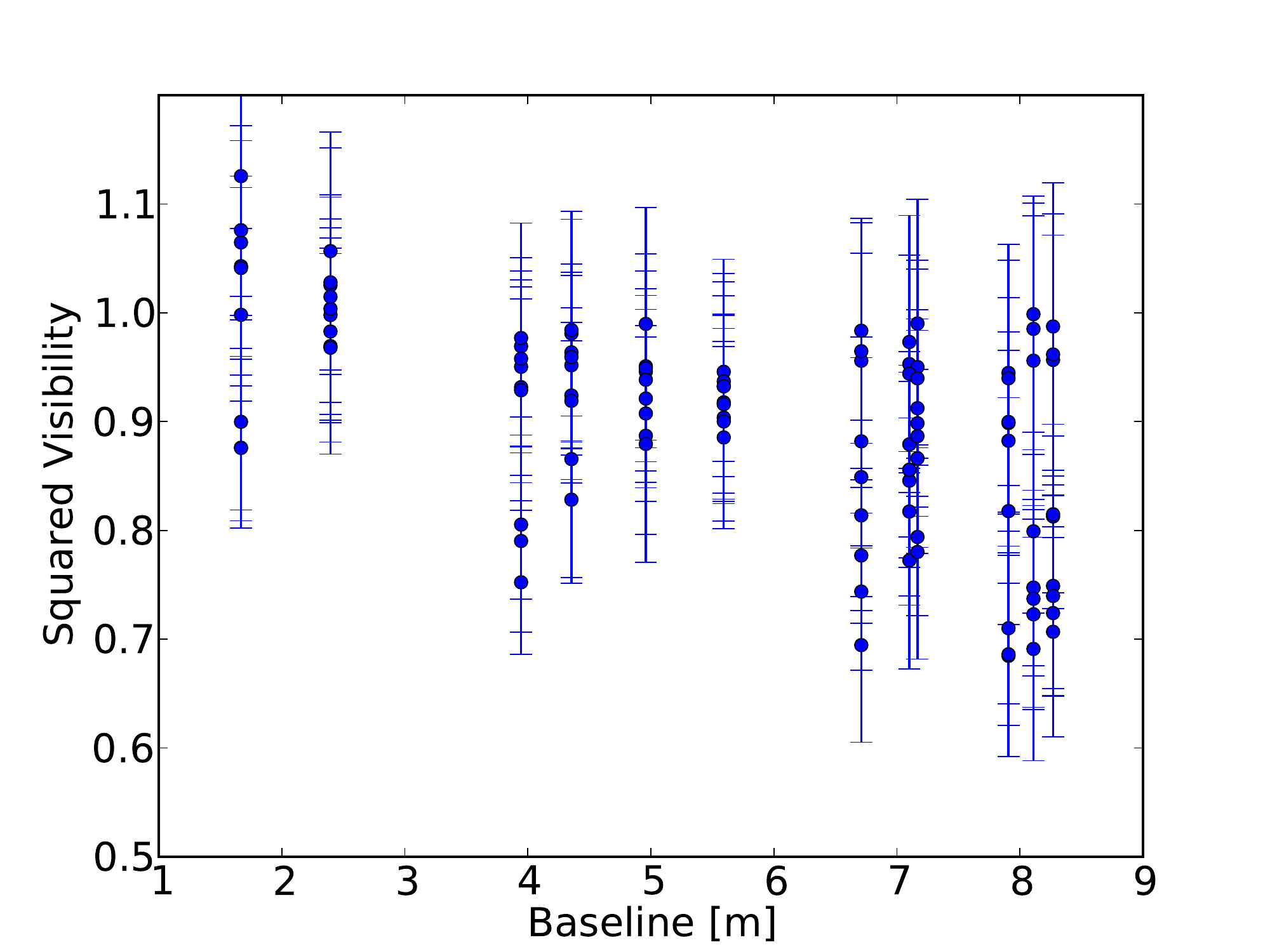}\\
        \end{array}$ 
\end{center}
\caption{Potential disc feature detections \textbf{Left:}  Reconstructed Image \textbf{Right:} Computed significance map \textbf{Right:} V$^2$ plot. FP Tau, K-band.}
\label{fig:PotDiskFeatures}
\end{figure*}

\begin{figure*}
 \begin{center}
\scriptsize
$\begin{array}{ @{\hspace{-1.0mm}} c @{\hspace{-4.7mm}} c @{\hspace{-5.0mm}} c }
    \includegraphics[height=4cm, angle=0]{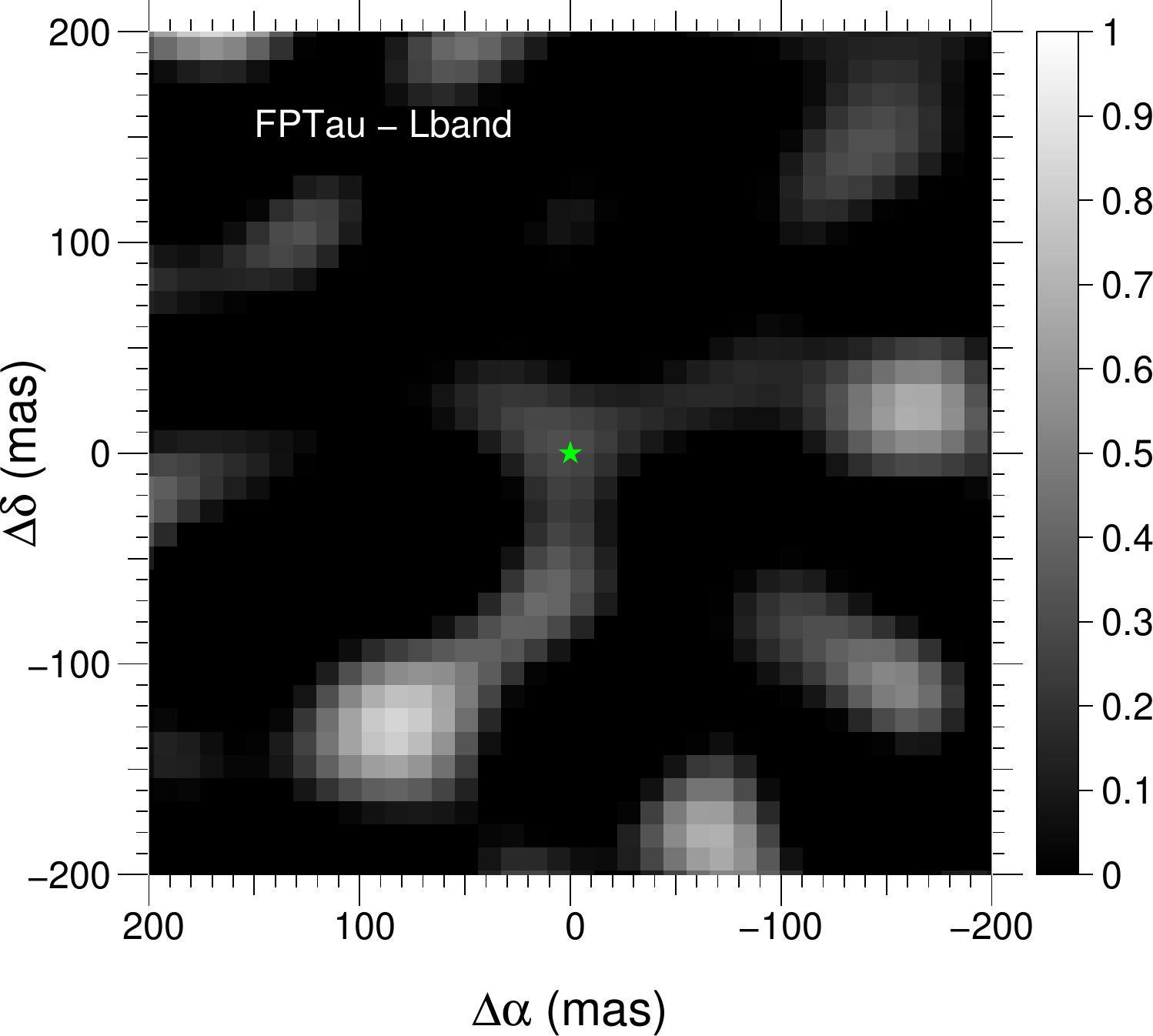} & \includegraphics[height=4cm, angle=0]{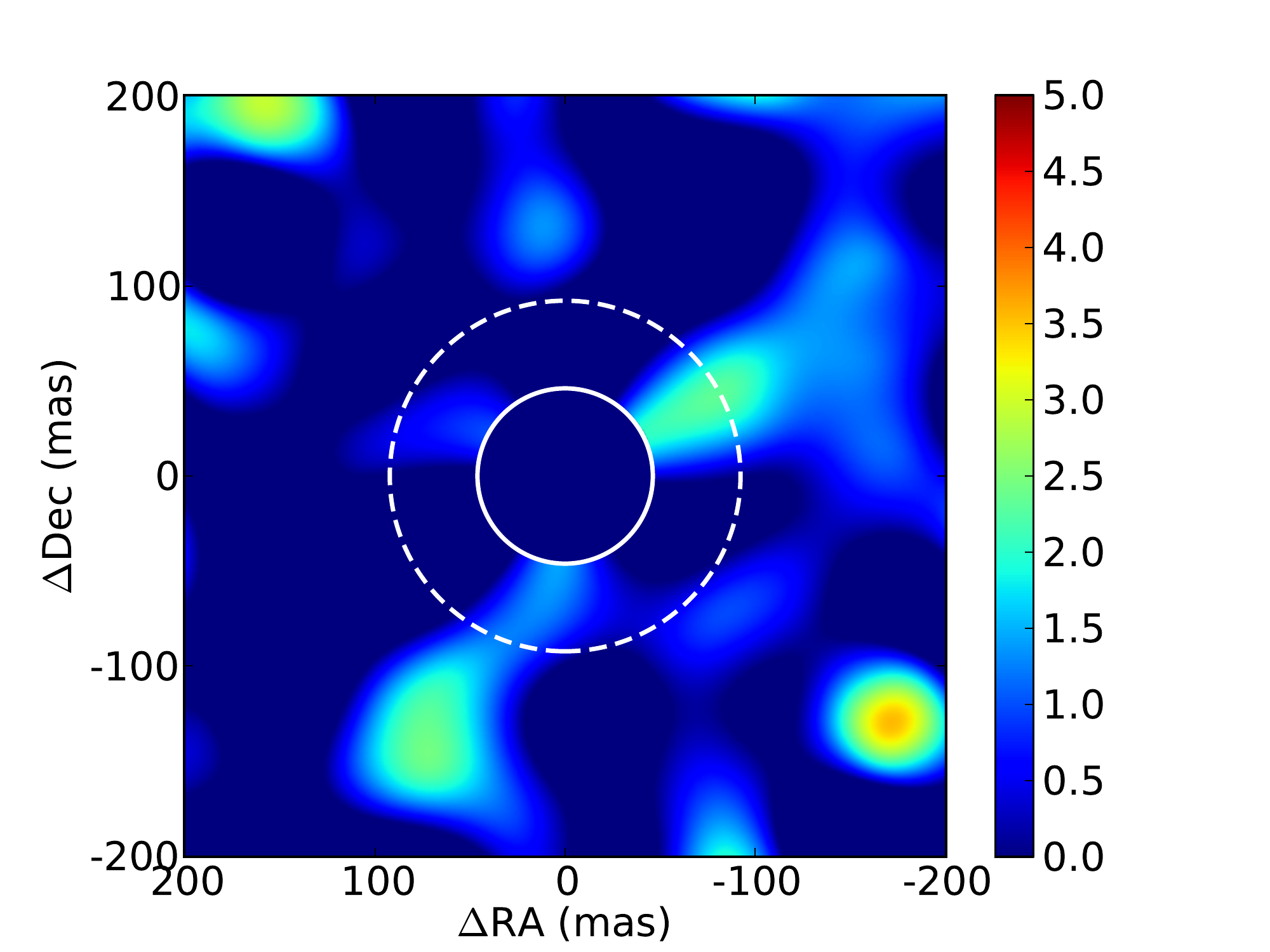} &
    \includegraphics[height=4cm, angle=0]{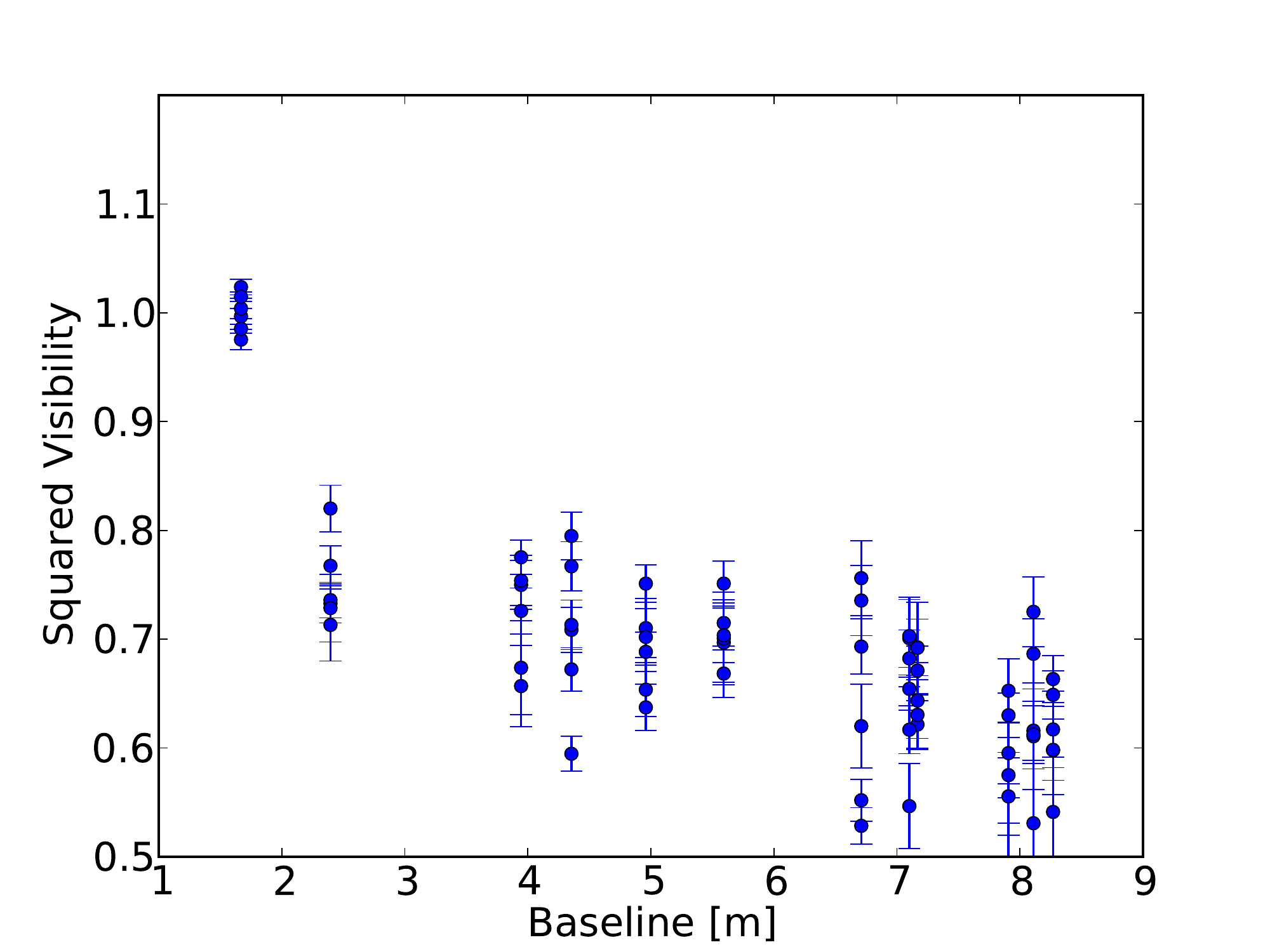}\\
    \includegraphics[height=4cm, angle=0]{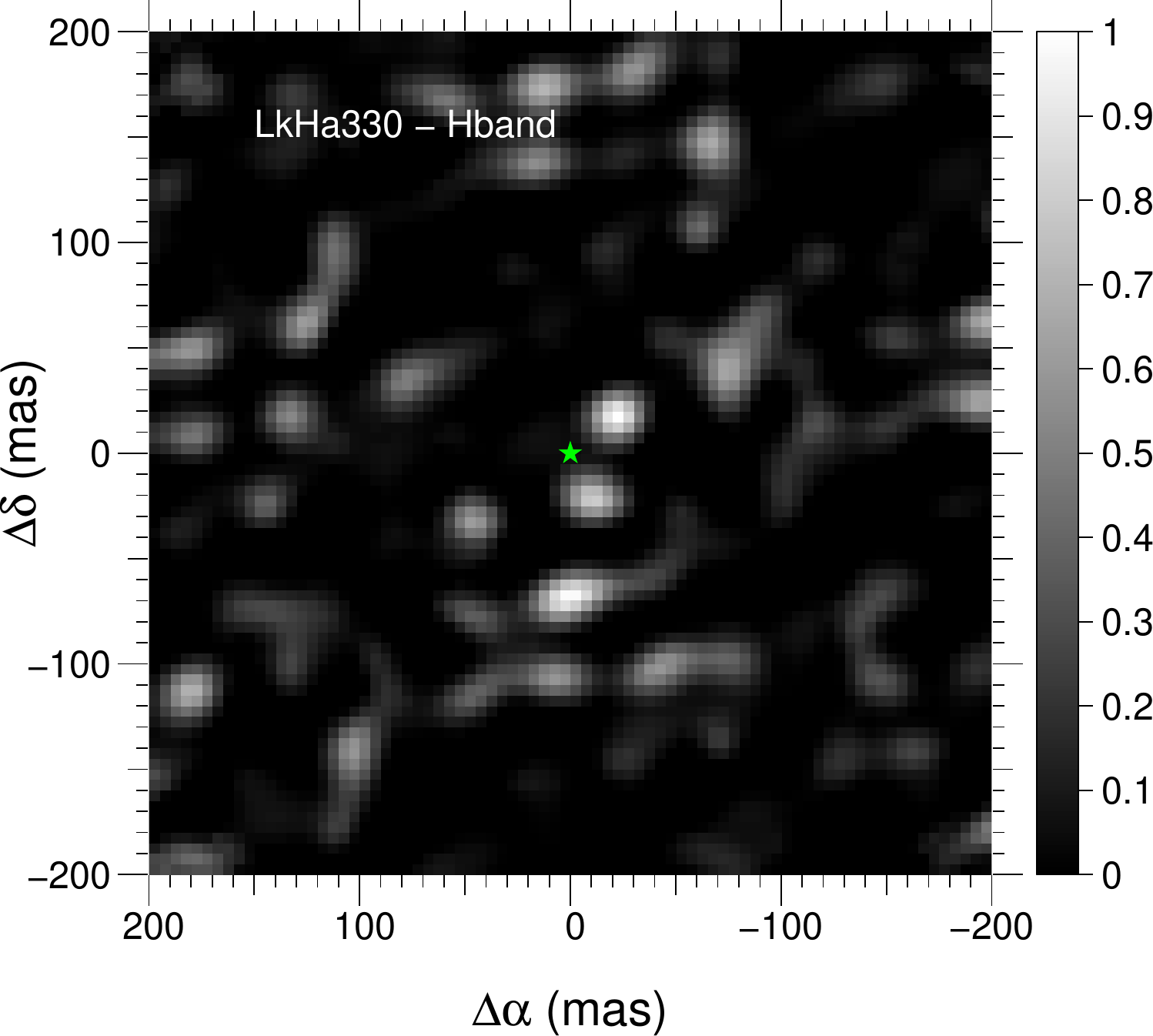} & \includegraphics[height=4cm, angle=0]{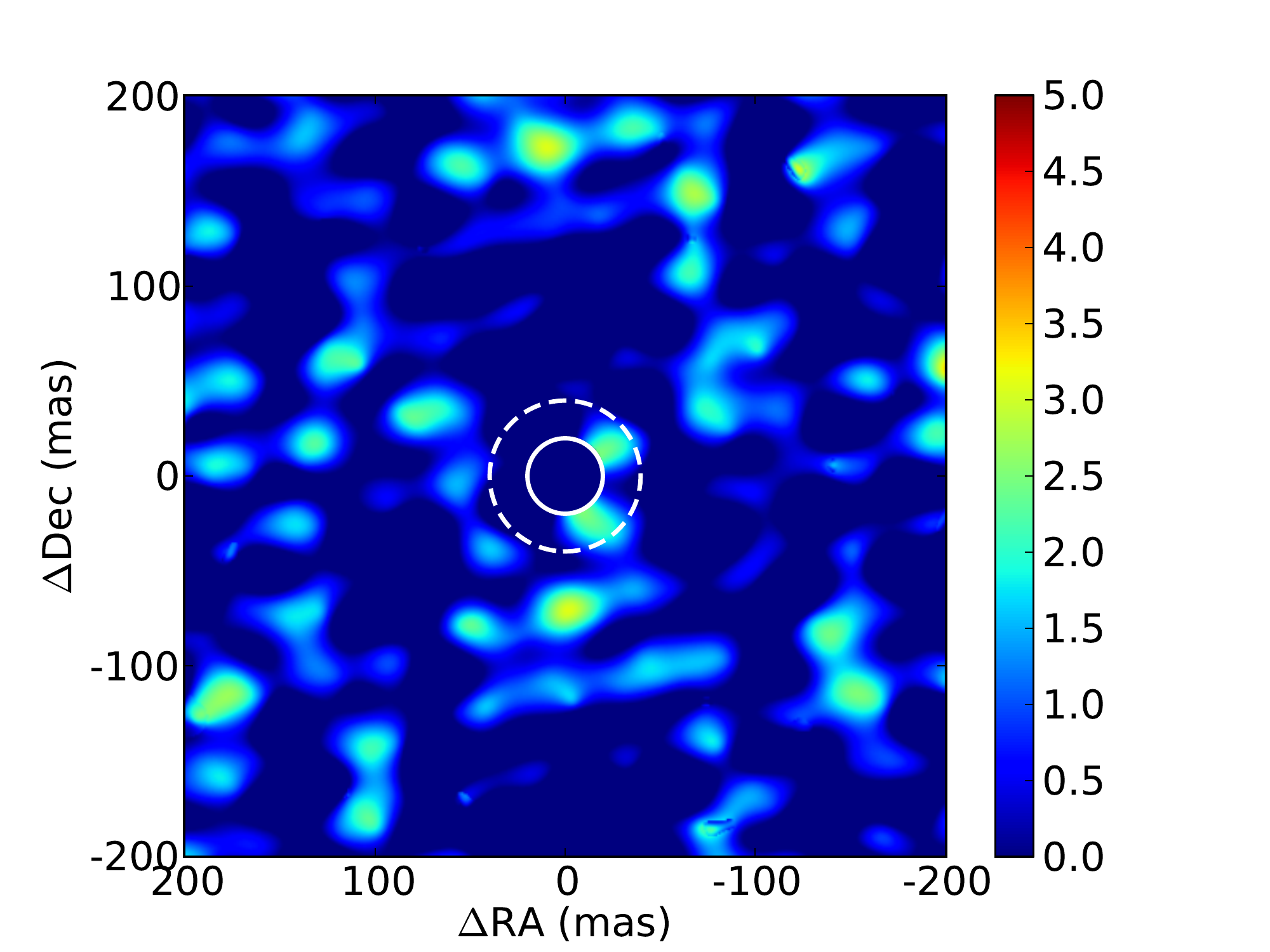} &
    \includegraphics[height=4cm, angle=0]{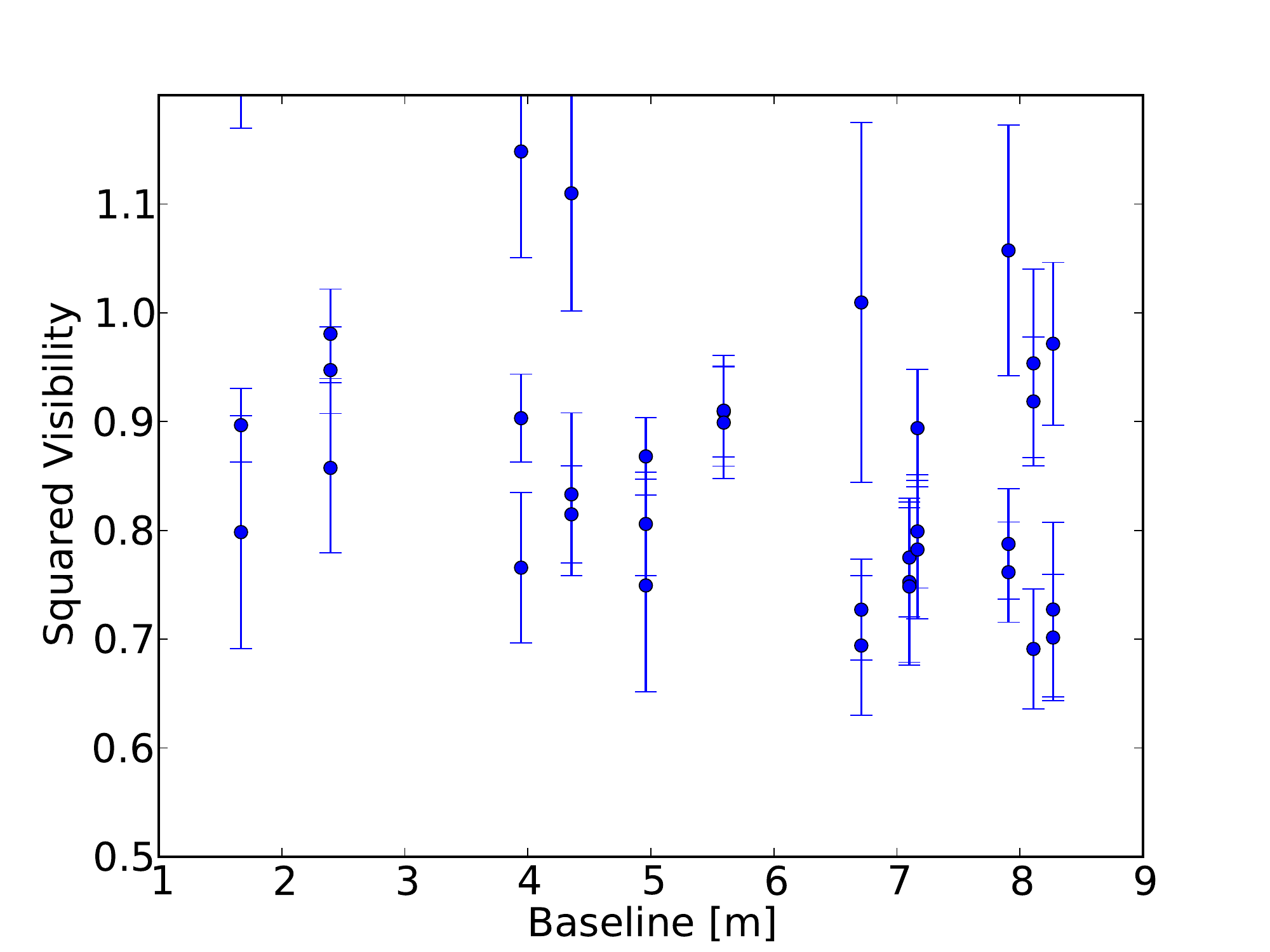}\\    
    \includegraphics[height=4cm, angle=0]{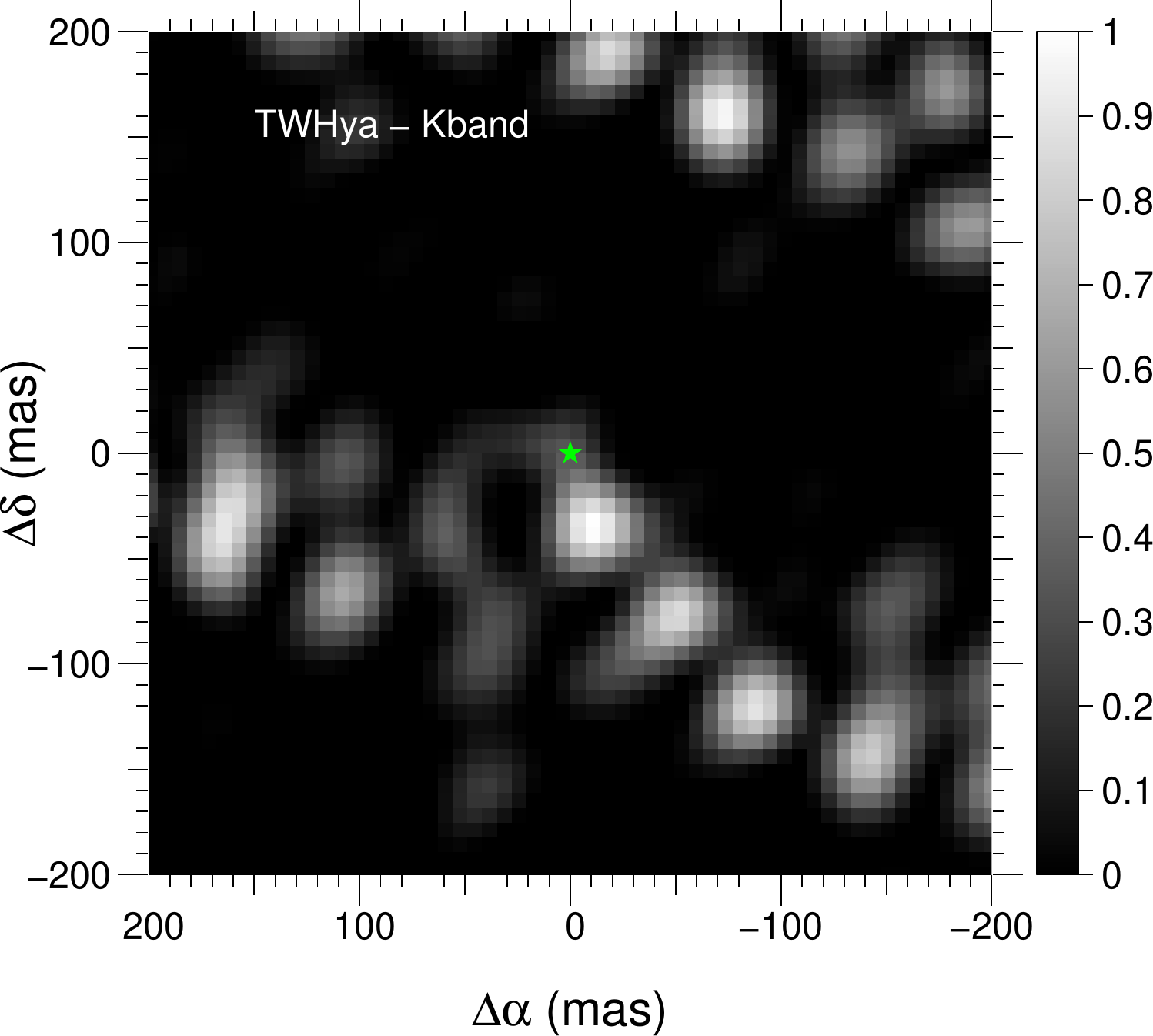} & 
    \includegraphics[height=4cm, angle=0]{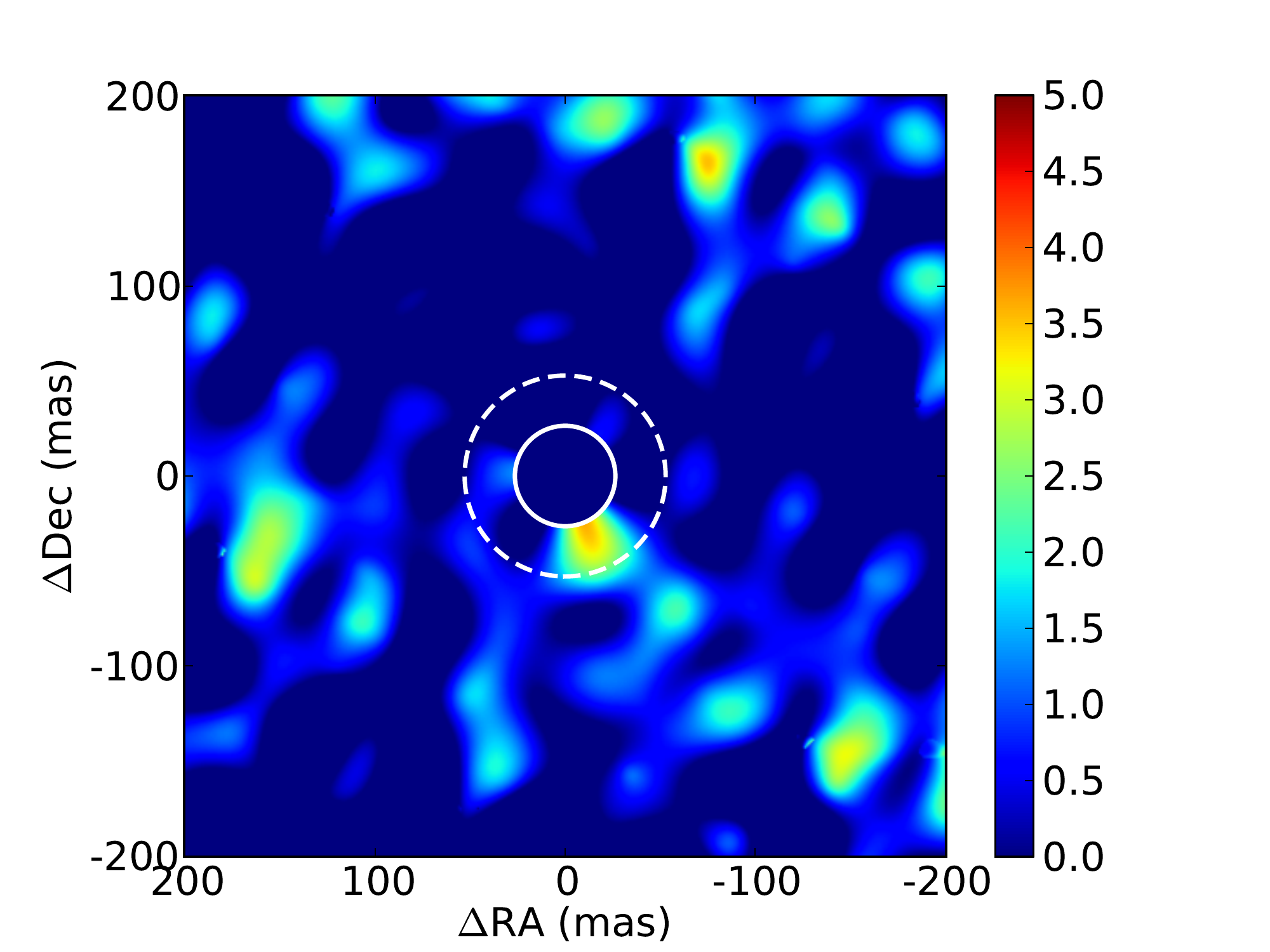} & 
    \includegraphics[height=4cm, angle=0]{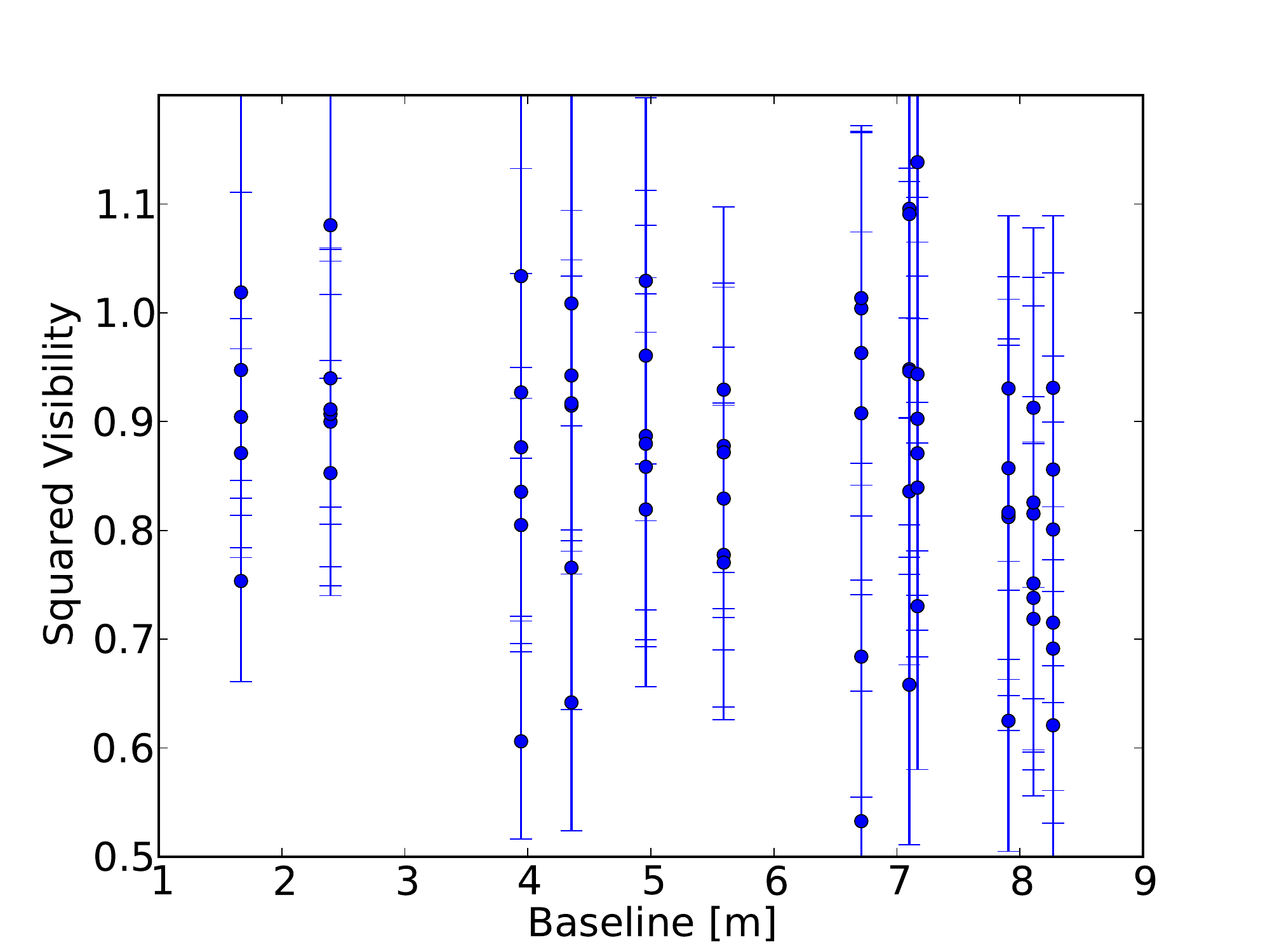}\\
    \includegraphics[height=4cm, angle=0]{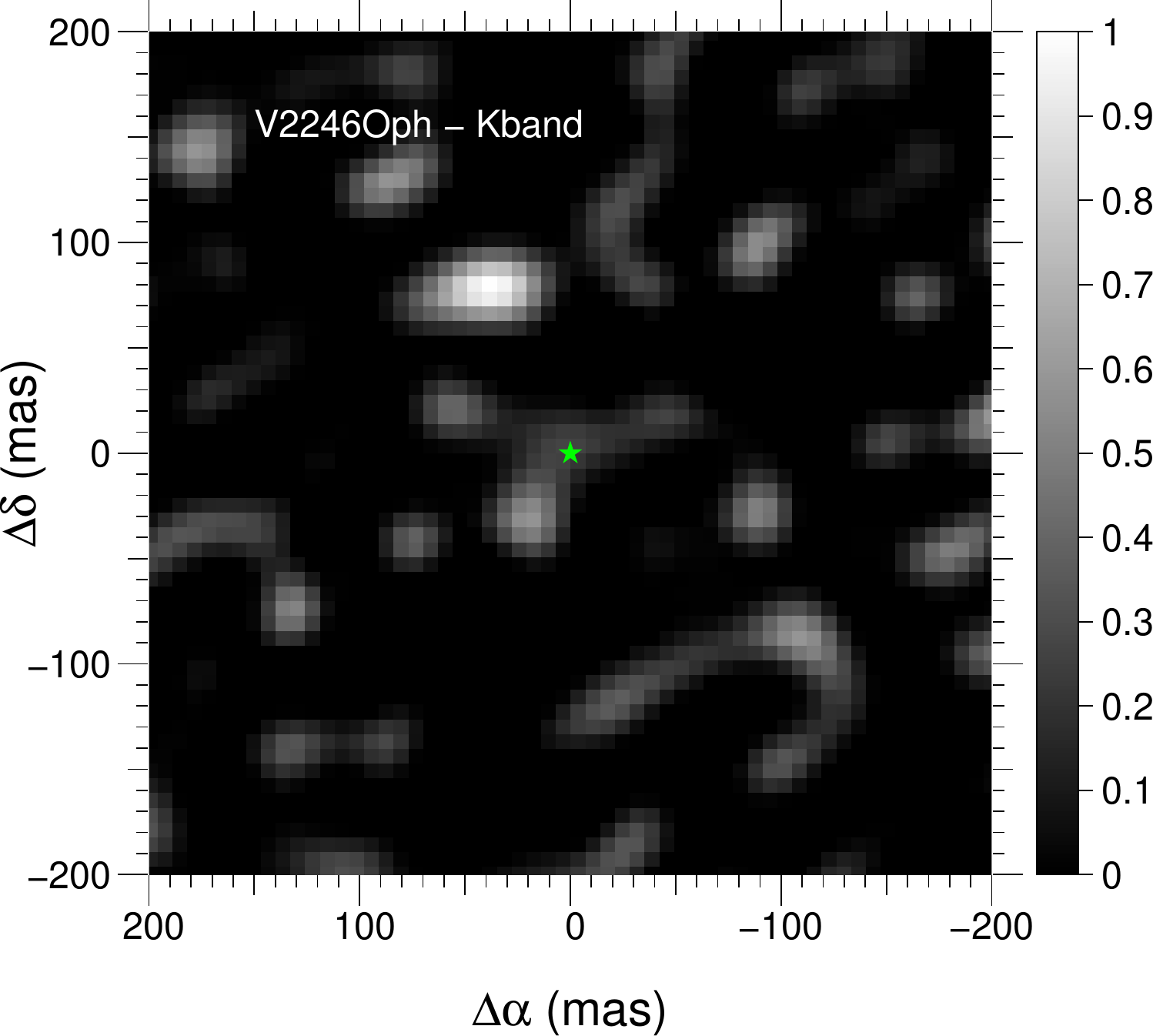} & \includegraphics[height=4cm, angle=0]{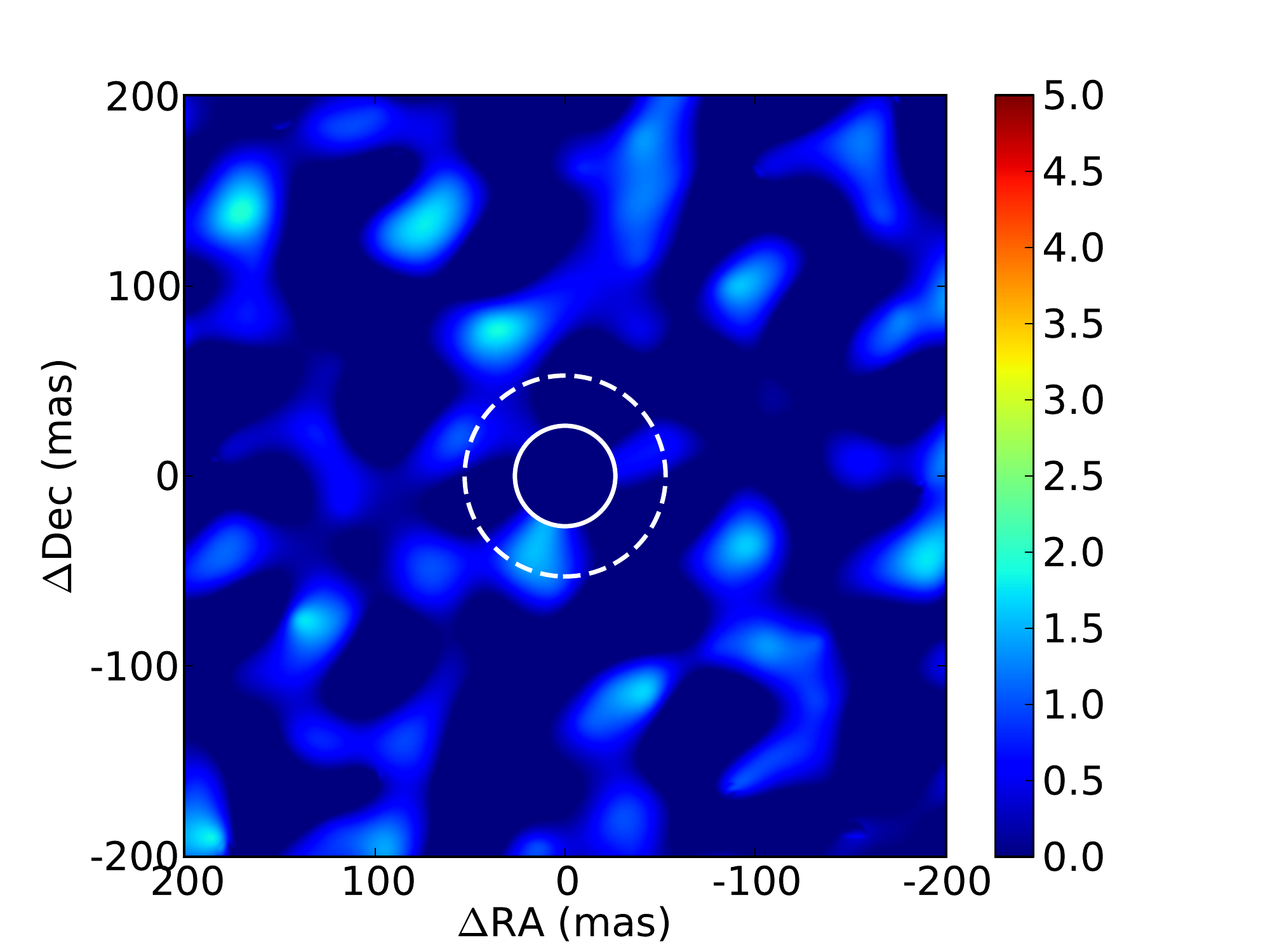} &
    \includegraphics[height=4cm, angle=0]{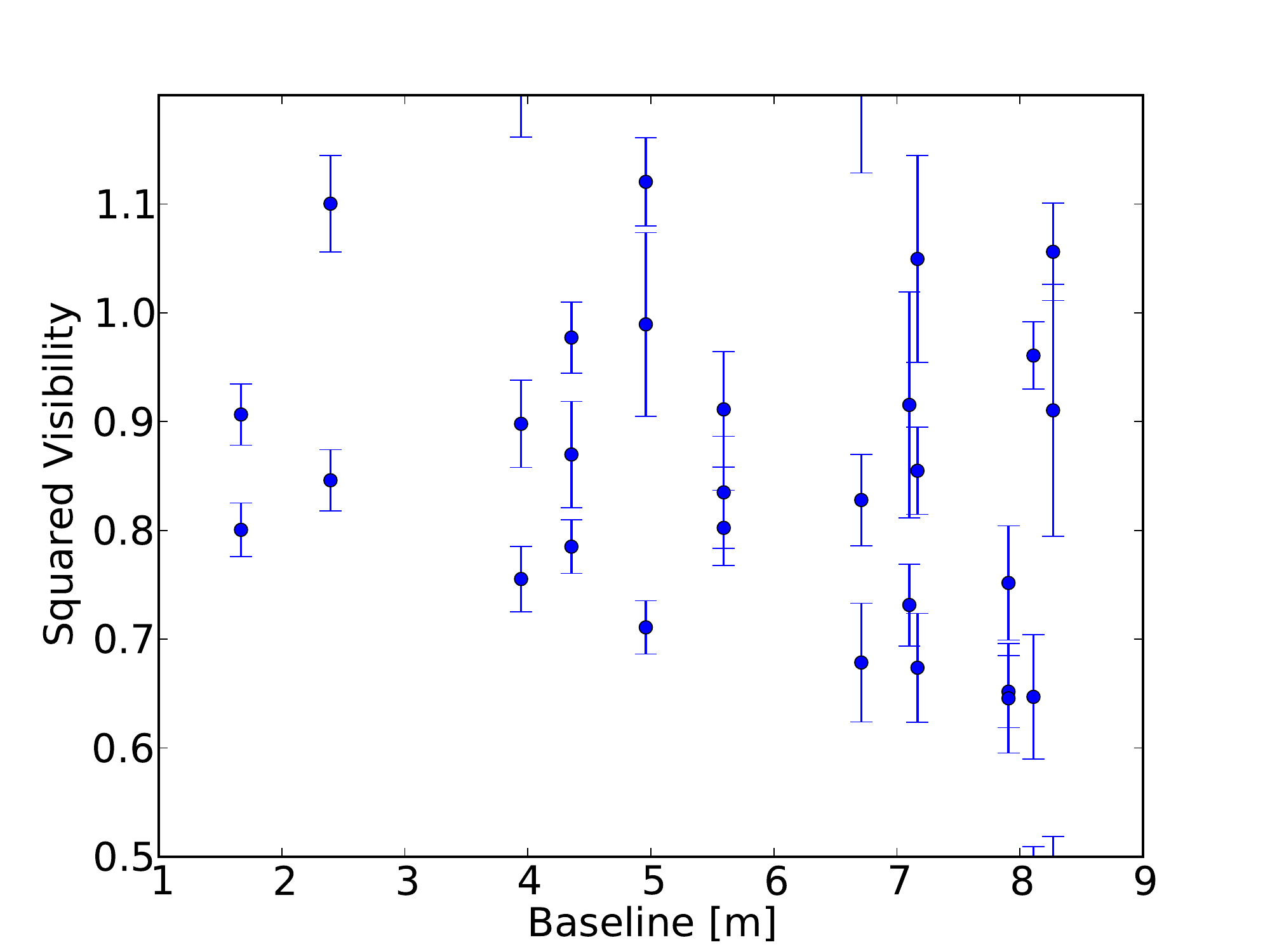}\\ 
 \end{array}$
 
\end{center}
\caption{Data sets where we see no significant emission \textbf{Left:}  Reconstructed Images \textbf{Middle:} Computed significance maps \textbf{Right:} V$^2$ plot. \textbf{First row:}FP Tau, L-band, \textbf{Second row:} LkH$\alpha$330, H-band, \textbf{Third row:}, TW\,Hya, K'-band, \textbf{Fourth row:}, V2246 Oph, K'-band. In these cases we set limits on the contrast of a potential candidate or disc feature.}
\label{fig:NonDetections}
\end{figure*}

\begin{figure*}
 \begin{center}
\scriptsize
$\begin{array}{ @{\hspace{-1.0mm}} c @{\hspace{-4.7mm}} c @{\hspace{-5.0mm}} c}
    \includegraphics[height=4cm, angle=0]{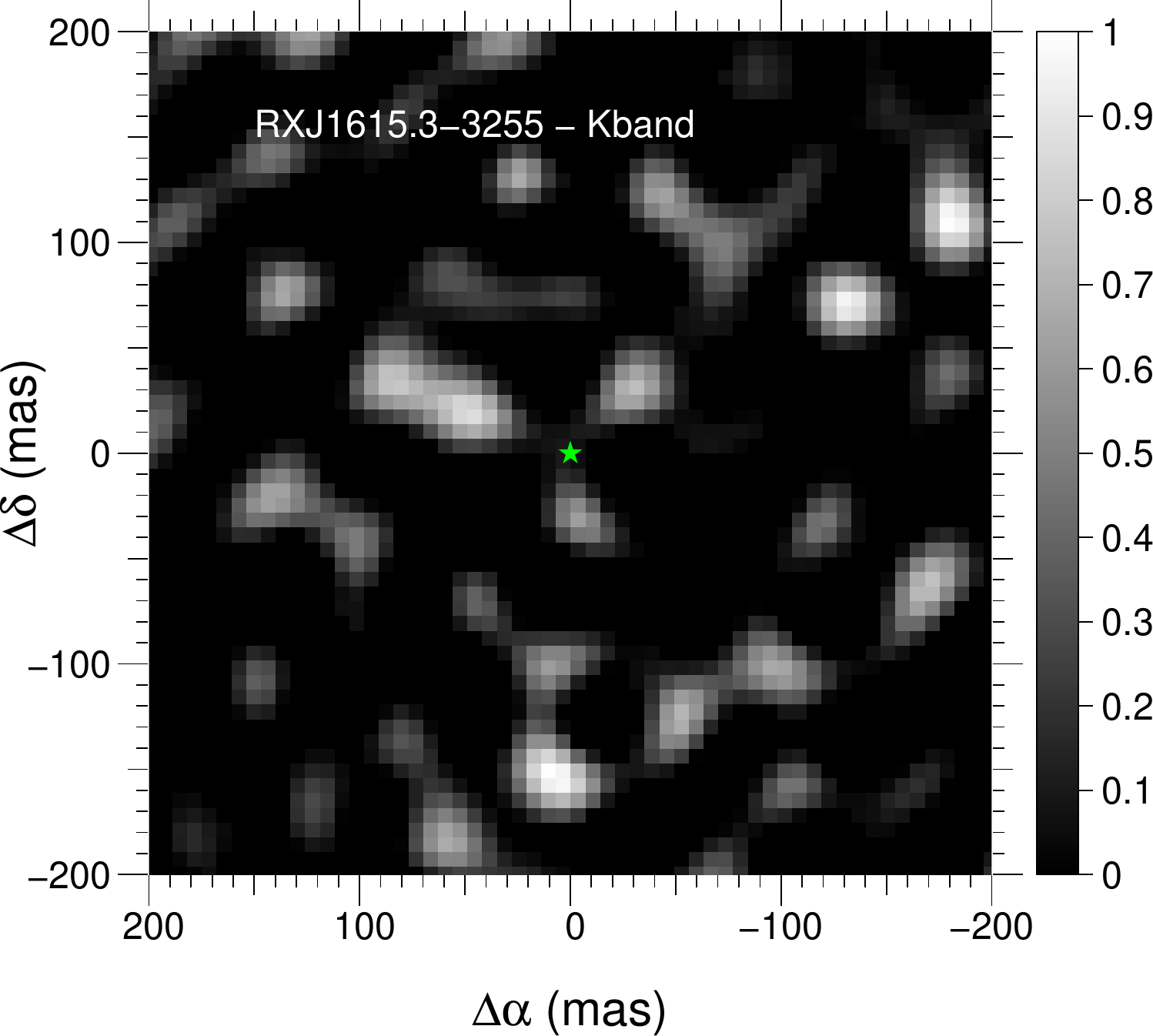} & \includegraphics[height=4cm, angle=0]{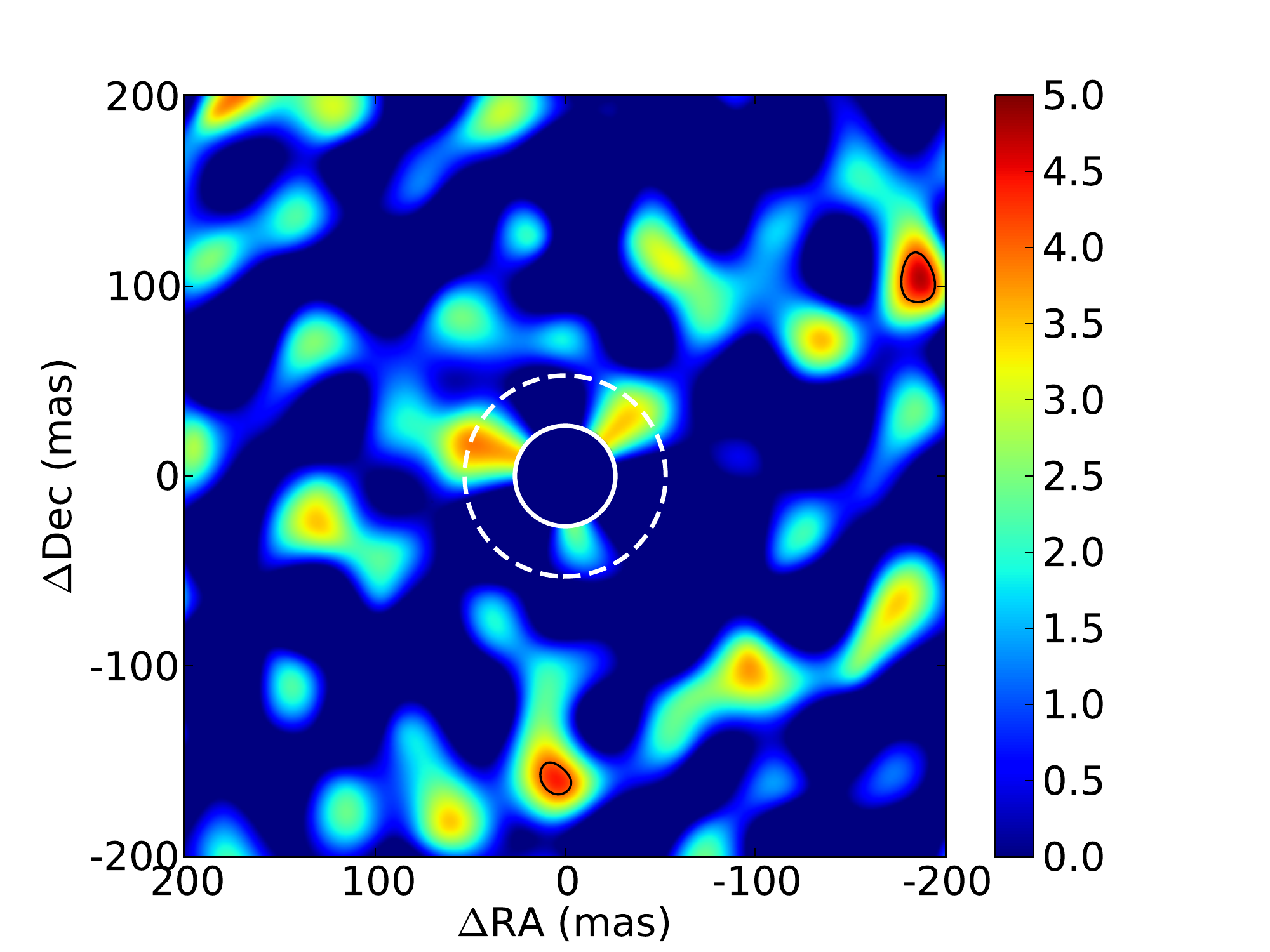} &
    \includegraphics[height=4cm, angle=0]{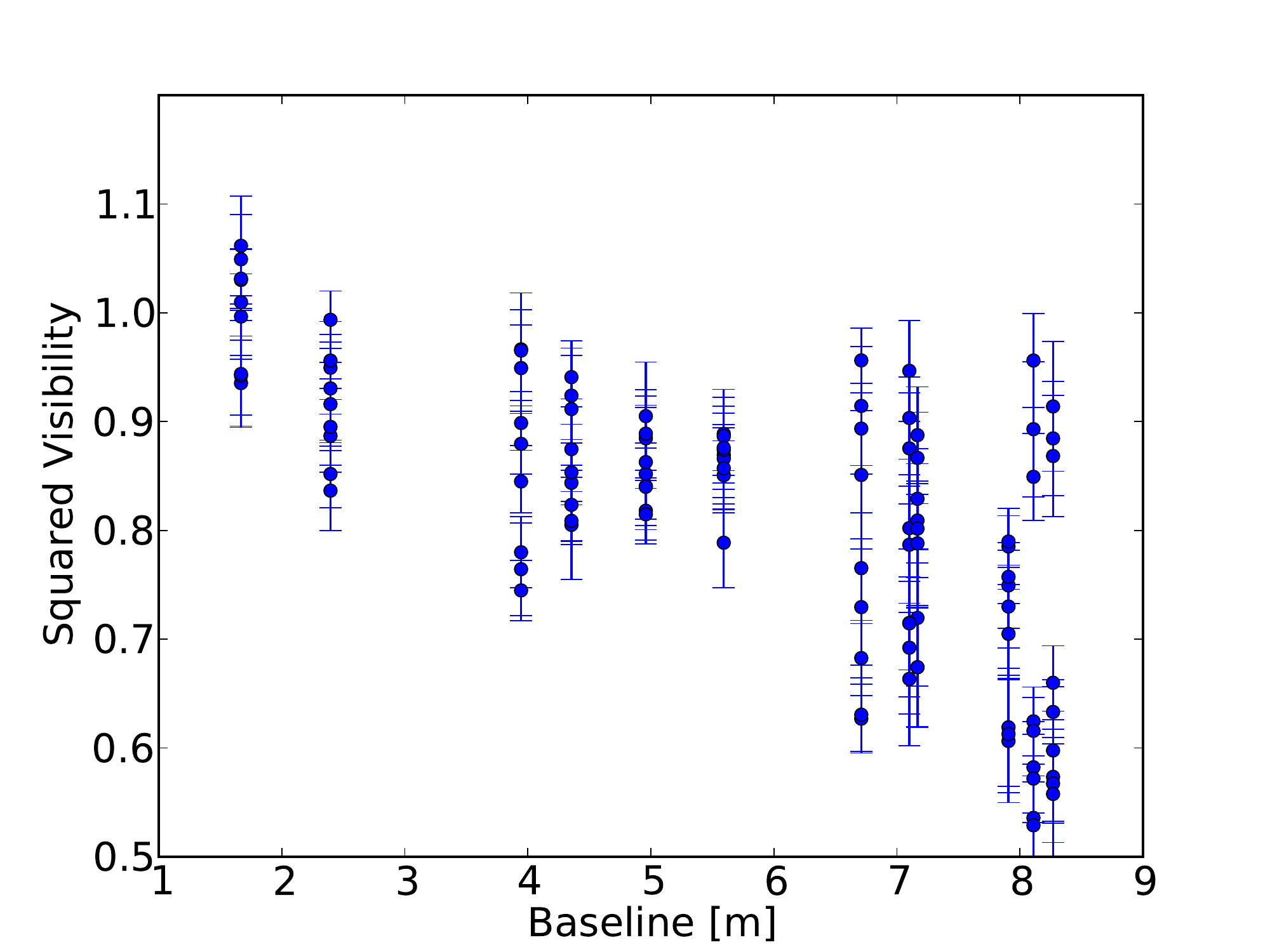}\\     
    \includegraphics[height=4cm, angle=0]{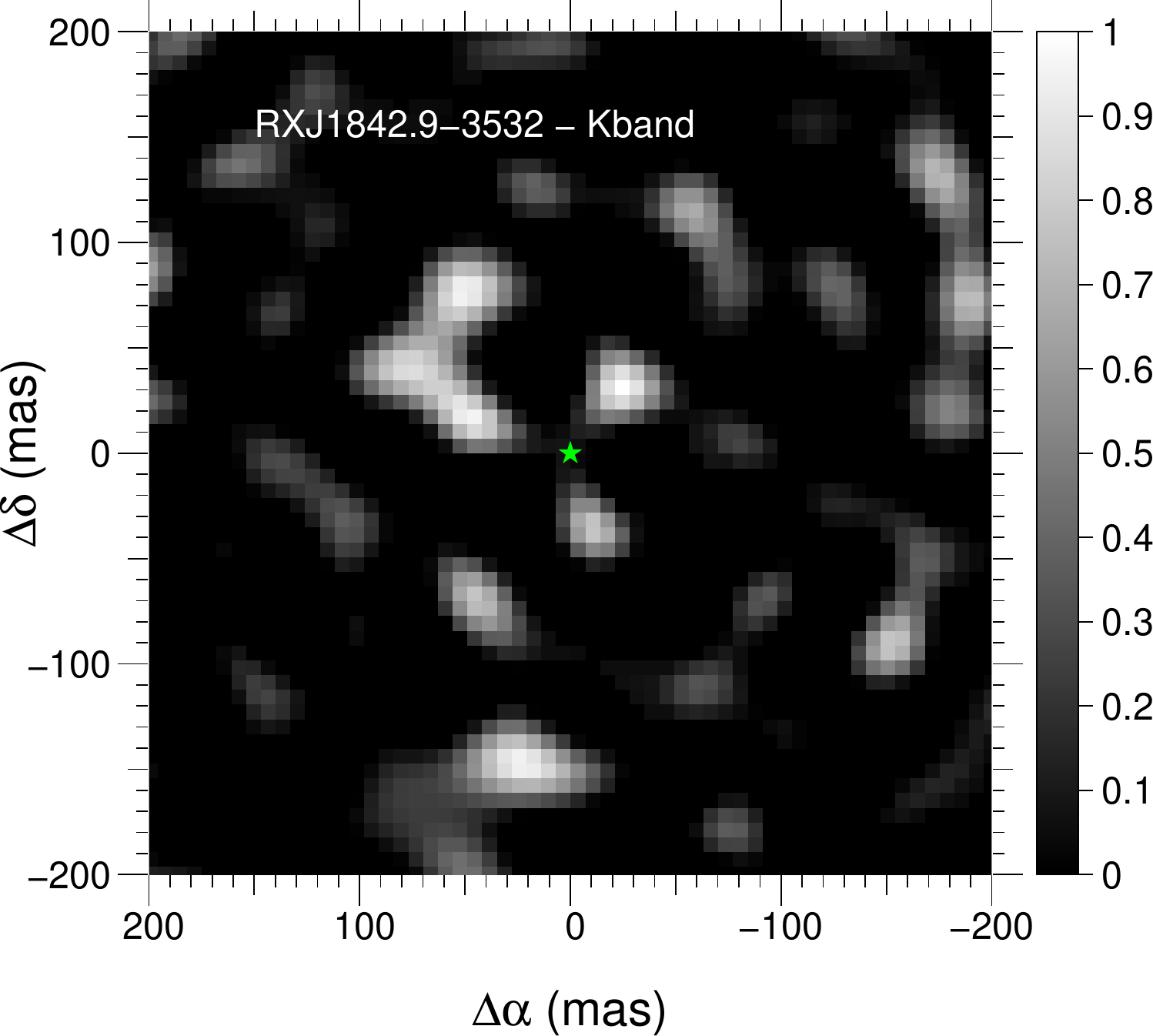} & 
    \includegraphics[height=4cm, angle=0]{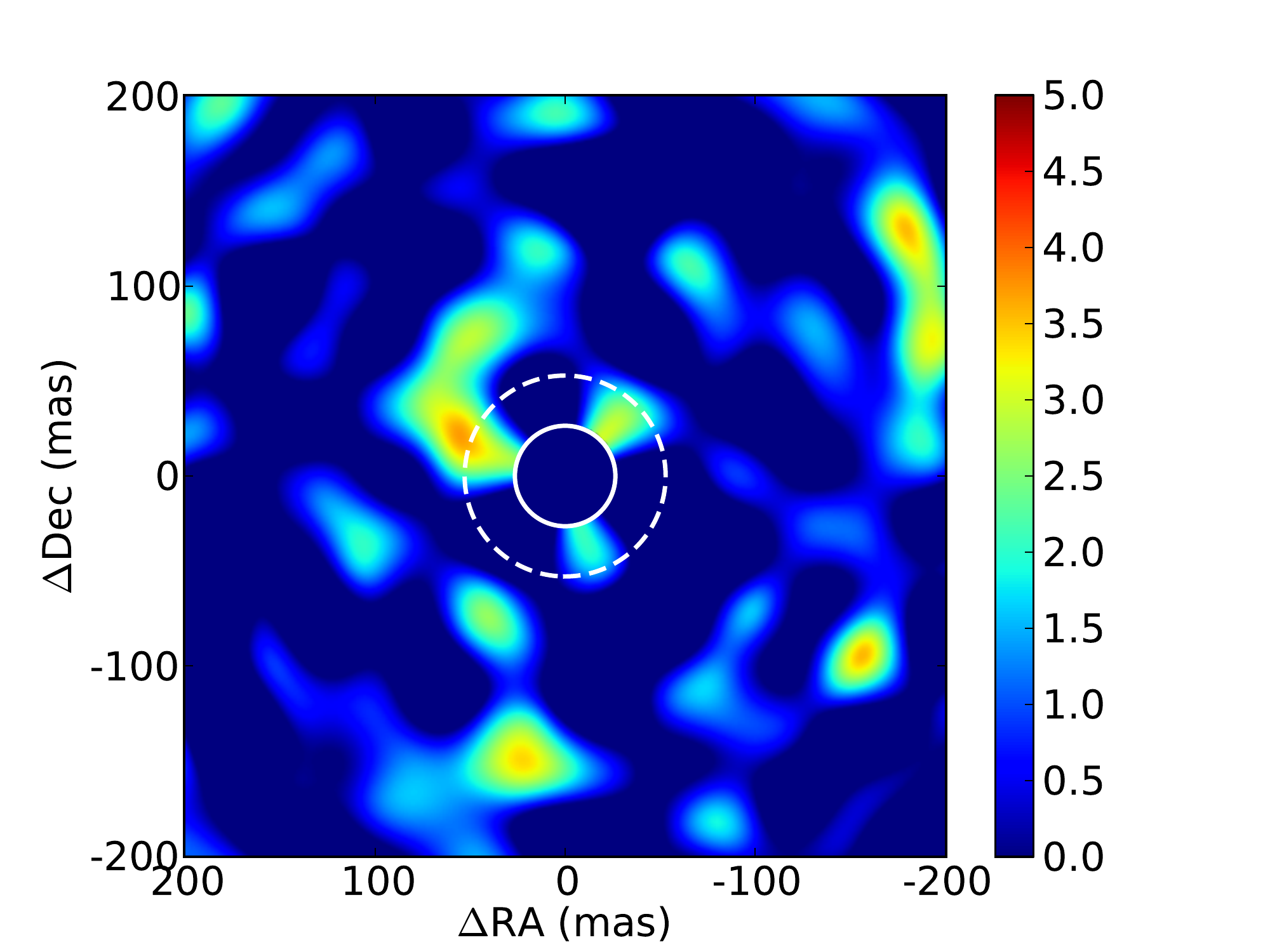} &
     \includegraphics[height=4cm, angle=0]{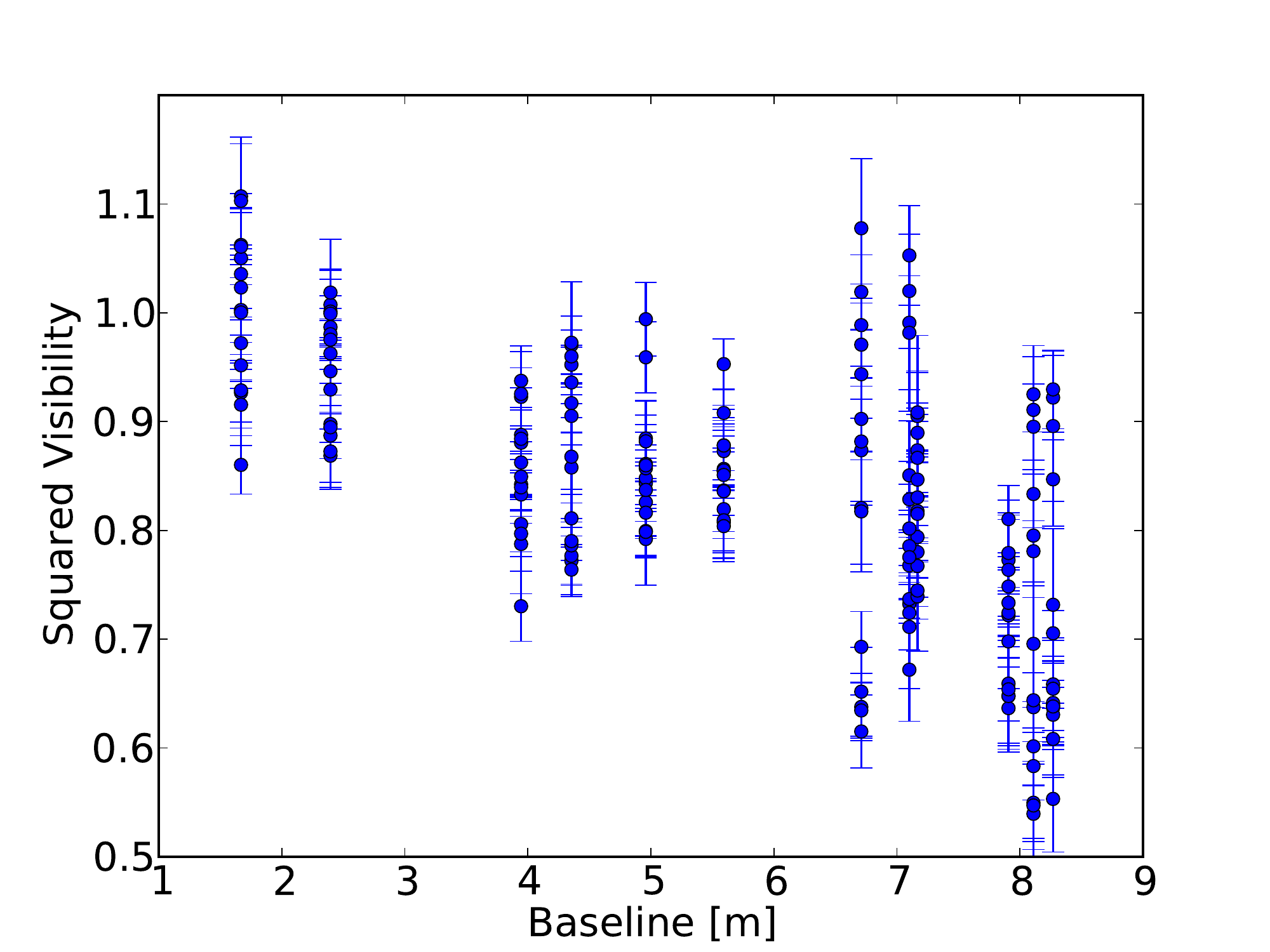}\\
    \includegraphics[height=4cm, angle=0]{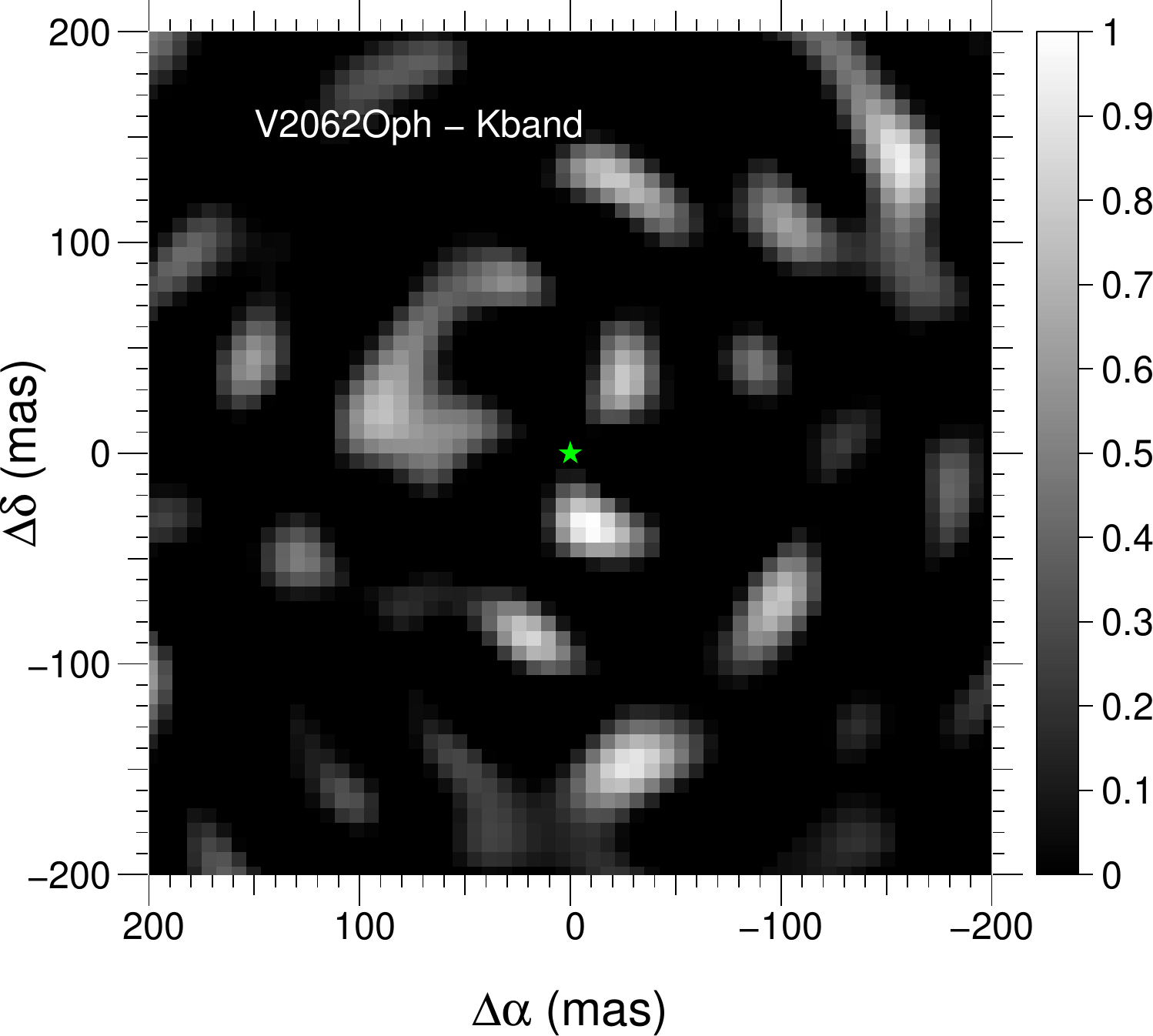} & 
    \includegraphics[height=4cm, angle=0]{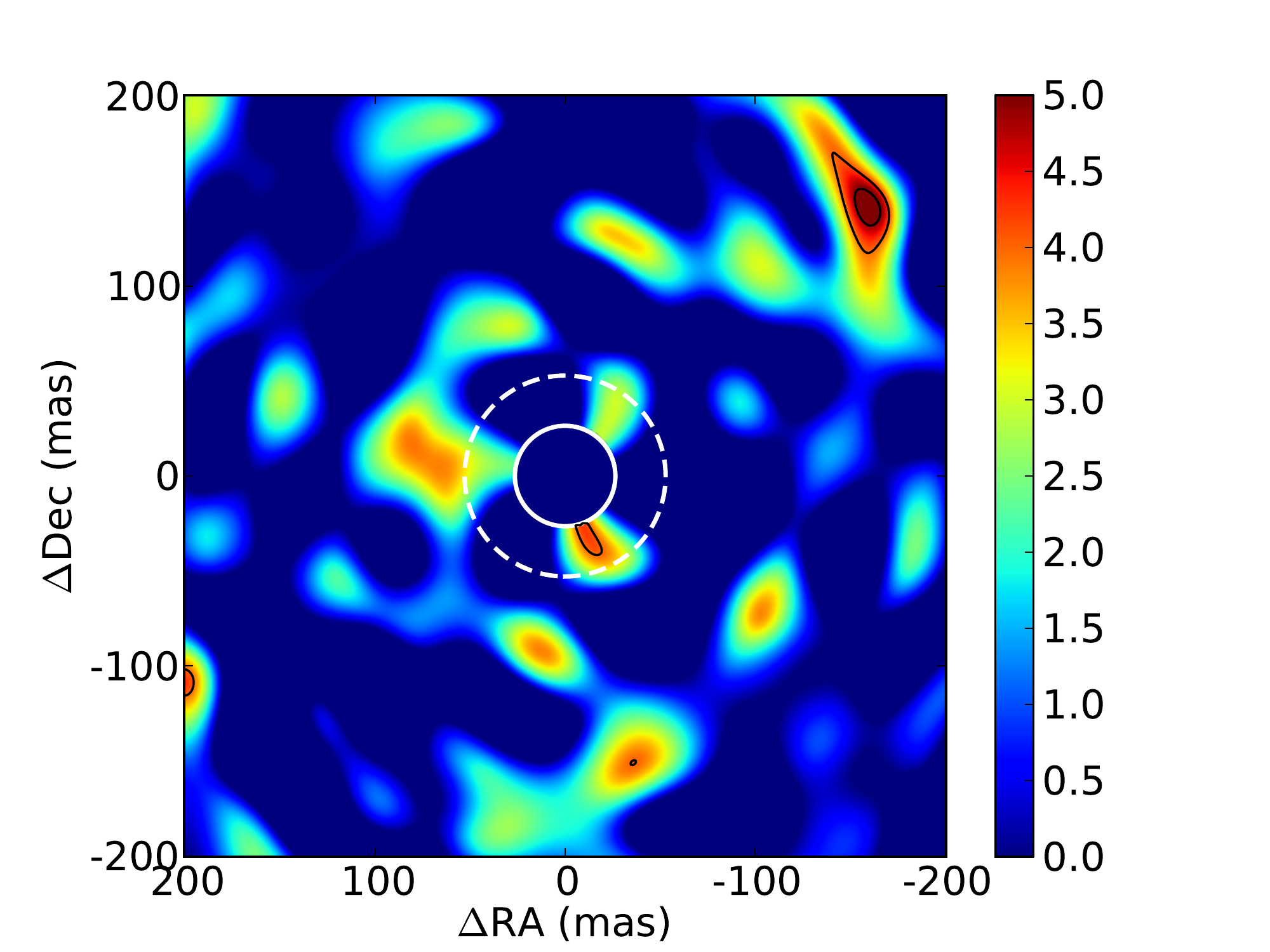} &
     \includegraphics[height=4cm, angle=0]{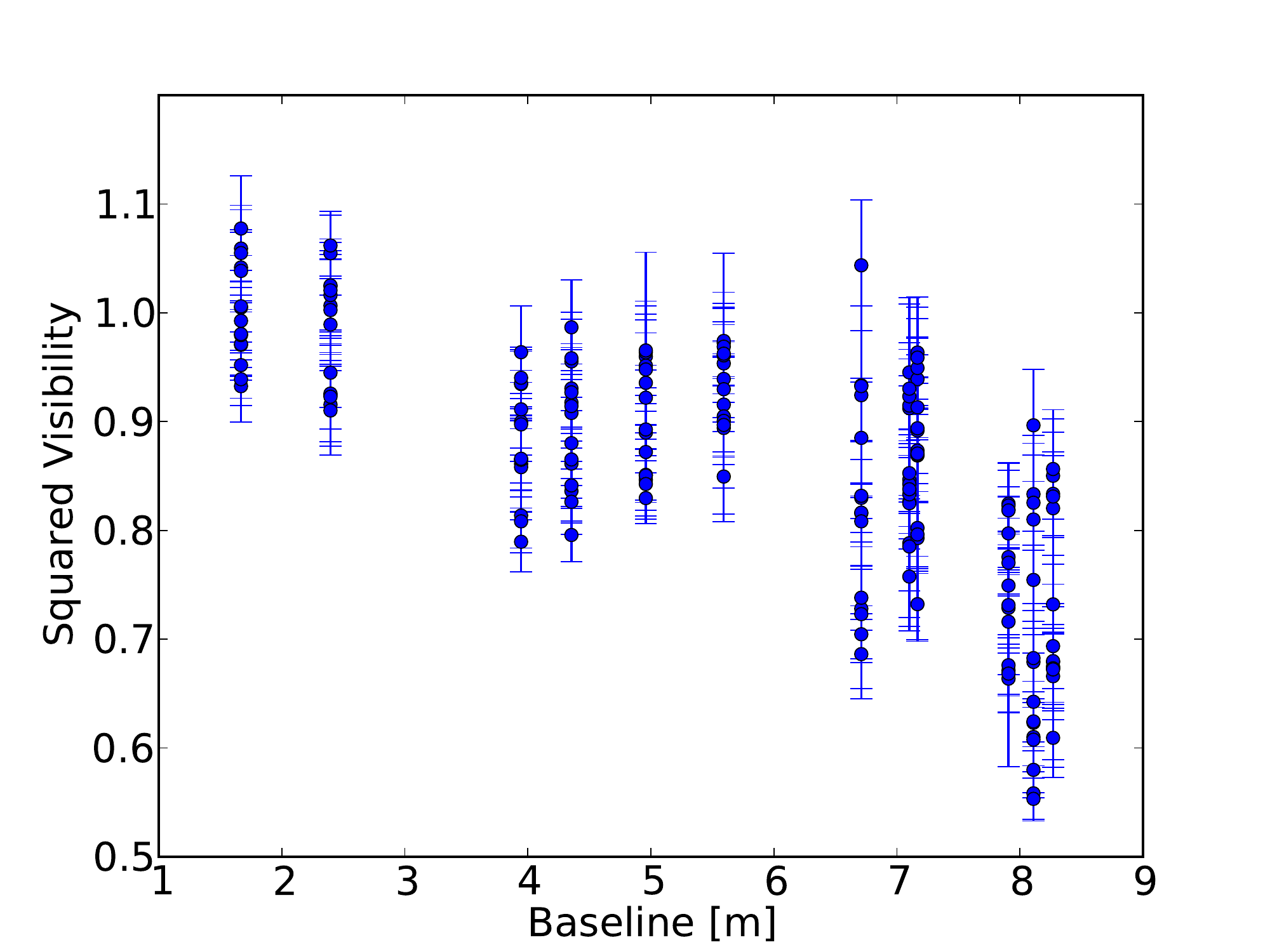}\\ 
 \end{array}$
 
\end{center}
\caption{Data sets where we detect significant asymmetric signal but rule it to be a false positive through similarity in structure to other data sets taken during the same night. \textbf{Left:}  Reconstructed Images \textbf{Middle:} Computed significance maps. \textbf{Right:} V$^2$ plots. \textbf{First row:} RXJ1615.3-3255, K-band, \textbf{Second row:} RXJ1842.9-3532, K-band, \textbf{Third row:}, V2062\,Oph, K'-band. We rule out these asymmetries because of the similarity in the structure of their artefacts. All three targets were observed on the same night and the same structure is also observed in some of the calibrators. The position angle of the structures differ by the same angle as the on sky rotation of the 9-hole mask (See Sections \ref{sec:resultsRXJ1842} and \ref{sec:resultsV2062Oph}). The cause of this effect is unknown.}
\label{fig:FalsePositives}
\end{figure*}

In addition, we place lower limits on the contrast at the 99\% confidence level. For the data sets that were recorded under detrimental conditions, our sensitivity is reduced from typical K-band contrasts of $\Delta m_{\lambda} =$\,5.5 to values of $\Delta m_{\lambda} > $\,5.0\,. 

In the following sections we discuss each object individually. Quoted disc masses represent gas+dust masses and the disc position angles are measured East-of-North along the major axis. 

Besides the significance maps derived from closure phase fitting, we also show the visibility amplitudes derived from our observations.

\subsection{DM\,Tau}

The structure of the inner 10s of au around DM\,Tau is complex and difficult to constrain with SED-based models alone. Studying the \textit{Spitzer} IRS spectrum, \citet{2005ApJ...630L.185C} modelled the SED of DM\,Tau inferring the presence of a 3\,au inner cavity in the disc. In contrast, \citet{2011ApJ...732...42A} used SMA data to spatially resolve an inner disc cavity with a radius of 19$\pm$2\,au in 880\,$\mu$m observations. Neither model however explains simultaneously the IR and sub-mm spectra suggesting the inner disc is potentially populated by a species of small dust grains \citep{2005ApJ...630L.185C}.
\citet{2011ApJ...732...42A} additionally estimated the total disc mass to be 0.04\,M$_{\odot}$ and measured the inclination and position angles of the disc to be 35$^{\circ}$ and 155$^{\circ}$ respectively.

The result of our simple binary model indicates a companion at 43\,mas\,($\approx$6\,au) with an absolute magnitude of $M_K\,=$\,11.0$\pm$0.3\,mag and a significance of 4.27$\sigma$ (93\% confidence level) (see Figure \ref{fig:PotCompanions}). This places the companion candidate within the disc cavity resolved by \citet{2011ApJ...732...42A} and outwards of the ring of small dust grains suggested by \citet{2005ApJ...630L.185C}. We find a value of $M_{c} \dot M_{c}\,=\,10^{-5}$\,M$^2 _J$yr$^{-1}$.
The source of the asymmetric signal is located within the partially resolved region nevertheless the SNR in the closure phases is sufficient to constrain the separation through our binary fitting. The net result is an inflation in the uncertainties within the separation and contrast (see Table \ref{tab:companions}).

We set limits on the contrast of a companion within 200\,mas at $\Delta m_K>$\,4.66\,mag and are most sensitive between 40-160\,mas where we set lower limits $\Delta m_K>$\,5.49\,mag.

We see a systematic reduction in the visibilities at longer baselines but these remain consistent with an unresolved target.

Strong caveats on this detection are placed owing to the small on sky rotation ($\sim$\,$1^{\circ}$) and low strength of the detection. The small on-sky rotation in particular makes this case vulnerable to systematics which may mimic a detection. Multiple visits to the target however should aid in reducing such effects but the confidence levels established by comparison to the sample collected by \citet{2016MNRAS.457.2877G} are likely to be more applicable to this case ($\sim$\,$70\%$) than the confidence levels established through our sample of calibrators so label this detection as only a possible detection.

\subsection{FP\,Tau}

\citet{2005ApJ...628L..65F} classified FP\,Tau as a Class\,II object based on \textit{Spitzer} mid-infrared spectra and inferred the presence of an extended gap within the disc from the lack of near-infrared excess flux. This was further supported by later analysis by \citet{2011ApJ...732...24C} but neither characterised the spatial extent of the gap. They did however measure the disc mass to be 2.5$\times 10^{-4}$\,M$_{\odot}$.

We see clear disc structures within FP\,Tau, with the K-band data set displaying both "dual lobing" in the significance map and dual point source-like emission in the image (Figure \ref{fig:PotDiskFeatures}), which we identified in Section \ref{sec:disk_feature_simulations} as likely indicators for disc-related asymmetries. From this we conclude the inner edge of the outer disc of FP\,Tau to be likely moderately inclined and located between 26-52\,mas. 
The SNR of the closure phase is insufficient to find a solution for the separation and we are forced to fix the separation to $\lambda$/2D within our fit. This corresponds to $\sim$26\,mas and leads to an inner edge located at 10.0$\pm$2.0\,au where the unknown inclination dominates the uncertainty. Values for the position of the inner rim of the disc can be as large as 20\,au however with an equally good fit to the data. 
Between 50-200\,mas we set a lower limit on the contrast of a companion at $\Delta m_{K}>$\,5.0\,mag.

The L-band data set on FP\,Tau does not show any significant asymmetries, as can be seen within the significance map. No signal reaches the 4$\sigma$ significance threshold, which is consistent with the reconstructed images where we see little off-centre flux (see Figure \ref{fig:NonDetections}). Between 50-200\,mas we set an lower limit on the contrast of a companion as $\Delta m_{L} > 3.4$\,mag.

We see a strong drop in the visibilities in the L-band. We fit simple geometric models to the visibilities to estimate the size and orientation of the extended emission. We fit both a ring model and a gaussian profile to the data and find consistent position and inclination angles in both models of $350 \pm 20^{\circ}$ and $25 \pm 5^{\circ}$ respectively.
We find a semi-major axis of 60\,mas in the case of the ring model and a FWHM of 80\,mas in the gaussian profile case. We find the K-band visibilities to be consistent with an unresolved object within the measurement uncertainties, indicating a more compact emitting region.

\subsection{LkH$\alpha$\,330}

LkH$\alpha$\,330 has been extensively studied in unresolved spectroscopy and through interferometry in the millimetre. \citet{2007ApJ...664L.107B} inferred a disc gap between 0.7-50\,au through SED modelling. This was in agreement with later modelling of the SED by \citet{2011ApJ...732...42A}. They resolved the gap cavity in sub-mm SMA observations, inferring in the process that the infrared emission had its origin within the cleared region gap. They attribute this infrared emission to an additional population of small dust grains located in the gap. The disc was found to have an inclination of 35$^{\circ}$ and to be oriented along position angle 80$^{\circ}$. They estimated the disc mass to be 0.025\,M$_{\odot}$.
\citet{2013ApJ...775...30I} carried on further study of the outer disc through the SMA data. They identified a "lopsided" ring in the 1.3\,mm thermal dust emission at a radius of 100\,au. Through hydrodynamic simulations they find this asymmetric ring to be consistent with perturbations in the surface density of the disc caused by an unseen companion. They set limits on the mass and orbital radius of this companion to $>$ 1\,M$_J$ and $<$ 70\,au respectively.

Our observations of LkH$\alpha$\,330 were performed in K- and H-band at two epochs separated by 678\,days. We see no significant signal within the H-band data set but do see a strong asymmetric signal within the K-band closure phases indicative of a companion detection. 

The contrast of the best-fit companion candidate was found to be $\Delta m_{K}= 5.5 \pm 0.2$\,mag with a significance level of $\sigma=4.88$. We estimate $M_{c} \dot M_{c}$ to be 10$^{-3}$\,M$^2 _J$yr$^{-1}$. 

Amongst our companion candidate detections, the K-band observations of LkH$\alpha$\,330 display the most pronounced visibility drop ($\sim$\,$0.8-0.9$). A strong extended component likely exists around this target, making a considerable contribution to the total flux observed in the K-band. The extended component contributes to the strong periodic patterns seen in the significance maps and reconstructed images (see Figure \ref{fig:PotCompanions}) caused by holes within our uv-coverage.
With the existing data set, we cannot rule out that the aforementioned asymmetric signal may be associated with these artefacts.  Additionally our calculated upper limits on the contrast of a companion were found to be $\Delta m_{K} < 5.5$\,mag, which are comparable to the contrast of our most significant detection.

Within the H-band observations we do not see the best fit position found in the K-band data set, however to reproduce the observed K-band M$^2 _J$yr$^{-1}$ values we would expect contrasts of 6.0-6.2\,mag in H-band, well below the 99$\%$ confidence level preventing us from ruling out the K-band detections using the H-band observations. We place upper limits on a companion contrast at $\Delta m_{H} < 4.5$\,mag.

\subsection{RXJ1615.3-3255}
\label{sec:resultsRXJ1615}

Previous, resolved observations of RXJ1615.3-3255 are limited. \citet{2007ApJ...658..480M} linked the object kinematically to the Lupus association at a distance of approximately 185 pc.  \citet{1976ApJS...30..491H} and \citet{1997A&AS..123..329K} classified RXJ1615.3-3255 as a weak-line T\,Tauri star, whereas \citet{2010ApJ...718.1200M} classified it as a potential transitional disc based on \textit{Spitzer} spectra. 

\citet{2011ApJ...732...42A} resolved the disc at 880\,$\mu m$ with SMA observations and found that the emission from the disc is highly extended suggesting a large disc extending out to 115\,au, and they measure a particularly low-density cavity extending to 30\,au. The low density of the cavity forced them to remove all dust from their models from within 0.5\,au of the star. The low far-infrared flux of the source was interpreted by them to be as a result of the effects of dust settling in the outer regions of the disc. This leads to a high estimate for the disc mass of $~$0.13M$_\odot$, that is $\sim 12$\% of the stellar mass. They estimated the disc inclination to be 4$^{\circ}$ with position angle 143$^{\circ}$.

We observed RXJ1615.3-3233 at a single epoch in the K-band and detected a significant asymmetry in the closure phases. However, inspecting the significance maps we see strong similarity between RXJ1615.3-3233, RXJ1842.9-3532 and V2062\,Oph. All three targets were observed on the same night (09/06/2014) with the same filter and appear to suffer from an systematic effect that results in close to identical structure. The rotation of the structure is equal to the on-sky rotation of the mask. We are not able to identify the precise cause of this systematic effect, but note that the night suffered from poor atmospheric conditions and variable wind speeds, which might have induced vibrations and degraded the AO performance (these poor conditions also reflect in a high variance in the individual uncalibrated closure phase; see Table~\ref{tab:cpFWHM}). The visibilities are also strongly affected by this systematic, showing similar strong drops and structure.

We set a lower limit for the contrast of a potential binary to $\Delta m_{K} >$\,4.0\,mag between 20-40\,mas and $\Delta m_{K} >$\,4.6\,mag between 40-200\,mas (see Figure \ref{fig:PotCompanions}). The systematics previously mentioned may affect adversely the accuracy of the limits we set in these cases.

\subsection{RXJ1842.9-3532}
\label{sec:resultsRXJ1842}

\citet{2010AJ....140..887H} used a combination of resolved SMA observations and SED modelling to infer the presence of an optically thin region inwards from 5\,au with a narrow ring of optically thick material at $\sim 0.01-0.2$\,au. Their models suggest little to no evidence for shadowing from the inner on the outer disc. They estimate the disc mass to be 0.01\,M$_\odot$ and measure the inclination to be 54$^{\circ}$ with a position angle of 32$^{\circ}$.

We detect no significant asymmetric signal in the closure phases but see the same systematic structure in the significance maps as in RXJ1615.3-3255 and V2062\,Oph (see Section \ref{sec:resultsRXJ1615}).

We set lower limits on the contrast of a companion at $\Delta m_{K} > $5.0\,mag between 40-200\,mas.

\subsection{TW\,Hya}
\label{sec:resultsTWHya}

Estimates by \citet{2002ApJ...568.1008C} found that the optically-thick disc of TW\,Hya extends from 4\,to\,140\,au, with a mass of 0.06\,$M_{\odot}$ for a 10\,Myr old disc. They additionally found that the inner region of the disc is not fully cleared. A population of 1\,$\mu m$ dust grains is required within the optically thin inner 4\,au to properly fit the SED in agreement with observed continued accretion onto TW\,Hya. This interpretation is supported by recent ALMA observations by \citet{2016arXiv160309352A} which probed, through 870\,$\mu m$ emission, the distribution of millimeter-sized grains to spatial scales on the order of an au. They observed ring structures suggestive of ongoing planet formation, in particular an unresolved inner disc within 0.5\,au and a bright ring at 2.4\,au separated by a dark annulus centred at 1\,au. 

Radial velocity studies of this object performed by \citet{2008Natur.451...38S} provided evidence for the presence of a 9.8$\pm3.3$\,$M_J$ planet on an orbit with a semi-major axis of 0.041$\pm$0.002\,au. This body could be responsible for the clearing of the inner regions of the disc. This interpretation of the RV data was disputed by \citet{2008A&A...489L...9H}, who attributed the signal to the presence of a cool stellar spot.

TW\,Hya was also observed as part of the AO imaging survey with Keck\,II by \citet{2003AJ....126.2009B}. They detected no companion in the H-band down to contrasts of $\sim 1$\,mag at 0$\farcs$05, increasing approximately linearly to 4\,mag at 0$\farcs$2, corresponding to distances of 2.75 to 11\,au.

Using VLT/NACO, \citet{2011A&A...533A.135V} searched for a potential companion in 1.75$\mu$m and 2.12$\mu$m. They employed the LOCI PSF removal algorithm and detected no companion more massive than 0.11\,$M_\odot$ outward of 5.5\,au (0$\farcs$1) or brown dwarf companion outward of 7\,au (0$\farcs$13) or planetary mass outward of 13\,au (0$\farcs$24) at a contrast of 2\,mag. Outward of 87\,au they achieve their maximum contrast sensitivity of 8\,mag allowing them to rule out companions above 7\,M$_J$. \citet{2012ApJ...744..120E} observed TW\,Hya with Keck-II/CONICA in L-band in March, 2009 and observed no significant asymmetric signal within 200\,mas and set lower limits on the contrast of a companion.

We observed TW\,Hya twice in the K-band on non-consecutive nights in the same epoch. In the data from the first night, we see a significant asymmetric signal corresponding to a contrast of $\Delta m_{K} =$\,5.6$\pm$0.3\,mag (see Figure \ref{fig:PotCompanions}). Assuming that the potential companion orbits co-planar to the disc, we find a corresponding semi major axis of $\sim$\,6\,au with $M_{c} \dot M_{c} = 10^{-5}$ M$^2 _J$yr$^{-1}$. We set limits on the contrast of a companion at $\Delta m_{K} >$\,5.4\,mag between 20-200\,mas. We additionally see no significant drop in the visibilities.

On the second night we see no significant asymmetries. The poor atmospheric conditions lead to large uncertainties in the closure phases as a result of frequent loss of AO lock during observation. The result can be seen in the significance map and in particular the reconstructed image where strong bands of artefacts are visible  (See Figure \ref{fig:NonDetections}). The contrast limits from the second night are also adversely affected. Between 80-160\,mas we set contrast limits of $\Delta m_{K} >$\,3.5\,mag, preventing us from ruling out the detection from the first night.

Comparing our result from the first night to the limits in L-band set by \citet{2012ApJ...744..120E}, we use the circumplanetary disc models in \citet{0004-637X-799-1-16} to estimate the L-band absolute magnitude an accreting companion of this absolute magnitude would display. We find the expected contrast to be $\Delta m_L \approx 4$\,mag compared to the limit imposed by \citet{2012ApJ...744..120E} of $\Delta m_L\,>\,6$\,mag. 
Assuming the scaling described within the  circumplanetary disc models to be accurate, to account for both observations the accretion rate onto the potential companion would be required to increase by at least an order of magnitude during the three years separating the observations. The value of $\dot M$ onto TW\,Hya is known to be highly variable with values fluctuating at least by an order of magnitude \citep{2002ApJ...571..378A}. This variation occurs on a time scale of years and we would expect the accretion rate onto a companion to be related to the amount of material flowing through the disc so we cannot rule out this companion candidate based on previous SAM observations. 

Another possible cause is that the origin of this asymmetry is not protoplanetary in nature but instead from a another potential source of asymmetry such as an accretion stream or disc asymmetry. ALMA observations carried out by \citet{2016arXiv160309352A} at $\sim$350GHz found no non-axisymmetric structures on these scales within the distribution of sub-mm particles but this does not rule out a disc asymmetry in our data set as our K-band observations probe the surface layer of the disc while their sub-mm observations probe the middle of the disc \citep{2015MNRAS.451.1147J}. We lack the required signal to noise to be sensitive to any companion within the 1\,au gap seen in their sub-mm data but find no significant asymmetry in the bright ring at 2.4\,au. We additionally note that if confirmed, our companion candidate would lie immediately outside the bright ring seen in the sub-mm data where the intensity distribution flattens out at $\sim$6\,au. ALMA observations taken in 138 and 230GHz \citep{2016arXiv160500289T} revealed a shallow gap of a few percent centred at $\sim$6\,au in agreement with the location of our companion candidate.

\subsection{V2062\,Oph}
\label{sec:resultsV2062Oph}

\citet{2010ApJ...717..441E} modelled the \textit{Spitzer} SED and found a disc cavity extending to 36\,au containing some optically thin dust consistent with other resolved observations of V2062\,Oph.  \citet{2011ApJ...732...42A} finds a cavity in the disc extending out to 30\,au. They additionally constrained the inclination and position angle to 35$^{\circ}$ and 80$^{\circ}$ respectively and they estimate the disc mass to be 0.007\,M$_{\odot}$.

We observed V2062\,Oph in the K-band within a single epoch. Observing conditions were not ideal, but we detect a significant asymmetry signal in the closure phases. As mentioned previously in Section \ref{sec:resultsRXJ1615}, the produced structures are also reproduced within the RXJ1615.3-3255 and RXJ1842.9-3532 data sets leading to the conclusion that these are false positives along with V2062\,Oph. 

We set lower limits on the contrast of a companion at $\Delta m_{K} >$ 4.9\,mag between 20-40\,mas and $>$ 5.1\,mag between 40-200\,mas.

\subsection{V2246\,Oph}

Mid-infrared 9-18\,$\mu$m Gemini observations by \citet{2009ApJ...703..252J} resolved V2246\,Oph at subarcsecond resolution and found very little mid-infrared excess within 100\,au. Beyond this region they observed strongly extended and asymmetric emission out to 100s of au. The asymmetric emission forms a half ring structure to the north west, at an angular separation of $1\farcs 1$. 

\citet{2011A&A...533A.135V} observed V2246\,Oph as part of their VLT/NACO high resolution observations. They reached sensitivities of 15\,M$_J$ and 6\,M$_J$ past separations of 3\,au and 192\,au respectively. They found no evidence for a companion within these limits.

We observed V2246 Oph in the K-band in a single epoch. Poor observing conditions severely limited the sensitivity of our observations. We place limits on the contrast of a potential companion at $\Delta m_{K} >$\,2.3\,mag between 20-40\,mas and $>$ 3.1\,mag between 40-200\,mas.

%

\section{Conclusions}
\label{sec:conclusions}

In this paper we presented results of five nights of Keck sparse aperture masking observations on eight targets in K-band, one in L-band (FP\,Tau), one in H-band (LkH$\alpha$\,330). Within this data set we find significant non-zero closure phases for six targets, indicating asymmetries in the brightness distribution on scales of few au. We however rule three of these to be false positives caused by a systematic effect that affected one of our observing nights. 
The remaining detected asymmetries indicate either the presence of complex disc structures and/or the presence of companions. We conducted detailed simulations in order to understand the signature that these different scenarios produce in our phase measurements and investigated the degeneracies that occur between the derived separation and contrast parameters in the case of marginally resolved companions. 

Using both modelling and image reconstruction methods, we investigated the likely origin of the asymmetries for each target star. We estimate confidence levels for our companion detections through fitting companion models to a sample of 24 calibrators stars known to be point source-like. We use the resultant ditribution to form our confidence levels. We report companion detections at a confidence level of $>99\%$ ($> 4.5\sigma$) in LkH$\alpha\,330$ and detections in two further stars (TW\,Hya and DM\,Tau) at the lower confidence level of $>95\%$ ($> 4.0\sigma$). For the detections, we derive $M_{c} \dot M_{c}$ values of 10$^{-3}$\,M$^2 _J$yr$^{-1}$ (LkH$\alpha 330$), 10$^{-5}$\,M$^2 _J$yr$^{-1}$ (DM\,Tau) and, 10$^{-5}$\,M$^2 _J$yr$^{-1}$ (TW\,Hya). Additionally we infer through comparison to limits previously set on the contrast of a companion in L-band that the origin of the asymmetry signal within the TW\,Hya data set would require an increase in the accretion rate of an order of magnitude within a few years for it to be consistent with an accreting protoplanet, assuming accurate scaling from L- to K-band. Observations by \citet{2002ApJ...571..378A} indicate TW\,Hya to be a highly variable disc with values of $\dot M$ varying by an order of magnitude over time scales of a year, adding support to this scenario.

In LkH$\alpha$\,330 and DM\,Tau the gap properties have been characterised by earlier observations and we find the companion candidates to be located within the disc gaps, suggesting that they are orbiting within the cleared regions of the disc. In the case of TW\,Hya we find the companion candidate to be located on the outer edge of the bright annulus located at 2.4\,au in recent 350GHz ALMA observations by \citet{2016arXiv160309352A}. Furthermore we find that separation of our companion candidate to lie within the shallow gap at 6\,au observed by \citet{2016arXiv160500289T} in 138 and 230GHz ALMA observations.

We interpret the asymmetries in FP\,Tau be associated with disc emission, most likely a disc wall between 20-40\,mas, similar to the asymmetries seen in T Cha \citep{2015MNRAS.450L...1C} and FL Cha \citep{2013ApJ...762L..12C}. This is supported through strong drops in the visibilties in both the K- and L-band observations of this target. Fitting geometric disc models to the data sets we find find visibilities consistent with a compact emitting region in K-Band and an extended component in L-band with a position angle of 350$\pm$20$^\circ$ and an inclination of 25$\pm$5$^\circ$.
Finally, for the remaining data sets we detect no significant asymmetries and set lower limits on the contrast of potential companions.  

With the detection of significant asymmetries in four out of eight target stars, our detection frequency is relatively high (50\%).  This is higher than the detection rate that was found in surveys of other object classes (14\%: \citealt{2008ApJ...679..762K}; 20\%: \citealt{2011ApJ...731....8K}) conducted with Keck/NIRC2 SAM interferometry with a same observational setup and a similar data analysis scheme. This demonstrates that transitional discs indeed trace a particularly interesting phase in disc evolution and highlights the need for further studies on these object classes with the unique observational window that SAM provides, both with the current-generation telescopes and the upcoming generation of Extremely Large Telescopes. Besides further continuum imaging, it is promising to image these objects in accretion tracing spectral lines such as H$\alpha$, in order to confirm that these objects are sites of continued accretion and to ultimately establish their classification as protoplanets.

\begin{acknowledgements}
We acknowledge support from a STFC Rutherford Fellowship and Grant (ST/J004030/1, ST/K003445/1), Marie Sklodowska-Curie CIG grant (Ref.\ 618910), and Philip Leverhulme Prize (PLP-2013-110). We additionally acknowledge support from NASA KPDA grants (JPL-1452321, 1474717, 1485953, 1496788). The authors wish to recognise and acknowledge the very significant cultural role and reverence that the summit of Mauna Kea has always had within the indigenous Hawaiian community. We are most fortunate to have the opportunity to conduct observations from this mountain. Some of the data presented herein were obtained at the W.M.\ Keck Observatory, which is operated as a scientific partnership among the California Institute of Technology, the University of California and the National Aeronautics and Space Administration. The Observatory was made possible by the generous financial support of the W.M.\ Keck Foundation.

\end{acknowledgements}

\bibliography{2015SAM.bib}{}

\begin{thebibliography}{68}
\expandafter\ifx\csname natexlab\endcsname\relax\def\natexlab#1{#1}\fi

\bibitem[{{Alencar} \& {Batalha}(2002)}]{2002ApJ...571..378A}
{Alencar}, S.~H.~P. \& {Batalha}, C. 2002, \apj, 571, 378

\bibitem[{{Alencar} {et~al.}(2010){Alencar}, {Teixeira}, {Guimar{\~a}es},
  {McGinnis}, {Gameiro}, {Bouvier}, {Aigrain}, {Flaccomio}, \&
  {Favata}}]{2010A&A...519A..88A}
{Alencar}, S.~H.~P., {Teixeira}, P.~S., {Guimar{\~a}es}, M.~M., {et~al.} 2010,
  \aap, 519, A88

\bibitem[{{Andrews} {et~al.}(2011){Andrews}, {Wilner}, {Espaillat}, {Hughes},
  {Dullemond}, {McClure}, {Qi}, \& {Brown}}]{2011ApJ...732...42A}
{Andrews}, S.~M., {Wilner}, D.~J., {Espaillat}, C., {et~al.} 2011, \apj, 732,
  42

\bibitem[{{Andrews} {et~al.}(2016){Andrews}, {Wilner}, {Zhu}, {Birnstiel},
  {Carpenter}, {Perez}, {Bai}, {Oberg}, {Hughes}, {Isella}, \&
  {Ricci}}]{2016arXiv160309352A}
{Andrews}, S.~M., {Wilner}, D.~J., {Zhu}, Z., {et~al.} 2016, ArXiv e-prints

\bibitem[{{Armitage}(2011)}]{2011ARA&A..49..195A}
{Armitage}, P.~J. 2011, \araa, 49, 195

\bibitem[{{Ayliffe} \& {Bate}(2012)}]{2012MNRAS.427.2597A}
{Ayliffe}, B.~A. \& {Bate}, M.~R. 2012, \mnras, 427, 2597

\bibitem[{{Biller} {et~al.}(2012){Biller}, {Lacour}, {Juh{\'a}sz}, {Benisty},
  {Chauvin}, {Olofsson}, {Pott}, {M{\"u}ller}, {Sicilia-Aguilar}, {Bonnefoy},
  {Tuthill}, {Thebault}, {Henning}, \& {Crida}}]{2012ApJ...753L..38B}
{Biller}, B., {Lacour}, S., {Juh{\'a}sz}, A., {et~al.} 2012, \apjl, 753, L38

\bibitem[{{Birnstiel} {et~al.}(2011){Birnstiel}, {Ormel}, \&
  {Dullemond}}]{2011A&A...525A..11B}
{Birnstiel}, T., {Ormel}, C.~W., \& {Dullemond}, C.~P. 2011, \aap, 525, A11

\bibitem[{{Bouvier} {et~al.}(2007){Bouvier}, {Alencar}, {Boutelier},
  {Dougados}, {Balog}, {Grankin}, {Hodgkin}, {Ibrahimov}, {Kun}, {Magakian}, \&
  {Pinte}}]{2007A&A...463.1017B}
{Bouvier}, J., {Alencar}, S.~H.~P., {Boutelier}, T., {et~al.} 2007, \aap, 463,
  1017

\bibitem[{{Bouvier} \& {Appenzeller}(1992)}]{1992A&AS...92..481B}
{Bouvier}, J. \& {Appenzeller}, I. 1992, \aaps, 92, 481

\bibitem[{{Brandeker} {et~al.}(2003){Brandeker}, {Jayawardhana}, \&
  {Najita}}]{2003AJ....126.2009B}
{Brandeker}, A., {Jayawardhana}, R., \& {Najita}, J. 2003, \aj, 126, 2009

\bibitem[{{Brauer} {et~al.}(2008){Brauer}, {Dullemond}, \&
  {Henning}}]{2008A&A...480..859B}
{Brauer}, F., {Dullemond}, C.~P., \& {Henning}, T. 2008, \aap, 480, 859

\bibitem[{{Brown} {et~al.}(2007){Brown}, {Blake}, {Dullemond}, {Mer{\'{\i}}n},
  {Augereau}, {Boogert}, {Evans}, {Geers}, {Lahuis}, {Kessler-Silacci},
  {Pontoppidan}, \& {van Dishoeck}}]{2007ApJ...664L.107B}
{Brown}, J.~M., {Blake}, G.~A., {Dullemond}, C.~P., {et~al.} 2007, \apjl, 664,
  L107

\bibitem[{{Brown} {et~al.}(2009){Brown}, {Blake}, {Qi}, {Dullemond}, {Wilner},
  \& {Williams}}]{2009ApJ...704..496B}
{Brown}, J.~M., {Blake}, G.~A., {Qi}, C., {et~al.} 2009, \apj, 704, 496

\bibitem[{{Calvet} {et~al.}(2002){Calvet}, {D'Alessio}, {Hartmann}, {Wilner},
  {Walsh}, \& {Sitko}}]{2002ApJ...568.1008C}
{Calvet}, N., {D'Alessio}, P., {Hartmann}, L., {et~al.} 2002, \apj, 568, 1008

\bibitem[{{Calvet} {et~al.}(2005){Calvet}, {D'Alessio}, {Watson},
  {Franco-Hern{\'a}ndez}, {Furlan}, {Green}, {Sutter}, {Forrest}, {Hartmann},
  {Uchida}, {Keller}, {Sargent}, {Najita}, {Herter}, {Barry}, \&
  {Hall}}]{2005ApJ...630L.185C}
{Calvet}, N., {D'Alessio}, P., {Watson}, D.~M., {et~al.} 2005, \apjl, 630, L185

\bibitem[{{Calvet} {et~al.}(2004){Calvet}, {Muzerolle}, {Brice{\~n}o},
  {Hern{\'a}ndez}, {Hartmann}, {Saucedo}, \& {Gordon}}]{2004AJ....128.1294C}
{Calvet}, N., {Muzerolle}, J., {Brice{\~n}o}, C., {et~al.} 2004, \aj, 128, 1294

\bibitem[{{Cardelli} {et~al.}(1989){Cardelli}, {Clayton}, \&
  {Mathis}}]{1989ApJ...345..245C}
{Cardelli}, J.~A., {Clayton}, G.~C., \& {Mathis}, J.~S. 1989, \apj, 345, 245

\bibitem[{{Cheetham} {et~al.}(2015){Cheetham}, {Hu{\'e}lamo}, {Lacour}, {de
  Gregorio-Monsalvo}, \& {Tuthill}}]{2015MNRAS.450L...1C}
{Cheetham}, A., {Hu{\'e}lamo}, N., {Lacour}, S., {de Gregorio-Monsalvo}, I., \&
  {Tuthill}, P. 2015, \mnras, 450, L1

\bibitem[{{Chen} {et~al.}(1995){Chen}, {Myers}, {Ladd}, \&
  {Wood}}]{1995ApJ...445..377C}
{Chen}, H., {Myers}, P.~C., {Ladd}, E.~F., \& {Wood}, D.~O.~S. 1995, \apj, 445,
  377

\bibitem[{{Cieza} {et~al.}(2013){Cieza}, {Lacour}, {Schreiber}, {Casassus},
  {Jord{\'a}n}, {Mathews}, {C{\'a}novas}, {M{\'e}nard}, {Kraus}, {P{\'e}rez},
  {Tuthill}, \& {Ireland}}]{2013ApJ...762L..12C}
{Cieza}, L.~A., {Lacour}, S., {Schreiber}, M.~R., {et~al.} 2013, \apjl, 762,
  L12

\bibitem[{{Currie} \& {Sicilia-Aguilar}(2011)}]{2011ApJ...732...24C}
{Currie}, T. \& {Sicilia-Aguilar}, A. 2011, \apj, 732, 24

\bibitem[{{de Val-Borro} {et~al.}(2007){de Val-Borro}, {Artymowicz},
  {D'Angelo}, \& {Peplinski}}]{2007A&A...471.1043D}
{de Val-Borro}, M., {Artymowicz}, P., {D'Angelo}, G., \& {Peplinski}, A. 2007,
  \aap, 471, 1043

\bibitem[{{Espaillat} {et~al.}(2010){Espaillat}, {D'Alessio}, {Hern{\'a}ndez},
  {Nagel}, {Luhman}, {Watson}, {Calvet}, {Muzerolle}, \&
  {McClure}}]{2010ApJ...717..441E}
{Espaillat}, C., {D'Alessio}, P., {Hern{\'a}ndez}, J., {et~al.} 2010, \apj,
  717, 441

\bibitem[{{Espaillat} {et~al.}(2011){Espaillat}, {Furlan}, {D'Alessio},
  {Sargent}, {Nagel}, {Calvet}, {Watson}, \& {Muzerolle}}]{2011ApJ...728...49E}
{Espaillat}, C., {Furlan}, E., {D'Alessio}, P., {et~al.} 2011, \apj, 728, 49

\bibitem[{{Espaillat} {et~al.}(2014){Espaillat}, {Muzerolle}, {Najita},
  {Andrews}, {Zhu}, {Calvet}, {Kraus}, {Hashimoto}, {Kraus}, \&
  {D'Alessio}}]{2014prpl.conf..497E}
{Espaillat}, C., {Muzerolle}, J., {Najita}, J., {et~al.} 2014, Protostars and
  Planets VI, 497

\bibitem[{{Evans} {et~al.}(2012){Evans}, {Ireland}, {Kraus}, {Martinache},
  {Stewart}, {Tuthill}, {Lacour}, {Carpenter}, \&
  {Hillenbrand}}]{2012ApJ...744..120E}
{Evans}, T.~M., {Ireland}, M.~J., {Kraus}, A.~L., {et~al.} 2012, \apj, 744, 120

\bibitem[{{Fernandez} {et~al.}(1995){Fernandez}, {Ortiz}, {Eiroa}, \&
  {Miranda}}]{1995A&AS..114..439F}
{Fernandez}, M., {Ortiz}, E., {Eiroa}, C., \& {Miranda}, L.~F. 1995, \aaps,
  114, 439

\bibitem[{{Fouchet} {et~al.}(2010){Fouchet}, {Gonzalez}, \&
  {Maddison}}]{2010A&A...518A..16F}
{Fouchet}, L., {Gonzalez}, J.-F., \& {Maddison}, S.~T. 2010, \aap, 518, A16

\bibitem[{{Furlan} {et~al.}(2005){Furlan}, {Calvet}, {D'Alessio}, {Hartmann},
  {Forrest}, {Watson}, {Uchida}, {Sargent}, {Green}, \&
  {Herter}}]{2005ApJ...628L..65F}
{Furlan}, E., {Calvet}, N., {D'Alessio}, P., {et~al.} 2005, \apjl, 628, L65

\bibitem[{{Furlan} {et~al.}(2006){Furlan}, {Hartmann}, {Calvet}, {D'Alessio},
  {Franco-Hern{\'a}ndez}, {Forrest}, {Watson}, {Uchida}, {Sargent}, {Green},
  {Keller}, \& {Herter}}]{2006ApJS..165..568F}
{Furlan}, E., {Hartmann}, L., {Calvet}, N., {et~al.} 2006, \apjs, 165, 568

\bibitem[{{Gaidos} {et~al.}(2016){Gaidos}, {Mann}, {Kraus}, \&
  {Ireland}}]{2016MNRAS.457.2877G}
{Gaidos}, E., {Mann}, A.~W., {Kraus}, A.~L., \& {Ireland}, M. 2016, \mnras,
  457, 2877

\bibitem[{{Grady} {et~al.}(2013){Grady}, {Muto}, {Hashimoto}, {Fukagawa},
  {Currie}, {Biller}, {Thalmann}, {Sitko}, {Russell}, {Wisniewski}, {Dong},
  {Kwon}, {Sai}, {Hornbeck}, {Schneider}, {Hines}, {Moro Mart{\'{\i}}n},
  {Feldt}, {Henning}, {Pott}, {Bonnefoy}, {Bouwman}, {Lacour}, {Mueller},
  {Juh{\'a}sz}, {Crida}, {Chauvin}, {Andrews}, {Wilner}, {Kraus}, {Dahm},
  {Robitaille}, {Jang-Condell}, {Abe}, {Akiyama}, {Brandner}, {Brandt},
  {Carson}, {Egner}, {Follette}, {Goto}, {Guyon}, {Hayano}, {Hayashi},
  {Hayashi}, {Hodapp}, {Ishii}, {Iye}, {Janson}, {Kandori}, {Knapp}, {Kudo},
  {Kusakabe}, {Kuzuhara}, {Mayama}, {McElwain}, {Matsuo}, {Miyama}, {Morino},
  {Nishimura}, {Pyo}, {Serabyn}, {Suto}, {Suzuki}, {Takami}, {Takato},
  {Terada}, {Tomono}, {Turner}, {Watanabe}, {Yamada}, {Takami}, {Usuda}, \&
  {Tamura}}]{2013ApJ...762...48G}
{Grady}, C.~A., {Muto}, T., {Hashimoto}, J., {et~al.} 2013, \apj, 762, 48

\bibitem[{{Harries}(2014)}]{2014ascl.soft04006H}
{Harries}, T. 2014, {TORUS: Radiation transport and hydrodynamics code},
  Astrophysics Source Code Library

\bibitem[{{Hashimoto} {et~al.}(2011){Hashimoto}, {Tamura}, {Muto}, {Kudo},
  {Fukagawa}, {Fukue}, {Goto}, {Grady}, {Henning}, {Hodapp}, {Honda},
  {Inutsuka}, {Kokubo}, {Knapp}, {McElwain}, {Momose}, {Ohashi}, {Okamoto},
  {Takami}, {Turner}, {Wisniewski}, {Janson}, {Abe}, {Brandner}, {Carson},
  {Egner}, {Feldt}, {Golota}, {Guyon}, {Hayano}, {Hayashi}, {Hayashi}, {Ishii},
  {Kandori}, {Kusakabe}, {Matsuo}, {Mayama}, {Miyama}, {Morino}, {Moro-Martin},
  {Nishimura}, {Pyo}, {Suto}, {Suzuki}, {Takato}, {Terada}, {Thalmann},
  {Tomono}, {Watanabe}, {Yamada}, {Takami}, \& {Usuda}}]{2011ApJ...729L..17H}
{Hashimoto}, J., {Tamura}, M., {Muto}, T., {et~al.} 2011, \apjl, 729, L17

\bibitem[{{Henize}(1976)}]{1976ApJS...30..491H}
{Henize}, K.~G. 1976, \apjs, 30, 491

\bibitem[{{Hu{\'e}lamo} {et~al.}(2008){Hu{\'e}lamo}, {Figueira}, {Bonfils},
  {Santos}, {Pepe}, {Gillon}, {Azevedo}, {Barman}, {Fern{\'a}ndez}, {di Folco},
  {Guenther}, {Lovis}, {Melo}, {Queloz}, \& {Udry}}]{2008A&A...489L...9H}
{Hu{\'e}lamo}, N., {Figueira}, P., {Bonfils}, X., {et~al.} 2008, \aap, 489, L9

\bibitem[{{Hu{\'e}lamo} {et~al.}(2011){Hu{\'e}lamo}, {Lacour}, {Tuthill},
  {Ireland}, {Kraus}, \& {Chauvin}}]{2011A&A...528L...7H}
{Hu{\'e}lamo}, N., {Lacour}, S., {Tuthill}, P., {et~al.} 2011, \aap, 528, L7

\bibitem[{{Hughes} {et~al.}(2010){Hughes}, {Andrews}, {Wilner}, {Meyer},
  {Carpenter}, {Qi}, {Hales}, {Casassus}, {Hogerheijde}, {Mamajek}, {Wolf},
  {Henning}, \& {Silverstone}}]{2010AJ....140..887H}
{Hughes}, A.~M., {Andrews}, S.~M., {Wilner}, D.~J., {et~al.} 2010, \aj, 140,
  887

\bibitem[{{Ireland} \& {Kraus}(2008)}]{2008ApJ...678L..59I}
{Ireland}, M.~J. \& {Kraus}, A.~L. 2008, \apjl, 678, L59

\bibitem[{{Isella} {et~al.}(2013){Isella}, {P{\'e}rez}, {Carpenter}, {Ricci},
  {Andrews}, \& {Rosenfeld}}]{2013ApJ...775...30I}
{Isella}, A., {P{\'e}rez}, L.~M., {Carpenter}, J.~M., {et~al.} 2013, \apj, 775,
  30

\bibitem[{{Jensen} {et~al.}(2009){Jensen}, {Cohen}, \&
  {Gagn{\'e}}}]{2009ApJ...703..252J}
{Jensen}, E.~L.~N., {Cohen}, D.~H., \& {Gagn{\'e}}, M. 2009, \apj, 703, 252

\bibitem[{{Juh{\'a}sz} {et~al.}(2015){Juh{\'a}sz}, {Benisty}, {Pohl},
  {Dullemond}, {Dominik}, \& {Paardekooper}}]{2015MNRAS.451.1147J}
{Juh{\'a}sz}, A., {Benisty}, M., {Pohl}, A., {et~al.} 2015, \mnras, 451, 1147

\bibitem[{{Kenyon} {et~al.}(1998){Kenyon}, {Brown}, {Tout}, \&
  {Berlind}}]{1998AJ....115.2491K}
{Kenyon}, S.~J., {Brown}, D.~I., {Tout}, C.~A., \& {Berlind}, P. 1998, \aj,
  115, 2491

\bibitem[{{Kluska} {et~al.}(2014){Kluska}, {Malbet}, {Berger}, {Baron},
  {Lazareff}, {Le Bouquin}, {Monnier}, {Soulez}, \&
  {Thi{\'e}baut}}]{Kluska2014}
{Kluska}, J., {Malbet}, F., {Berger}, J.-P., {et~al.} 2014, \aap, 564, A80

\bibitem[{{Kraus} \& {Ireland}(2012)}]{2012ApJ...745....5K}
{Kraus}, A.~L. \& {Ireland}, M.~J. 2012, \apj, 745, 5

\bibitem[{{Kraus} {et~al.}(2011){Kraus}, {Ireland}, {Martinache}, \&
  {Hillenbrand}}]{2011ApJ...731....8K}
{Kraus}, A.~L., {Ireland}, M.~J., {Martinache}, F., \& {Hillenbrand}, L.~A.
  2011, \apj, 731, 8

\bibitem[{{Kraus} {et~al.}(2008){Kraus}, {Ireland}, {Martinache}, \&
  {Lloyd}}]{2008ApJ...679..762K}
{Kraus}, A.~L., {Ireland}, M.~J., {Martinache}, F., \& {Lloyd}, J.~P. 2008,
  \apj, 679, 762

\bibitem[{{Kraus} {et~al.}(2013){Kraus}, {Ireland}, {Sitko}, {Monnier},
  {Calvet}, {Espaillat}, {Grady}, {Harries}, {H{\"o}nig}, {Russell},
  {Swearingen}, {Werren}, \& {Wilner}}]{2013ApJ...768...80K}
{Kraus}, S., {Ireland}, M.~J., {Sitko}, M.~L., {et~al.} 2013, \apj, 768, 80

\bibitem[{{Krautter} {et~al.}(1997){Krautter}, {Wichmann}, {Schmitt}, {Alcala},
  {Neuhauser}, \& {Terranegra}}]{1997A&AS..123..329K}
{Krautter}, J., {Wichmann}, R., {Schmitt}, J.~H.~M.~M., {et~al.} 1997, \aaps,
  123, 329

\bibitem[{{Loinard} {et~al.}(2008){Loinard}, {Torres}, {Mioduszewski}, \&
  {Rodr{\'{\i}}guez}}]{2008ApJ...675L..29L}
{Loinard}, L., {Torres}, R.~M., {Mioduszewski}, A.~J., \& {Rodr{\'{\i}}guez},
  L.~F. 2008, \apjl, 675, L29

\bibitem[{{Makarov}(2007)}]{2007ApJ...658..480M}
{Makarov}, V.~V. 2007, \apj, 658, 480

\bibitem[{{Mer{\'{\i}}n} {et~al.}(2010){Mer{\'{\i}}n}, {Brown}, {Oliveira},
  {Herczeg}, {van Dishoeck}, {Bottinelli}, {Evans}, {Cieza}, {Spezzi},
  {Alcal{\'a}}, {Harvey}, {Blake}, {Bayo}, {Geers}, {Lahuis}, {Prusti},
  {Augereau}, {Olofsson}, {Walter}, \& {Chiu}}]{2010ApJ...718.1200M}
{Mer{\'{\i}}n}, B., {Brown}, J.~M., {Oliveira}, I., {et~al.} 2010, \apj, 718,
  1200

\bibitem[{{Muzerolle} {et~al.}(2009){Muzerolle}, {Flaherty}, {Balog}, {Furlan},
  {Smith}, {Allen}, {Calvet}, {D'Alessio}, {Megeath}, {Muench}, {Rieke}, \&
  {Sherry}}]{2009ApJ...704L..15M}
{Muzerolle}, J., {Flaherty}, K., {Balog}, Z., {et~al.} 2009, \apjl, 704, L15

\bibitem[{{Najita} {et~al.}(2007){Najita}, {Strom}, \&
  {Muzerolle}}]{2007MNRAS.378..369N}
{Najita}, J.~R., {Strom}, S.~E., \& {Muzerolle}, J. 2007, \mnras, 378, 369

\bibitem[{{Olofsson} {et~al.}(2013){Olofsson}, {Benisty}, {Le Bouquin},
  {Berger}, {Lacour}, {M{\'e}nard}, {Henning}, {Crida}, {Burtscher}, {Meeus},
  {Ratzka}, {Pinte}, {Augereau}, {Malbet}, {Lazareff}, \&
  {Traub}}]{2013A&A...552A...4O}
{Olofsson}, J., {Benisty}, M., {Le Bouquin}, J.-B., {et~al.} 2013, \aap, 552,
  A4

\bibitem[{{Pollack} {et~al.}(1996){Pollack}, {Hubickyj}, {Bodenheimer},
  {Lissauer}, {Podolak}, \& {Greenzweig}}]{1996Icar..124...62P}
{Pollack}, J.~B., {Hubickyj}, O., {Bodenheimer}, P., {et~al.} 1996, \icarus,
  124, 62

\bibitem[{{Reipurth} {et~al.}(1996){Reipurth}, {Pedrosa}, \&
  {Lago}}]{1996A&AS..120..229R}
{Reipurth}, B., {Pedrosa}, A., \& {Lago}, M.~T.~V.~T. 1996, \aaps, 120, 229

\bibitem[{{Renard} {et~al.}(2011){Renard}, {Thi{\'e}baut}, \&
  {Malbet}}]{Renard2011}
{Renard}, S., {Thi{\'e}baut}, E., \& {Malbet}, F. 2011, \aap, 533, A64

\bibitem[{{Sallum} {et~al.}(2015){Sallum}, {Follette}, {Eisner}, {Close},
  {Hinz}, {Kratter}, {Males}, {Skemer}, {Macintosh}, {Tuthill}, {Bailey},
  {Defr{\`e}re}, {Morzinski}, {Rodigas}, {Spalding}, {Vaz}, \&
  {Weinberger}}]{2015Natur.527..342S}
{Sallum}, S., {Follette}, K.~B., {Eisner}, J.~A., {et~al.} 2015, \nat, 527, 342

\bibitem[{{Setiawan} {et~al.}(2008){Setiawan}, {Henning}, {Launhardt},
  {M{\"u}ller}, {Weise}, \& {K{\"u}rster}}]{2008Natur.451...38S}
{Setiawan}, J., {Henning}, T., {Launhardt}, R., {et~al.} 2008, \nat, 451, 38

\bibitem[{{Silverstone} {et~al.}(2006){Silverstone}, {Meyer}, {Mamajek},
  {Hines}, {Hillenbrand}, {Najita}, {Pascucci}, {Bouwman}, {Kim}, {Carpenter},
  {Stauffer}, {Backman}, {Moro-Martin}, {Henning}, {Wolf}, {Brooke}, \&
  {Padgett}}]{2006ApJ...639.1138S}
{Silverstone}, M.~D., {Meyer}, M.~R., {Mamajek}, E.~E., {et~al.} 2006, \apj,
  639, 1138

\bibitem[{{Thi{\'e}baut}(2008)}]{Thiebaut2008}
{Thi{\'e}baut}, E. 2008, in Society of Photo-Optical Instrumentation Engineers
  (SPIE) Conference Series, Vol. 7013, Society of Photo-Optical Instrumentation
  Engineers (SPIE) Conference Series, 1

\bibitem[{{Tsukagoshi} {et~al.}(2016){Tsukagoshi}, {Nomura}, {Muto}, {Kawabe},
  {Ishimoto}, {Kanagawa}, {Okuzumi}, {Ida}, {Walsh}, \&
  {Millar}}]{2016arXiv160500289T}
{Tsukagoshi}, T., {Nomura}, H., {Muto}, T., {et~al.} 2016, ArXiv e-prints

\bibitem[{{Varni{\`e}re} {et~al.}(2006){Varni{\`e}re}, {Blackman}, {Frank}, \&
  {Quillen}}]{2006ApJ...640.1110V}
{Varni{\`e}re}, P., {Blackman}, E.~G., {Frank}, A., \& {Quillen}, A.~C. 2006,
  \apj, 640, 1110

\bibitem[{{Vicente} {et~al.}(2011){Vicente}, {Mer{\'{\i}}n}, {Hartung}, {Bouy},
  {Hu{\'e}lamo}, {Artigau}, {Augereau}, {van Dishoeck}, {Olofsson}, {Oliveira},
  \& {Prusti}}]{2011A&A...533A.135V}
{Vicente}, S., {Mer{\'{\i}}n}, B., {Hartung}, M., {et~al.} 2011, \aap, 533,
  A135

\bibitem[{{White} {et~al.}(2007){White}, {Gabor}, \&
  {Hillenbrand}}]{2007AJ....133.2524W}
{White}, R.~J., {Gabor}, J.~M., \& {Hillenbrand}, L.~A. 2007, \aj, 133, 2524

\bibitem[{Zhu(2015)}]{0004-637X-799-1-16}
Zhu, Z. 2015, ApJ, 799, 16

\end{thebibliography}
\bibliographystyle{aa.bst}
\end{document}